%% file: TeVLHC_review.tex
\pdfoutput=1
\documentclass[rmp,aps]{revtex4-1}

\usepackage{graphicx}
\usepackage{epsfig}
\usepackage{rotating}
\usepackage{xspace}
\usepackage{lineno}
\usepackage{hyperref}
\usepackage{placeins}
\usepackage[utf8]{inputenc} 

\input{mycommands}

\newcommand\pubnumber{CP3-17-45}
\newcommand\pubblock{\rightline{\begin{tabular}{l} \pubnumber\\
         \today  \end{tabular}}}

\begin{document}

\pubblock

\title{Single top-quark production at the Tevatron and the LHC}

\begin{abstract}
This paper provides a review of the experimental studies of processes with a single top quark at the Tevatron proton-antiproton collider and the LHC proton-proton collider. Single top-quark production in the $t$-channel process has been measured at both colliders. 
The $s$-channel process has been observed at the Tevatron, and its rate has been also measured at the center-of-mass energy of 8 TeV at the LHC in spite of the comparatively harsher background contamination. LHC data also brought the observation of the associated production of a single top quark with a $W$ boson as well as with a $Z$~boson.
 The Cabibbo-Kobayashi-Maskawa matrix element $|V_{tb}|$ is extracted from the single-top-quark production cross sections, and $t$-channel events are used to measure several properties of the top quark and set constraints on models of physics beyond the Standard Model. Rare final states with a single top quark are searched for, as enhancements in their production rates, if observed, would be clear signs of new physics.
\end{abstract}

\author{Andrea~Giammanco}
\email{andrea.giammanco@uclouvain.be}
\affiliation{Centre for Cosmology\justacomma Particle Physics and Phenomenology, Universit\'e catholique de Louvain, Louvain-la-Neuve, B-1348, Belgium}

\author{Reinhard Schwienhorst}
\email{schwier@pa.msu.edu}
\affiliation{Michigan State University, East Lansing, MI 48823, USA}

\maketitle
\tableofcontents

\input{tex/intro}

\input{tex/experiments}

\input{tex/xs}

\input{tex/extraction}

\input{tex/searches}

\input{tex/conclusions}

\begin{acknowledgments}
The authors acknowledge Nikolaos Kidonakis for his kindness in providing the theory curves for $t$-channel, $s$-channel, and $W$-associated production 
and J\'er\'emy Andrea for calculating the theory curve for the $Z$-associated process in Figure~\ref{fig:sqrts}. 
We acknowledge the CMS Collaboration for an earlier version of that figure. 
J\'er\'emy Andrea, together with Mara Senghi Soares, also helped to clarify the differences between the ATLAS and CMS modeling choices for the \tZ signal. 
Figure~\ref{fig:Rt} was produced starting from a macro 
from the CMS Collaboration and the theory predictions for that figure have been calculated by Wajid Ali Khan and Dominic Hirschb\"uhl. We also thank the latter, as well as Alberto Orso Maria Iorio, for clarifying the methodology followed in ATLAS and CMS, respectively, to compute the impact of PDF uncertainties on their \Rt measurements.
Some paragraphs in Sections~\ref{sec:intro}, \ref{sec:tchannel}, \ref{sec:vtb}, \ref{sec:fcnc}, and~\ref{sec:tHq} are adapted from ~\textcite{Giammanco:2015bxk}.
The authors wish to thank C.-P. Yuan, Muhammad Alhroob, Regina Moles Valls, Rebeca Gonzalez Suarez, Thorsten Chwalek and Tom Junk for their comments on a preliminary draft of this paper.
The Feynman diagrams in this paper have been created with {\textsc JaxoDraw}~\cite{Binosi:2003yf}. All the other original figures have been created with {\textsc ROOT}~\cite{Brun:1997pa}. The work of Schwienhorst was supported in part by NSF grants PHY-1410972 and PHY-1707812.
\end{acknowledgments}

\clearpage 
\bibliography{biblio}

\end{document}

%% file: mycommands.tex
\newcommand{\justacomma}{,\xspace} 
\newcommand{\stat}{\ensuremath{{\rm (stat.)}}}
\newcommand{\syst}{\ensuremath{{\rm (syst.)}}}
\newcommand{\pb}{\ensuremath{{\rm pb}^{-1}}\xspace}
\newcommand{\fb}{\ensuremath{{\rm fb}^{-1}}\xspace}

\newcommand{\pp}{\ensuremath{pp}\xspace}
\newcommand{\ppbar}{\ensuremath{p\bar{p}}\xspace}
\newcommand{\ttbar}{\ensuremath{t\bar{t}}\xspace}
\newcommand{\Wt}{\ensuremath{tW}\xspace}
\newcommand{\tW}{\Wt}
\newcommand{\tZ}{\ensuremath{tZq}\xspace}

\newcommand{\ttH}{\ensuremath{t\bar{t}H}\xspace}
\newcommand{\tH}{\ensuremath{tH}\xspace}
\newcommand{\tHq}{\ensuremath{tHq}\xspace}
\newcommand{\tHW}{\ensuremath{tHW}\xspace}
\newcommand{\tWH}{\tHW}
\newcommand{\pt}{\ensuremath{p_T}\xspace}
\newcommand{\mtop}{\ensuremath{m_t}\xspace}
\newcommand{\thetaL}{\ensuremath{\theta_{\ell}}\xspace}
\newcommand{\costheta}{\ensuremath{\cos\theta_{\ell}}\xspace}

\newcommand{\thetaW}{\ensuremath{\theta^*_W}\xspace}

\newcommand{\etalj}{\ensuremath{\eta_{j'}}\xspace}
\newcommand{\fLV}{\ensuremath{f_{L}}\xspace}
\newcommand{\fLR}{\ensuremath{f_{R}}\xspace}
\newcommand{\fLVVtb}{\ensuremath{|f_{L}\cdot V_{tb}|}\xspace}
\newcommand{\vtb}{\ensuremath{|V_{tb}|}\xspace}
\newcommand{\vts}{\ensuremath{|V_{ts}|}\xspace}
\newcommand{\vtd}{\ensuremath{|V_{td}|}\xspace}
\newcommand{\vtq}{\ensuremath{|V_{tq}|}\xspace}
\newcommand{\vtbprime}{\ensuremath{|V_{tb\prime}|}\xspace}
\newcommand{\MET}{\ensuremath{{E\!\!\!/}_{T}}\xspace}
\newcommand{\etmiss}{\MET}

\newcommand{\Rb}{\ensuremath{R_{b}}\xspace}
\newcommand{\RbDefOne}{\ensuremath{\frac{BR({t\to Wb})}{BR({t\to Wq})}}\xspace}
\newcommand{\RbDefTwo}{\ensuremath{\frac{\vtb^2}{\vtd^2+\vts^2+\vtb^2}}\xspace}
\newcommand{\Rt}{\ensuremath{R_{t}}\xspace}
\newcommand{\RtDef}{\ensuremath{\sigma_{t}/\sigma_{\bar t}}\xspace}
\newcommand{\tbW}{\ensuremath{tWb}\xspace}
\newcommand{\tWb}{\tbW}
\newcommand{\utgamma}{$ut\gamma$\xspace}
\newcommand{\ctgamma}{$ct\gamma$\xspace}

\newcommand{\xB}{$x_B$\xspace}

\newcommand{\mT}{\ensuremath{m_T^{W}}\xspace}
\newcommand{\yt}{\ensuremath{y_{t}}\xspace}
\newcommand{\BRHgg}{\ensuremath{{\rm BR}(H\to\gamma\gamma)}\xspace}

\newcommand{\tprime}{\ensuremath{t^{\prime}}\xspace}
\newcommand{\POWHEG}{{\sc POWHEG}\xspace}
\newcommand{\HATHOR}{{\sc HATHOR}\xspace}
\newcommand{\PYTHIA}{{\sc Pythia}\xspace}
\newcommand{\HERWIG}{{\sc Herwig}\xspace}
\newcommand{\MCatNLO}{{\sc MC@NLO}\xspace}
\newcommand{\aMCatNLO}{{\sc aMC@NLO}\xspace}
\newcommand{\CompHEP}{{\sc CompHEP}\xspace}

%% file: tex/intro.tex
\section{Introduction}
\label{sec:intro}

The top quark is the heaviest elementary particle in the Standard Model (SM), having a mass of more than 170~GeV~\cite{PDG2016}. 
According to the description of the origin of fermion masses provided by the SM (also valid in many of its extensions)~\cite{Weinberg:1967tq}, we can relate the top-quark mass to the strength of the interaction between top-quark and Higgs-boson fields (a so called ``Yukawa coupling'', here indicated as $y_t$), obtaining a value of order unity. 
After the discovery of the Higgs boson~\cite{Aad:2012tfa,Chatrchyan:2012ufa} this has been confirmed by direct studies of its couplings~\cite{Khachatryan:2016vau}. 
The top quark therefore plays an outsized role in electroweak symmetry breaking due to its large mass, which also makes it a sensitive probe to physics beyond the SM (BSM). 

The relationship between the mass and the decay width of an elementary fermion allows to determine for the top quark a lifetime of order $10^{-25}$~s, a couple of orders of magnitude shorter than the timescale of the so called hadronization process, that ``dresses'' colored quarks into color-neutral hadrons. 
That a decay mediated by a weak interaction may be faster than a process mediated by the strong interaction is at first sight surprising; intuitively, this is due to the fact that the top-quark mass is larger than the sum of the $W$ and $b$ masses, therefore there is no barrier to overcome and we have a two-body decay $t\to Wb$ with a real $W$ boson, instead of the usual three-body decay mediated by a virtual $W$ boson. 
The top quark is the only quark to decay before it can hadronize~\cite{Bigi:1986jk}, providing the unique opportunity to study a ``naked'' quark. 

At hadron colliders, the predominant production process is top-quark pair production (\ttbar), mediated by the strong force. 
In contrast, this article is devoted to various mechanisms that produce single top quarks or antiquarks, mediated in the SM by electroweak interactions and possibly receiving contributions from BSM physics. 
While the pair-production process was discovered more than twenty years ago~\cite{Abe:1995hr,Abachi:1995iq} and entered the domain of precision physics many years ago, single top-quark production has been observed less than a decade ago at the Tevatron~\cite{Abazov:2009ii,Aaltonen:2009jj}. In comparison to \ttbar production, the single top-quark signal is small and difficult to separate from the backgrounds (including \ttbar itself), hence the measurement precision for its cross sections and other properties has generally been relatively modest until recently.  
Nevertheless, despite being mediated by the weak interaction, single top-quark production has a production cross-section that is within an order of magnitude of \ttbar production. This is due to the more copious bottom quark and gluon content of the proton at the smaller energy required to produce a single top quark ($\approx 200$~GeV) compared to two of them ($\approx 400$~GeV), as pointed out by \textcite{Willenbrock:1986cr} for the first time.

\begin{figure}[!htpb]
  \includegraphics[width=0.19\textwidth]{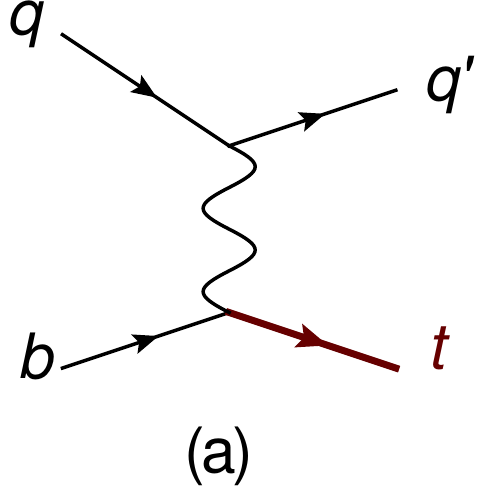}
  \hfill
  \includegraphics[width=0.2\textwidth]{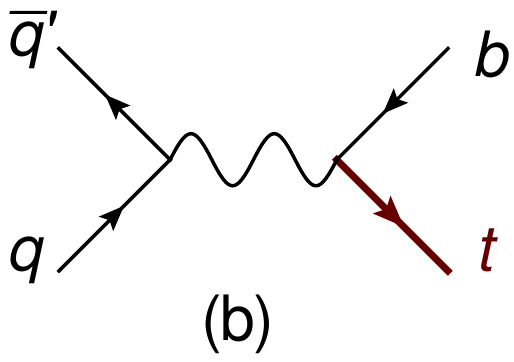}
  \hfill
  \includegraphics[width=0.24\textwidth]{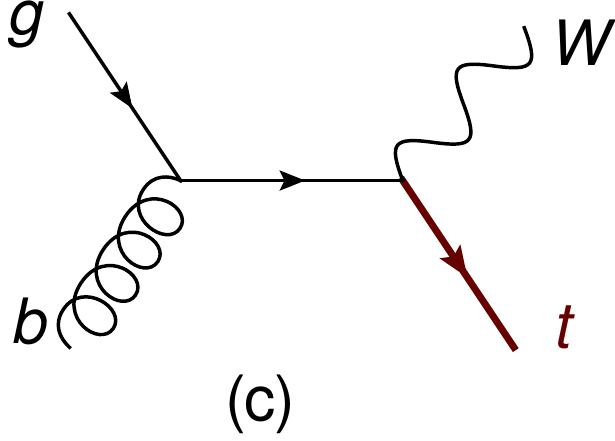}
  \caption{Representative diagrams for electroweak single top-quark production in the (a) $t$-channel, (b) $s$-channel, and (c) $W$-associated production (\Wt).}
  \label{fig:FG}
\end{figure}

In the SM, single top-quark production is a charged-current electroweak process that involves the \tWb vertex in the production of the top quark and in its decay, with only negligible contributions from $tWd$ and $tWs$ couplings, and even smaller contributions from Flavor-Changing Neutral Currents (FCNC). Precise measurements of single top-quark cross sections are motivated by their sensitivity to new physics that modifies either the production or the decay vertex or both~\cite{AguilarSaavedra:2008zc}. The single top-quark production cross section under the SM assumptions is proportional to the square of the Cabibbo-Kobayashi-Maskawa (CKM)~\cite{Cabibbo:1963yz,Kobayashi:1973fv} matrix element $V_{tb}$~\cite{Alwall:2006bx,Lacker:2012ek}.
 The three most abundant and most studied single top-quark processes are illustrated at Born level in Fig.~\ref{fig:FG}. Their production cross sections differ between the Tevatron proton-antiproton collider and the LHC proton-proton collider. The $t$-channel process proceeds through the exchange of a $W$~boson between a light-quark line and a heavy-quark line and has the largest production cross section at both colliders. The $s$-channel process is the production and decay of a heavy off-shell $W$~boson. Since it starts from a quark-antiquark initial state, this process has a comparatively large cross section in $p\bar p$ collisions (roughly half that of the $t$-channel, at the Tevatron) and a comparatively small cross section in $pp$ collisions at the LHC. The $W$-associated production, or \tW, has a top quark and a $W$~boson in the final state. Its initial state consists of a gluon and a $b$~quark, and its production cross section at the Tevatron center-of-mass (CM) energy is so small that this was never observed at that collider, while at LHC energies it is the second-largest production mechanism.

Being produced by parity-violating electroweak processes, the top quarks in single top-quark production are always polarized.
 The degree of polarization is close to 100\% in $t$- and $s$-channel production~\cite{Mahlon:1999gz,Jezabek:1994zv}, in striking difference to \ttbar production, where the SM expects them to be completely unpolarized. Both the timescales for production  ($\approx 1/\mtop$) and decay ($1/\Gamma$, where $\Gamma$ is about 2~GeV) of the top quark are smaller than the hadronization time scale ($\approx 1/\Lambda_{\rm QCD}$, where $\Lambda_{\rm QCD}\approx 0.2$~GeV) which, in turn, is an order of magnitude smaller than the spin decorrelation time ($\approx \mtop/\Lambda_{\rm QCD}^2$). Thus the top-quark polarization is transferred to its decay products and can be accessed through their angular distributionss, as described in Section~\ref{sec:wtb-vertex}. 

Different BSM scenarios predict different effects in the different production channels~\cite{Tait:2000sh}, and this motivates the study of all of them, in conjunction with \ttbar\ properties, to exploit their complementarity.
Some of these new-physics effects in $t$-channel and \Wt production might be mimicked by inaccuracies in the gluon or $b$-quark parton distribution functions (PDF) at large \xB~\footnote{The symbol \xB is used to indicate the quantity ``Bjorken $x$'', i.e. the fraction of the incoming proton's total momentum involved in the parton-level scattering.} and it is therefore necessary to rule out this possibility by additional dedicated inputs. 
 Precise measurements of the cross sections of the three main production modes may have a deep impact  on PDF constraints, with the three channels being complementary to each other and also to \ttbar\ production.
 For example, the $t$-channel and \Wt cross sections are sensitive to the $b$-quark PDF and anti-correlated with the $W/Z$ cross section, while the $s$-channel (essentially a Drell-Yan process, hence correlated with the $W/Z$ cross section) is insensitive to the $b$-quark PDF and can therefore act as a control process~\cite{Guffanti:2010yu}.
 Moreover, the integrated or differential charge asymmetry in $t$-channel production provides a  powerful input to constrain PDFs, similar to the case of $W$-boson production, in a region of \xB very relevant for several searches.
Examples of new physics that might influence $t$-channel production include a vector-like fourth generation quark with chromo-magnetic couplings~\cite{Nutter:2012an}, a color triplet~\cite{Drueke:2014pla}, and FCNC interactions of the top quark with the gluon and the charm quark~\cite{AguilarSaavedra:2008zc}.
The $s$-channel mode is also sensitive to new resonances decaying to a top quark~\cite{Drueke:2014pla}, while the \Wt mode is sensitive to vector-like quarks~\cite{AguilarSaavedra:2009es} and resonances decaying to a top quark and a $W$~boson~\cite{Nutter:2012an}. 

Experimentally, the study of top quarks proceeds by the reconstruction of its decay products. Almost all top quarks decay into a $W$ boson and a $b$ quark~\cite{Abazov:2011zk,Aaltonen:2013luz,Aaltonen:2014yua,Khachatryan:2014nda}. The former promptly decays either into a charged lepton and a neutrino, or into a light quark-antiquark pair. 
The presence of an isolated electron or muon, in particular, is used as a selection requirement in almost all single top-quark production studies, as those two particles are particularly easy to identify with large efficiency and low background contamination even in the busy particle environment created by hadron-hadron collisions. 
The neutrino is undetectable because of its negligible cross section of interaction with the detector material. But the large momentum that it carries, being boosted by the decay of the massive $W$~boson, which is in turn boosted by the decay of the even more massive top quark, is conspicuous by its absence: the large momentum imbalance of the system formed by all visible particles can be used to reconstruct the neutrino momentum. At hadron colliders, this quantity is meaningful only in the plane transverse to the beam directions (the fraction of proton or antiproton momentum carried by the interacting quarks or gluons is only known on a statistical basis via their PDF), and therefore it is customary to define a missing transverse momentum or missing transverse energy (\MET). 
The jets from $b$-quark hadronization can be separated on a statistical basis from those originating from lighter quarks (i.e., those jets can be ``$b$-tagged''). The heavier a quark is, the more asymmetric is the sharing of energy among the hadronization products~\cite{Bjorken:1977md}; in particular, a $b$-flavored hadron carries about 70\% of the original momentum of the corresponding $b$~quark~\cite{Heister:2001jg,DELPHI:2011aa,Abbiendi:2002vt,Abe:2002iq}. The long lifetime of this $b$-flavored hadron ($10^{-12}$~s) corresponds to a flight distance of the order of millimeters, which can be measured in the detectors. 
Charged leptons, \MET and $b$-tagged jets are among the tell-tale signs of the presence of top quarks in a collision event; to further identify the production mechanism, the presence or absence of accompanying objects is crucially exploited, as we will show in the following sections. The single top-quark signal is further separated from the backgrounds through the use of multi-variate analysis (MVA) algorithms that combine kinematic properties of the reconstructed objects into a powerful discriminant.

Ten years ago, \textcite{Gerber:2007xk} extrapolated the Tevatron single top-quark studies to LHC conditions; it was already clear, at the time of that report, that the large increase in cross section would make precision measurements possible. 
We recommend \textcite{Boos:2012hi} as reading material for the relevant theoretical issues, while \textcite{Husemann:2017eka} and~\textcite{Cristinziani:2016vif} provide recent overviews of the full LHC top-quark physics program.
~\textcite{Giammanco:2015bxk} wrote a previous experimental review of single top-quark studies, limited to the LHC experiments and written before the first measurements at 13~TeV were available. 

The theoretical cross section for single top-quark production in the $t$-channel has been computed at next-to-leading order (NLO) in quantum chromo-dynamics (QCD)~\cite{Harris:2002md,Campbell:2004ch,Schwienhorst:2010je,Cao:2005pq,Cao:2004ky}, including next-to-next-to-leading log (NNLL) corrections~\cite{Kidonakis:2011wy} and at next-to-next-to-leading order (NNLO)~\cite{Brucherseifer:2014ama,Berger:2016oht}. The cross section for the $s$-channel process has been computed at NLO~\cite{Harris:2002md,Campbell:2004ch,Cao:2004ap,Heim:2009ku}, and including NNLL corrections~\cite{Kidonakis:2010tc}. The cross section for the \Wt process has been computed at NLO~\cite{Campbell:2004ch}, and including NNLL corrections~\cite{Kidonakis:2010ux}. For each process, both total and differential cross sections are available.

This review is organized as follows: The Tevatron and LHC colliders and experiments are described in Section~\ref{sec:ColExp}, the cross section measurements are summarized and compared in Section~\ref{sec:xs}, the extraction of parameters from the cross-section measurements and searches for new physics are described in Section~\ref{sec:params}. We conclude in Section~\ref{sec:concl}, providing some thoughts on the future of this research direction.

%% file: tex/experiments.tex
\section{Hadron colliders and experiments}
\label{sec:ColExp}

Only two particle colliders have had sufficient CM energy and integrated enough luminosity to produce top quarks --- the Tevatron proton-antiproton collider at Fermilab~\cite{Wilson:1977nk,HolmesSD:1998aa,Lebedev:2014bza} and the LHC proton-proton collider at CERN~\cite{Evans:2008zzb}. The different initial states lead to different production processes: At the Tevatron, hard-scale processes (including all top-quark production mechanisms or processes involving the exchange of massive mediators) are dominated by quark-antiquark initial states, while at the LHC they are dominated by initial states with one or two gluons. In addition, the LHC has accumulated large amounts of proton-proton (\pp) collision data at three different CM energies, 7~TeV, 8~TeV, and 13~TeV, while the Tevatron accumulated a large amount of proton-antiproton data at 1.96~TeV. The Tevatron initially collected data at 1.8~TeV, with sufficient statistics to discover the top quark in pair production~\cite{Abe:1995hr,Abachi:1995iq}, but insufficient to measure single top-quark production~\cite{Acosta:2001un,Abbott:2000pa}. 

The algorithms for the identification and reconstruction of the so-called analysis objects (e.g., electrons, muons, hadronic jets) are similar though not identical at the different experiments, reflecting their complementary strengths. The focus in single top-quark selections is on identifying isolated high-\pt electrons or muons together with large \MET and one or more jets, at least one of which is required to be $b$-tagged to identify the $b$~quark from the top-quark decay. The Tevatron experiments, CDF and D0, use two different jet reconstruction algorithms with different cone sizes. The LHC experiments ATLAS and CMS use the same anti-$k_T$ algorithm~\cite{Salam:2007xv}, though during Run~1 different radius parameters were used. The \pt thresholds for leptons and jets at the Tevatron are typically lower (15~GeV to 20~GeV) than at the LHC (20~GeV to 30~GeV), giving higher acceptances for single top-quark events, compensated partially by the harder spectrum caused by the larger CM energies at the LHC. All $b$-tagging algorithms in these four experiments exploit information related to the lifetime of the $b$-flavored hadrons, in many cases combined with complementary information such as the mass and track multiplicity of the secondary vertices (when present) and/or by the observation of charged leptons inside the jet. The $b$-tagging efficiencies, for similar light-quark rejection factors, are smaller at the Tevatron (50\% to 65\%)~\cite{Acosta:2004hw,Abazov:2013gaa} compared to the LHC (65\% to 85\%)~\cite{Aad:2015ydr,Chatrchyan:2012jua}.

\subsection{Tevatron}
\label{sec:TeV}

The Tevatron was a proton-antiproton collider with two interaction regions that were surrounded by two multi-purpose experiments, CDF and D0, to record the collisions. Run~1 at the Tevatron lasted from 1992 to 1996 and delivered 0.12~fb$^{-1}$ of data at a CM energy of 1.8~TeV. That was sufficient to produce top-quark pairs via the strong interaction, leading to the top-quark discovery~\cite{Abe:1995hr,Abachi:1995iq}. Run~2 at the Tevatron lasted from 2002 to 2011, delivering 10~fb$^{-1}$ of data at a CM energy of 1.96~TeV and kicking off the single top-quark program.

\subsubsection{CDF}

The CDF (Collider Detector at Fermilab) experiment~\cite{Acosta:2004yw} in Run 2 at the Tevatron consisted of a magnetic spectrometer surrounded by calorimeters and muon detectors. The charged-particle tracking system was contained in a 1.4~T solenoid. CDF had a precision tracking system, with silicon microstrip detectors providing charged-particle tracking close to the beam pipe. 
It was surrounded by an open-cell drift chamber 
which covered a radial distance out to 137 cm and provided up to 96 measurements of the track position. The fiducial region of the silicon detector extended in pseudorapidity $|\eta|$ up to $|\eta| = 2$, while the drift chamber provided full radial coverage up to $|\eta| = 1$. Segmented electromagnetic 
and hadronic (iron-scintillator) 
sampling calorimeters surrounded the tracking system and measured the energy of interacting particles, covering the range $|\eta| < 3.6$.
The momentum of muons was measured by drift chambers and scintillation counters out to $|\eta| = 1.5$. 
The CDF trigger system selected events in a three-level architecture. The first (hardware-based) level accepted events at a rate of up to 30~kHz, while the second (firmware and software-based) level reduced the rate to less than 750~Hz, and the third (software-based) level reduced that rate to up to 200~Hz.

In the offline analyses of CDF data, jets were identified using a fixed-cone algorithm with a cone radius of 0.4. Heavy-flavor jets were $b$-tagged based on secondary vertex reconstruction. 
Electrons were reconstructed as charged particles in the tracking system that leave the majority of their energy in the electromagnetic section of the calorimeter. Muons were identified as charged particles in the tracker that leave hits in the muon chambers located outside the calorimeter. The \MET was measured from the imbalance of energy observed in the calorimeter, projected in the transverse plane of the detector, with corrections to take into account the calibration of the energy that could be attributed to analysis objects such as jets, electrons or muons.
CDF collected an integrated luminosity of 9.5~\fb in Run~2.

\subsubsection{D0}

The D0 detector~\cite{Abazov:2005pn} in Run 2 at the Tevatron had a central tracking system consisting of a silicon microstrip tracker and a central fiber tracker, both located within a 2~T superconducting solenoidal magnet. The central tracking system was designed to optimize tracking and vertexing at detector pseudorapidities of $|\eta| < 2.5$. A liquid-argon sampling calorimeter had a central section covering $|\eta| < 1.1$ and two endcap calorimeters that extended coverage to $|\eta| < 4.2$. An outer muon system, with pseudorapidity coverage of $|\eta| < 2$, consisted of a layer of tracking detectors and scintillation trigger counters in a magnetic field of 1.8~T provided by iron toroids. 
Events were selected by a three-level trigger system, with the first two (hardware-based and hardware/software-based) levels accepting an event rate of about 1~kHz, which was reduced to less than 100~Hz with the software-based third level.

In the offline analyses, jets were identified as energy clusters in the electromagnetic and hadronic parts of the calorimeter, reconstructed using an iterative mid-point cone algorithm with radius $R = 0.5$~\cite{Blazey:2000qt}. Heavy-flavor jets were $b$-tagged based on a multivariate analysis (MVA) algorithm that combines the information from the impact parameters
of tracks and from variables that characterize the properties of secondary vertices within jets. Electrons were identified as energy clusters in the calorimeter with a radius of 0.2, matched to a track. Muons were identified as segments in the muon system that are matched to tracks reconstructed in the central tracking system. The \MET was measured with the calorimeter and corrected for the presence of reconstructed objects. D0 collected an integrated luminosity of 9.7~\fb in Run~2.

\subsection{LHC}

The Large Hadron Collider (LHC) operates since 2009 as a proton-proton, proton-lead and lead-lead collider~\footnote{A short ``pilot run'' in October 2017 also provided few hours of xenon-xenon collisions.}, at CM energies ranging from 900~GeV to 13~TeV. 
Collisions happen at four beam-crossing points, and data are recorded by seven experiments: the multi-purpose experiments ATLAS~\cite{Aad:2008zzm} and CMS~\cite{Chatrchyan:2008aa}, the $b$-physics experiment LHCb~\cite{Alves:2008zz}, the heavy-ion experiment ALICE~\cite{Aamodt:2008zz}, the forward-physics experiments TOTEM (at the CMS collision point)~\cite{Berardi:2004bq,Berardi:2004ku} and LHCf (at the ATLAS collision point)~\cite{Adriani:2006jd}, and the MoEDAL experiment (at the LHCb collision point) optimized for the search of magnetic monopoles and other highly-ionizing hypothetical particles~\cite{Pinfold:2009oia}. 
The following run periods are of relevance for the studies reported in this review: 7~TeV runs in 2010 and 2011, with about 5~\fb of good data collected by each of the multi-purpose experiments; 8~TeV run in 2012, where about 20~\fb of data were collected per experiment; and 13~TeV runs since 2015, with around 40~\fb per experiment collected by the end of 2016. The LHC and the experiments continue to operate well at the time of writing, with much larger datasets expected to be collected.
Only the experiments that contribute to single top-quark studies (ATLAS, CMS, and LHCb) are described in this section.

\subsubsection{ATLAS}
\label{sec:atlas}

The ATLAS (A Toroidal LHC ApparatuS) experiment~\cite{Aad:2008zzm} is a multi-purpose particle detector with a forward-backward symmetric cylindrical geometry. ATLAS comprises an inner detector (ID) surrounded by a thin superconducting solenoid providing a 2~T axial magnetic field, a calorimeter system and a
muon spectrometer in a toroidal magnetic field. The ID tracking system covers the pseudorapidity
range $|\eta| < 2.5$ and consists of silicon pixel, silicon microstrip, and transition radiation tracking detectors. 
Lead/liquid-argon sampling EM and forward calorimeters and steel/scintillator-tile central hadronic calorimeters provide energy measurements with pseudorapidity coverage of $|\eta| < 4.9$. The muon spectrometer surrounds the calorimeters and consists of large air-core toroid superconducting magnets with trigger and tracking chambers out to $|\eta| < 2.7$. Events are selected in Run~1 in a three-level trigger system with the first (hardware-based) level accepting an event rate of less than 75~kHz and Level~2 and the event filter (both software-based) reducing the accepted rate to about 400~Hz. In Run~2, there are two trigger levels, accepting event rates of 100~kHz and 1~kHz, respectively.

Jets are reconstructed using the anti-$k_T$ jet clustering algorithm~\cite{Salam:2007xv} with a 
radius parameter of $R=0.4$. Heavy-flavor jets are $b$-tagged based on a combination of multivariate algorithms which  take advantage of the long lifetime of $b$-flavored hadrons and the topological properties of secondary and tertiary decay vertices reconstructed within the jet. Electrons are reconstructed from energy clusters in the calorimeter which are matched to inner detector tracks. Electrons are identified in the pseudorapidity region $|\eta| < 2.47$, excluding the transition region between barrel and endcap calorimeters of $1.37<|\eta|<1.52$. Muons are reconstructed by combining matching tracks reconstructed in both the inner detector and the muon spectrometer up to $|\eta|<2.5$.
An upgrade of the silicon pixel detector, with the addition of a fourth layer of pixel sensors closer to the beam pipe, was performed between Run~1 and Run~2, enhancing the ATLAS performances in tracking and vertexing and consequently improving $b$-tagging performances.

During the runs at 7~TeV, in 2010 and 2011, ATLAS accumulated respectively 35~\pb and about 
5~\fb of data usable for physics analysis. In 2012, 
about 20~\fb were accumulated at 8~TeV, while about 3~\fb and  33~\fb were collected at 13~TeV in 2015 and 2016, respectively.

\subsubsection{CMS}

The CMS (Compact Muon Solenoid) experiment is, similarly to ATLAS, a multi-purpose detector with cylindrical forward-backward symmetry. It features a superconducting solenoid of 6~m internal diameter, providing a magnetic field of 3.8~T. Within the solenoid volume are a silicon pixel and strip tracker, a lead tungstate crystal electromagnetic calorimeter (ECAL), and a brass and scintillator hadron calorimeter (HCAL), each composed of a barrel and two endcap sections. Forward calorimeters extend the pseudorapidity coverage provided by the barrel and endcap detectors. Muons are measured in gas-ionization detectors embedded in the steel flux-return yoke outside the solenoid. 
 A more detailed description of the CMS detector can be found in \textcite{Chatrchyan:2008aa}.
Events of interest are selected using a two-tiered trigger system~\cite{Khachatryan:2016bia}. The first level (L1), composed of custom hardware processors, uses information from the calorimeters and muon detectors to select events at a rate of around 100~kHz. The second level, known as the high-level trigger (HLT), consists of a farm of processors running a version of the full event reconstruction software optimized for fast processing, and reduces the event rate to less than 1~kHz before data storage. 

All single top-quark analyses published by the CMS collaboration have profited from the performances of the so called particle-flow (PF) algorithm~\cite{Sirunyan:2017ulk}. 
 The PF algorithm (also called global event reconstruction) reconstructs and identifies each individual particle with an optimized combination of information from the various elements of the CMS detector. The energy of photons is directly obtained from the ECAL measurement. The energy of electrons is determined from a combination of the electron momentum at the primary interaction vertex as determined by the tracker, the energy of the corresponding ECAL cluster, and the energy sum of all bremsstrahlung photons spatially compatible with originating from the electron track. The energy of muons is obtained from the curvature of the corresponding track. The energy of charged hadrons is determined from a combination of their momentum measured in the tracker and the matching ECAL and HCAL energy deposits, corrected for zero-suppression effects and for the response function of the calorimeters to hadronic showers. Finally, the energy of neutral hadrons is obtained from the corresponding corrected ECAL and HCAL energy. 
Jets and \MET are reconstructed using as input the list of particles provided by the PF algorithm. Jets are reconstructed with the the anti-$k_T$ jet clustering algorithm  with a radius parameter of $R=0.5$ in Run~1 and $R=0.4$ in Run~2. Heavy-flavor jets are $b$-tagged based on a combination of multivariate algorithms which  take advantage of the long lifetime of $b$-hadrons and the topological properties of secondary and tertiary decay vertices reconstructed within the jet.

During the runs at 7~TeV, in 2010 and 2011, CMS accumulated respectively 36~\pb and 5~\fb of certified data, defined as the data collected when all sub-detectors and the magnet are fully operational. In 2012, 20~\fb were accumulated at 8~TeV, while 2.3~\fb and 36~\fb of certified data were recorded at 13~TeV in 2015 and 2016, respectively.

\subsubsection{LHCb}

The LHCb detector~\cite{Alves:2008zz} is a single-arm forward spectrometer with pseudo-rapidity acceptance of $2 < \eta < 5$, designed for the study of particles containing $b$~or $c$~quarks. A warm dipole magnet provides an integrated field of 4~Tm and surrounds the tracking systems, which include a vertex locator and silicon microstrip tracker. Additional tracking stations are located outside the magnet, made of silicon microstrips and Ring
Imaging Cherenkov counters. The calorimeter has a preshower, electromagnetic, and hadronic part. Five muon stations based on multi-wire proportional chambers, one in front of and the rest behind the calorimeters, record the trajectory of muons. Events are recorded by a two-level triggering: a hardware-based Level 0 which accepts events at a rate of about 1~MHz and a software-based HLT that reduces the rate to about 2~kHz. Events passing the muon trigger have been used for top-quark analysis (Section~\ref{sec:top-lhcb}.)

 As the LHCb detector is not hermetic, a complete reconstruction of top-quark decay products is unfeasible as \MET, the usual proxy for the sum of transverse neutrino momenta, is not usable, and the visible decay products of the top quark are unlikely to be all directed to the same hemisphere in \ttbar events. For this reason, top-quark measurements can only be performed in a fiducial region that includes contributions to the $W+b$ and $W+b\bar b$ final states from single and pair production modes~\cite{Aaij:2015mwa,Aaij:2016vsy}. LHCb recorded 1.1~\fb at 7~TeV, 2.1~\fb at 8~TeV and about 2~\fb at 13 TeV in 2015 and 2016.

\FloatBarrier

%% file: tex/xs.tex
\FloatBarrier
\section{Cross section measurements}
\label{sec:xs}

The cross sections of four single top-quark production mechanisms have been measured at the hadron colliders. The cross section of $t$-channel production, Fig.~\ref{fig:FG}(a), is largest at both the Tevatron and LHC colliders, about $1/3$ of the top-quark pair production cross section. The production of $s$-channel single top quarks, Fig.~\ref{fig:FG}(b), is initiated at Born level by $q\bar q^\prime$ annihilation and the cross section is therefore larger in \ppbar than in \pp collisions (at the same CM energy), about half that of $t$-channel production at the Tevatron. The cross section of \Wt production, Fig.~\ref{fig:FG}(c), while being experimentally inaccessible at the Tevatron, is the second largest one at the LHC due to the higher CM energy and larger gluon PDF. The much rarer \tZ process has been observed only recently thanks to the large statistics accumulated by the LHC in Run~2.

\begin{figure}[!htbp]
  \includegraphics[width=0.42\textwidth]{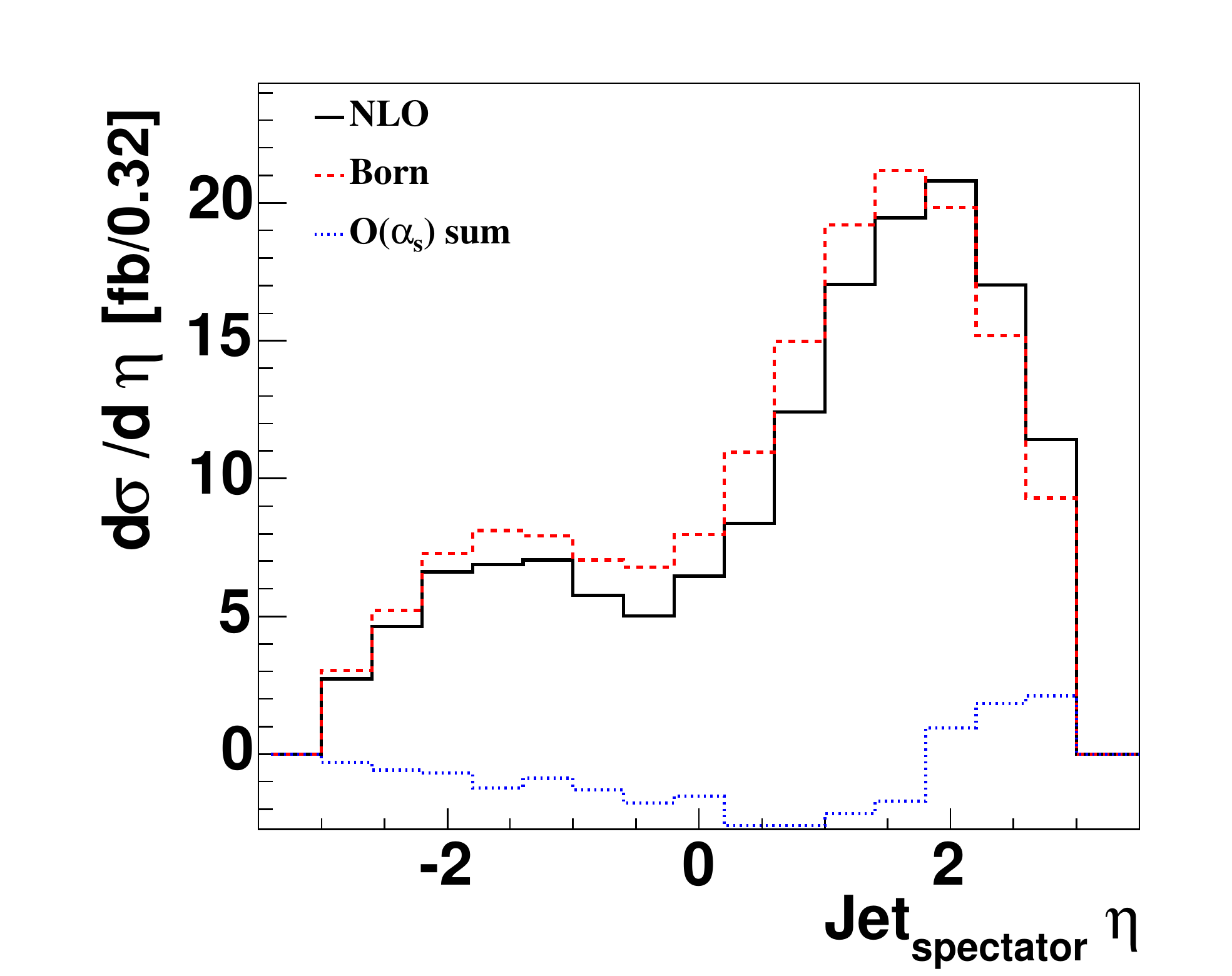}
\hspace{1cm} 
  \includegraphics[width=0.4\textwidth]{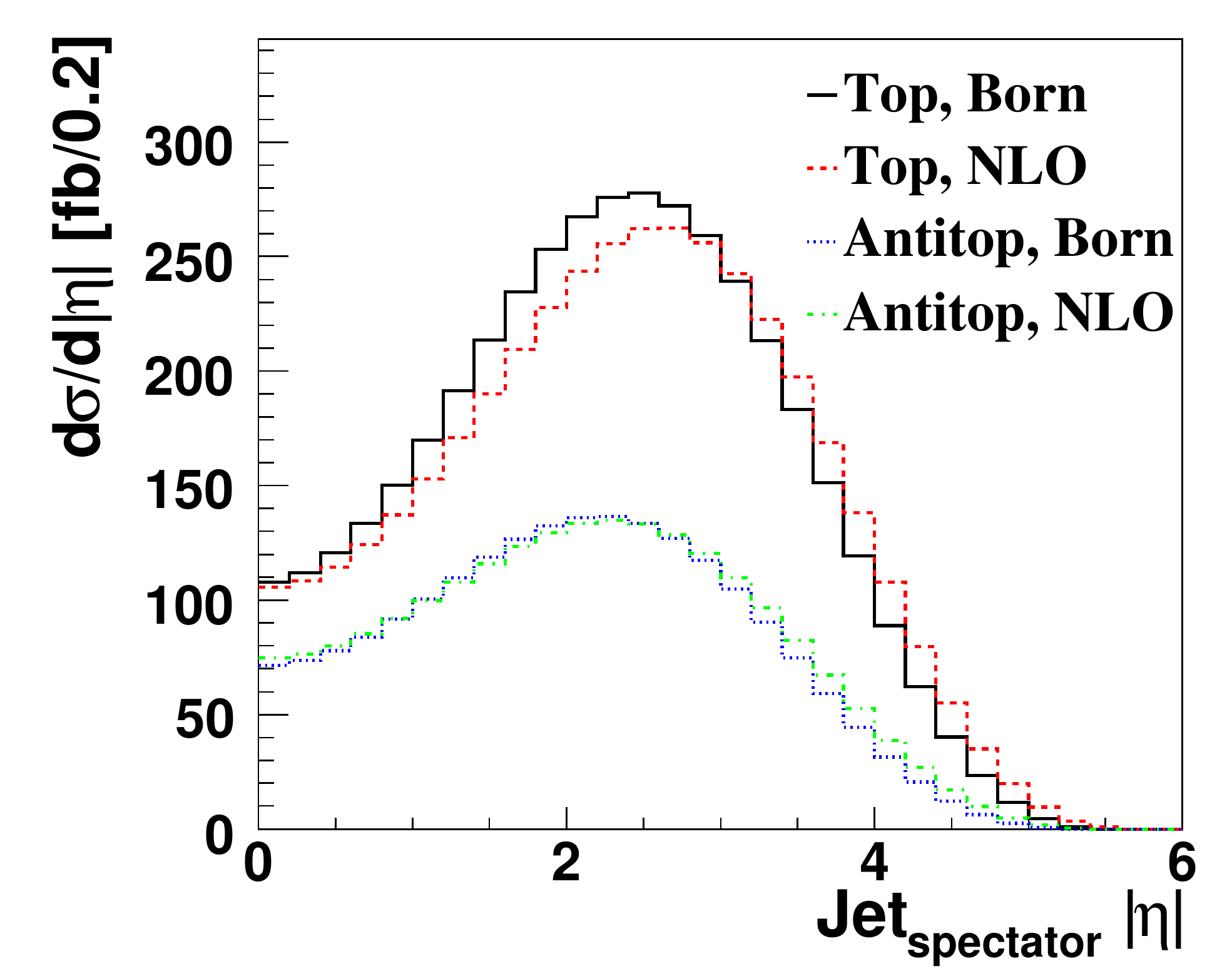}
  \caption{Spectator jet pseudorapidity distribution, corresponding to the light-quark line in Fig.~\ref{fig:FG}(a), comparing Born-level to NLO, (left) for $\eta$ at the Tevatron for top quark (not antiquark) production (from \textcite{Cao:2005pq} and (right) for $|\eta|$ at the LHC for top quark and antiquark $t$-channel production (from \textcite{Schwienhorst:2010je}).}
  \label{fig:lqeta}
\end{figure}

Figure~\ref{fig:lqeta} compares the pseudorapidity distributions of the light quark in the dominant $t$-channel production at Born level (LO) and NLO between the Tevatron and the LHC~\cite{Cao:2005pq,Schwienhorst:2010je}. At the Tevatron, the distribution is asymmetric due to the proton-antiproton initial state. The light quark that recoils against the top quark (antiquark), often called ``spectator'' quark, goes preferentially along the direction of the incoming proton (antiproton). At the LHC, the pseudorapidity distribution is symmetric, thus only $|\eta|$ is shown. For the same reason, the cross sections for the production of top quarks and antiquarks are different. The light quark distribution peaks more forward at the LHC than at the Tevatron due to the larger CM energy, and more forward for top quarks than top antiquarks because the incoming light quark is a valence quark for top-quark production.

The single top-quark analyses in the $t$-channel and $s$-channel at the Tevatron and the LHC select events in the lepton plus jets ($l+$jets) final state~\footnote{Here and anywhere in this article, symbol $l$ is used to refer to a charged lepton (electron or muon), $p_x$ and $p_y$ indicate momentum components along the $x$ and $y$ axis chosen as orthogonal directions to the beam axis, and $p_T \equiv \sqrt{p_x^2+p_y^2}$ (transverse momentum).}, which requires a high-\pt lepton and at least one $b$-tagged jet. The exception is one CDF analysis, which selects events with large $\etmiss$ and $b$-tagged jets. The \Wt measurements select events in the dilepton final state. 
The searches for \tZ production exploit the trilepton final state, where the price paid in terms of leptonic branching fractions of the $Z$ boson and of the top quark gets compensated in terms of purity.

In this article we follow the usual convention in the High-Energy Physics community~\footnote{The authors are aware of the shortcomings of this convention, especially in cases where the signal expectation is precisely determined in the SM; see discussion in~\textcite{Dorigo:2015baa}.}  
of indicating with the words ``evidence'' and ``observation'' a significance of the signal with respect to the background-only hypothesis that surpasses three and five standard deviations, respectively. 

\subsection{Tevatron}
\label{sec:xstev}

At the Tevatron, the $t$-channel process has the largest predicted production cross section of $2.10 \pm 0.13$~pb~\cite{Kidonakis:2011wy} and is easiest to separate from the backgrounds due to the unique signature of a forward light-quark jet, see Figs.~\ref{fig:FG}(a) and~\ref{fig:lqeta}. The $s$-channel process has a smaller predicted production cross section of $1.05\pm 0.06$~pb~\cite{Kidonakis:2010tc}. Both theory predictions have been computed at NLO, including NNLL corrections, and for a top-quark mass of 172.5~GeV. The \tW cross section is $0.10\pm 0.01$~pb~\cite{Kidonakis:2016sjf}, too small to disentangle from other processes with similar final states, and it is therefore neglected in all Tevatron analyses. Due to the challenge of separating the signal from the background and the two signals from each other, the Tevatron experiments report both combined $s+t$-channel measurements, where the ratio between the two processes is assumed to take the SM value, and individual measurements for $t$-channel and $s$-channel. 
The SM ratio assumption is suitable for the early measurements that aim to establish the existence of this signal and provide the first \vtb extraction. It does limit the sensitivity to new physics~\footnote{This approach is only rigorous as a test for models that coherently modify the cross section of both channels, such as an anomalous $tWb$ coupling.}, for which a two-dimensional cross-section fit is more appropriate as presented in Section~\ref{sec:TevComb}.

\subsubsection{Observation of single top-quark production}

The amount of data collected in Run~1 at the Tevatron at a CM energy of 1.8~TeV was not sufficient to accumulate a measurable sample of single top-quark events and only upper limits on the production cross section were set~\cite{Acosta:2001un,Abbott:2000pa,Abazov:2001ns}. 
In Run~2, Tevatron delivered collisions at a CM energy of 1.96~TeV. Tighter constraints were set~\cite{Abazov:2005zz}, then evidence for single top-quark production was reported by D0 in 2006~\cite{Abazov:2006gd,Abazov:2008kt} and by CDF in 2008~\cite{Aaltonen:2008sy}. The production of single top-quark events was first observed in 2009 by CDF~\cite{Aaltonen:2009jj,Aaltonen:2010jr} and D0~\cite{Abazov:2009ii}. The two measurements were also combined~\cite{Group:2009qk}.

Two approaches are critical in the Tevatron single top-quark discovery. First, no attempt is made to separate the $t$-channel and $s$-channel production modes, though the analyses are mostly sensitive to $t$-channel production due to its larger expected cross section and distinct kinematic properties, in particular the forward light-quark jet, the pseudorapidity of which is shown in Fig.~\ref{fig:lqeta}. The number of expected signal events with two jets and one $b$-tag in 3.2/2.3~fb$^{-1}$ for CDF/D0 was 85/77 for the $t$-channel and 62/45 for the $s$-channel. 

\begin{figure}[!htbp]
  \includegraphics[width=0.4\textwidth]{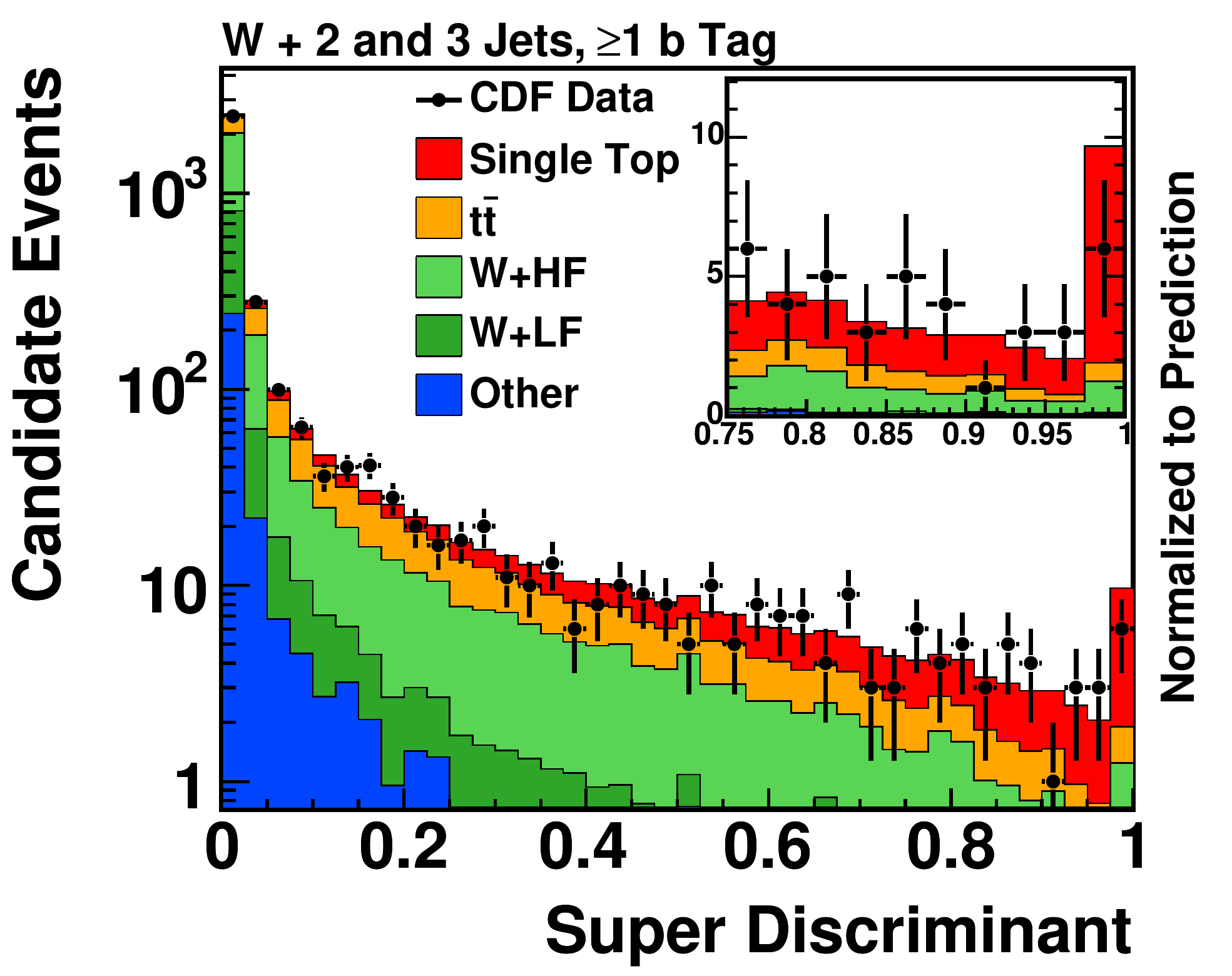}
\hspace{1cm} 
  \includegraphics[width=0.4\textwidth]{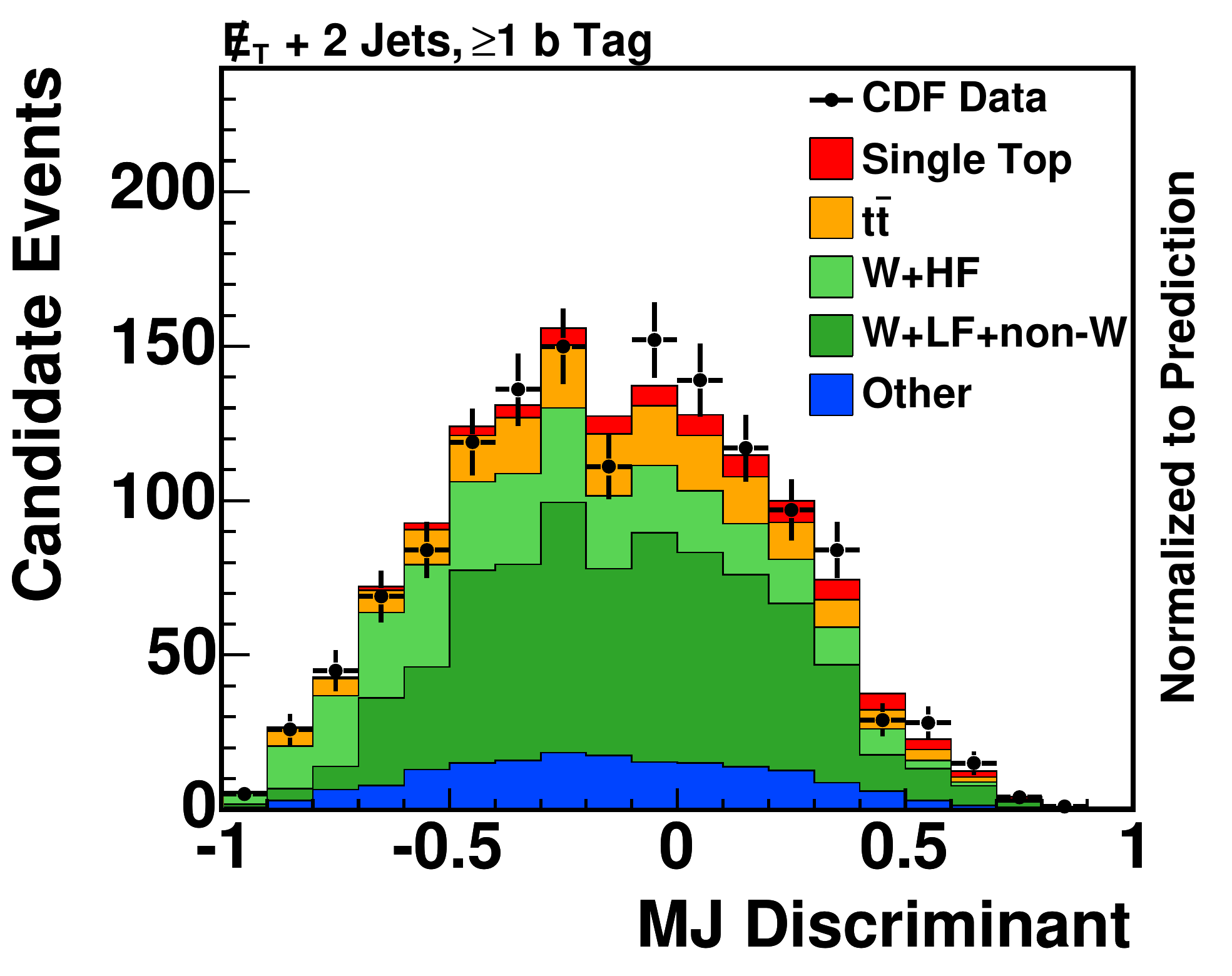}
  \caption{(Left) Combination discriminant distribution and (right) $\MET$+jets analysis discriminant distribution for the CDF single top-quark observation analysis (from \textcite{Aaltonen:2010jr}).}
  \label{fig:CDF3.2}
\end{figure}

\begin{figure}[!htbp]
  \includegraphics[width=0.35\textwidth]{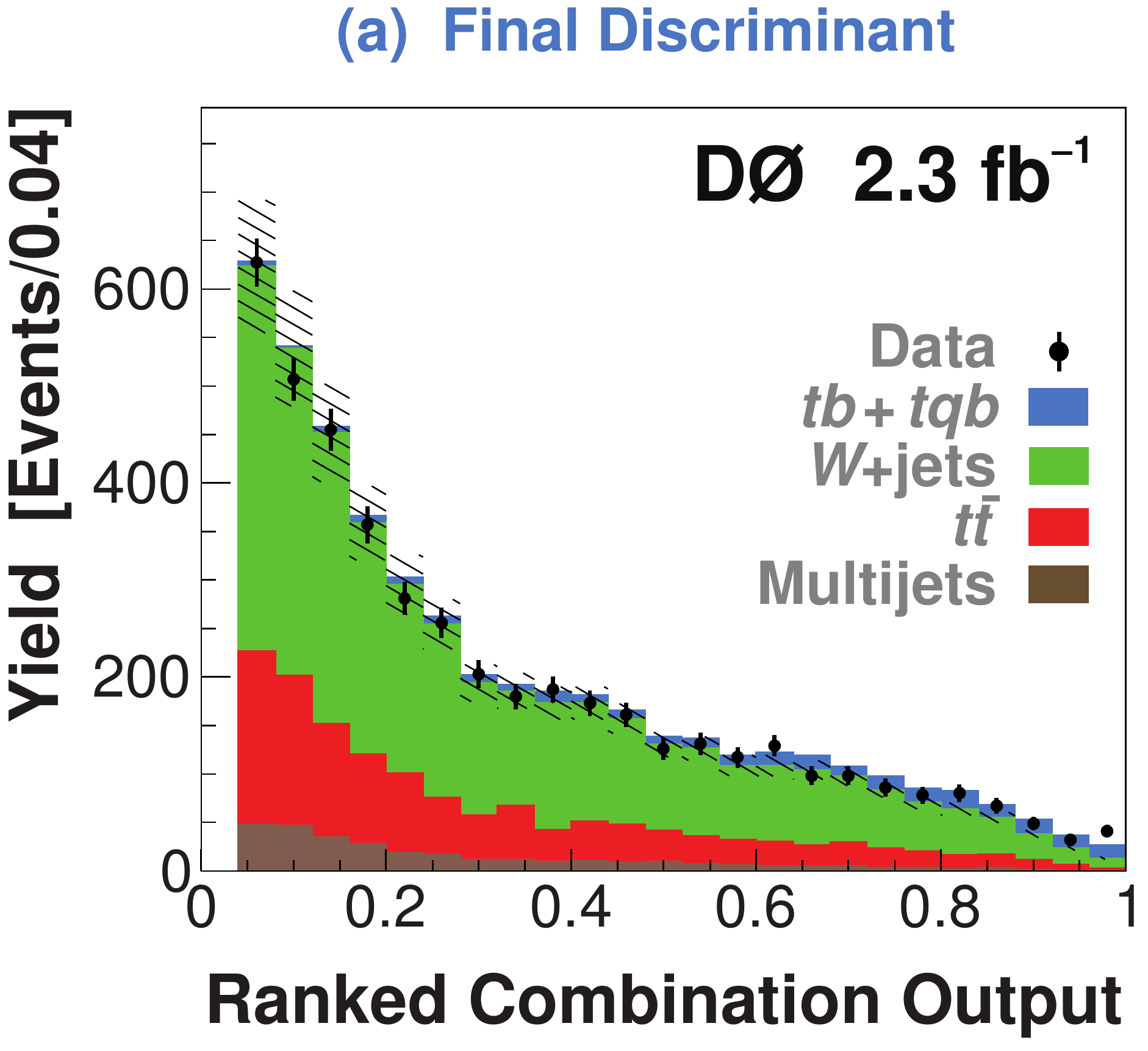}
\hspace{1cm} 
  \includegraphics[width=0.35\textwidth]{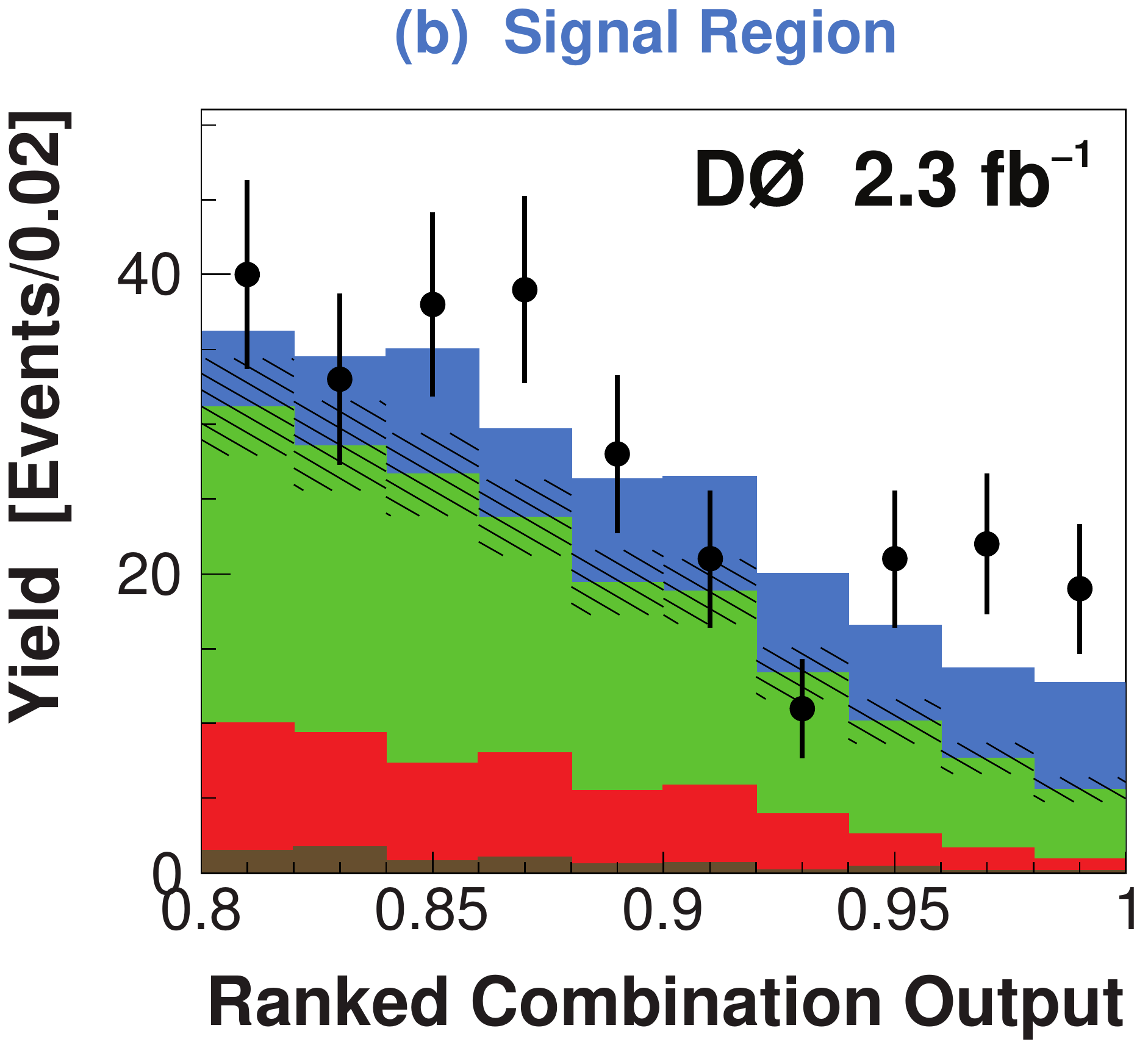}
  \caption{Combination discriminant distribution for the D0 single top-quark observation analysis for (a) the full range and (b) zoomed in on the signal region (from \textcite{Abazov:2009ii}).}
  \label{fig:D02.3}
\end{figure}

Second, the Tevatron single top-quark searches and measurements rely on MVA techniques to separate the small signal from the large backgrounds with large systematic uncertainties. And not just MVAs, but the discovery sensitivity is only reached when multiple MVAs are combined in another MVA. Figure~\ref{fig:CDF3.2} shows the discriminant distributions in the two CDF analyses that enter the observation: The super discriminant, from a combination of multiple $l$+jets analyses, and the MVA discriminant from the $\MET$+jets (MJ) analysis which vetoes isolated leptons~\cite{Aaltonen:2010jr}. The super discriminant only has a single bin with more than 5 signal events expected, and the MJ discriminant also has very few signal events in the signal-enriched region.
Figure~\ref{fig:D02.3} shows the combination discriminant for the D0 analysis. Even in the signal-enriched region close to an MVA output of 1, there are only about 8 expected signal events for an expected background of about 10 events.
The combined cross section for $t$-channel and $s$-channel production is obtained in a Bayesian likelihood analysis, assuming the SM ratio of the two processes. The same approach is also used to combine the two measurements, and the combined $t$-channel plus $s$-channel ($t+s$) cross section is $2.76^{+0.58}_{-0.47}$ pb~\cite{Group:2009qk}.

\begin{figure}[!htpb]
  \includegraphics[width=0.2\textwidth]{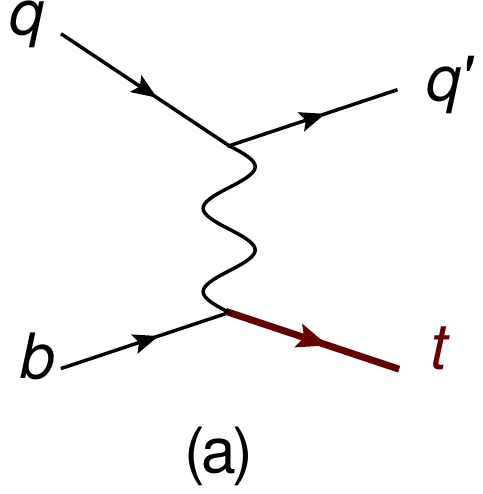}
\hspace{1cm}
  \includegraphics[width=0.25\textwidth]{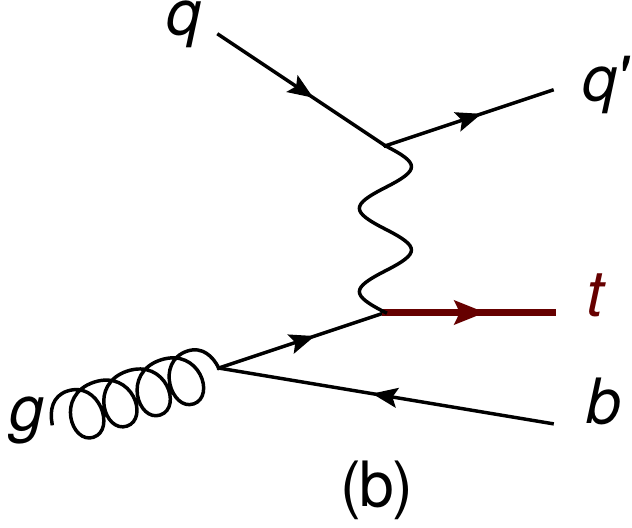}
  \caption{Representative diagrams for electroweak single top-quark $t$-channel production in (a) the $2\rightarrow 2$ mode, corresponding to the 5-flavor-number scheme and (b) the $2\rightarrow 3$ mode, corresponding to the 4-flavor-number scheme. }
  \label{fig:FG23}
\end{figure}

CDF required a data sample about 50\% larger than D0 to observe single top-quark production due to a downward fluctuation in the data, as can be seen in Fig.~\ref{fig:CDF3.2}(left), while D0 had an upward fluctuation in data in the signal region, see Fig.~\ref{fig:D02.3}. An additional reason was the limited accuracy of single top-quark theory modeling. Only leading order (LO) generators existed at the time, while the production cross section receives contributions from both the $2\rightarrow 2$ process shown in Fig.~\ref{fig:FG23}(a) and the $2\rightarrow 3$ process shown in Fig.~\ref{fig:FG23}(b). The $2\rightarrow 2$ process corresponds to the 5-flavor-number scheme (5FNS) where the parton distribution functions include $b$~quarks. The $2\rightarrow 3$ process is a part of the real corrections in QCD to the $2\rightarrow 2$ process in this scheme. However, this diagram actually contributes a large fraction of the selected single top-quark events~\cite{Cao:2005pq}. Alternatively, when generating events in the 4-flavor-number scheme (4FNS) where the parton distribution functions do not include $b$~quarks, the $2\rightarrow 3$ process in Fig.~\ref{fig:FG23} is the LO process~\cite{Frederix:2012dh}. Consequently, LO generators need to employ a matching scheme that includes both diagrams. D0 employs the SingleTop generator~\cite{Boos:2006af}, based on \CompHEP~\cite{Boos:2004kh}, which matches the kinematics of the scattered $b$~quark to NLO prediction. This approach gives reasonable agreement with NLO distribution~\cite{Binoth:2010nha,Campbell:2009ss}. This is not the case for the CDF signal model, which was tuned by comparing the LO parton-level distribution to NLO~\cite{Aaltonen:2010jr}. For the analysis with the full Tevatron Run~2 dataset, the CDF signal model was updated to NLO using \POWHEG generator~\cite{Alioli:2009je, Re:2010bp}.

\subsubsection{Tevatron legacy measurements and s-channel observation}
\label{sec:tev-legacy}

\begin{figure}[!htbp]
  \includegraphics[width=0.4\textwidth]{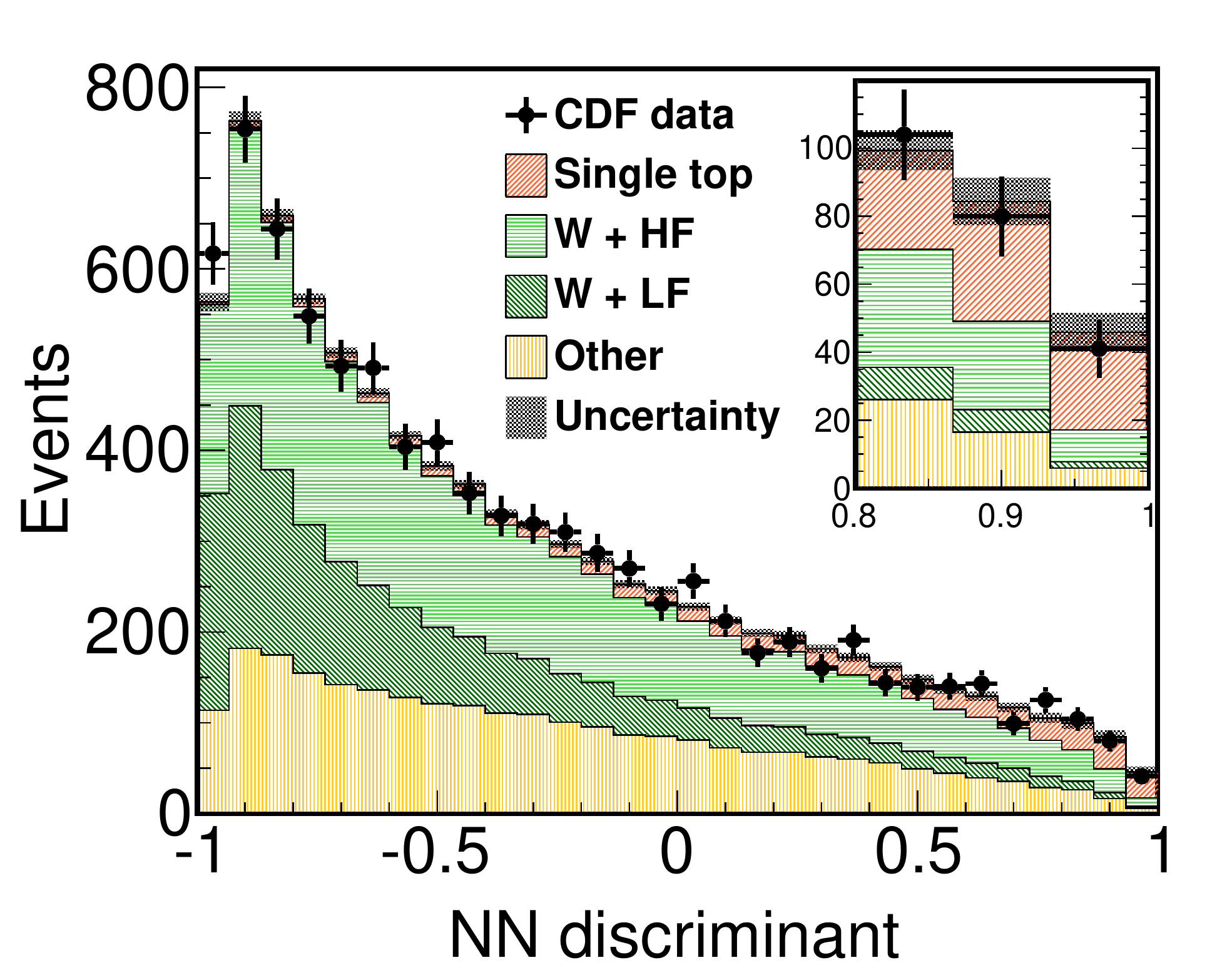}
\hspace{1cm} 
\includegraphics[width=0.38\textwidth]{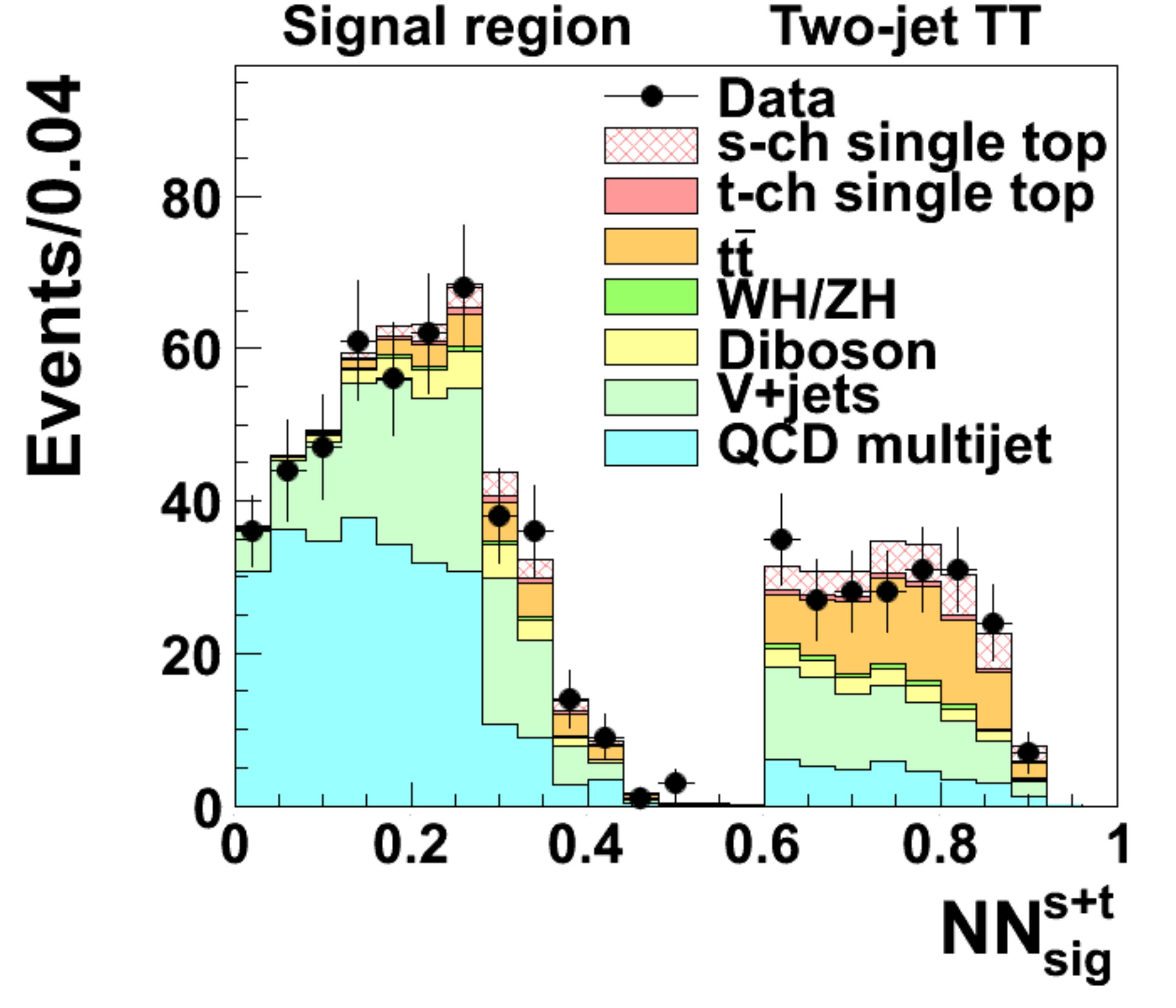}
  \caption{Multivariate discriminant for (left) the CDF $l$+jets analysis for events with 1 $b$-tag  (from \textcite{Aaltonen:2014ura}) and (right) the CDF \etmiss+jets analysis for events with two tight $b$-tags (from \textcite{Aaltonen:2014mza}).}
  \label{fig:NNCDFRun2}
\end{figure}

The CDF and D0 analyses with the full Tevatron dataset of about 10~\fb utilize the same analysis techniques as the observation analyses described above. 
CDF combines two measurements, one in the $l$+jets channel, and one in the MJ channel. The first measurement selects events with a lepton (electron or muon), jets and large \etmiss in 7.5~\fb of data~\cite{Aaltonen:2014ura}. 
The data events are separated into four categories by jet multiplicity (2-jet and 3-jet) and $b$-tag multiplicity (1-tag and 2-tag). The single top-quark signal is separated from the backgrounds using a Neural Network (NN) discriminant, trained separately in each analysis region, using only $s$-channel events as the signal in the training for 2-jet, 2-tag events, and only $t$-channel events as the signal in the training for all other events. This dedicated training enhances the separate sensitivity to $s$-channel and $t$-channel. 
In addition, simulated samples with variations related to the main systematic uncertainties (jet energy scale, factorization and renormalization scales) are included in the training in order to reduce the sensitivity to these sources of uncertainty. The NN discriminant for 1-tag events is shown in Fig.~\ref{fig:NNCDFRun2}. 

The second measurement selects events containing large \etmiss, $b$-tagged jets, but no identified leptons~\cite{Aaltonen:2014mza} in 9.5~\fb of data. 
Events are separated into six regions by jet multiplicity (2 or 3) and $b$-tag categories (exactly one tight, one tight and one loose, and two tight tags). In total, 22,700 events are selected in data, of which 530 are expected to be from single top-quark production. This amount of signal is similar to the $l$+jets analysis, but the background here is much larger. The signal is separated from the large background from QCD multijet events with a NN. The $t$-channel ($s$-channel) signal is isolated from the background in 1~$b$-tag (2~$b$-tag) events with a separate NN. The resulting NN output for events with two $b$-tagged jets is shown in Fig.~\ref{fig:NNCDFRun2}. The \etmiss+jets analysis has less sensitivity than the $l$+jets one, but still contributes in the combination and enhances the single-top sensitivity.

\begin{figure}[!htbp]
  \includegraphics[width=0.4\textwidth]{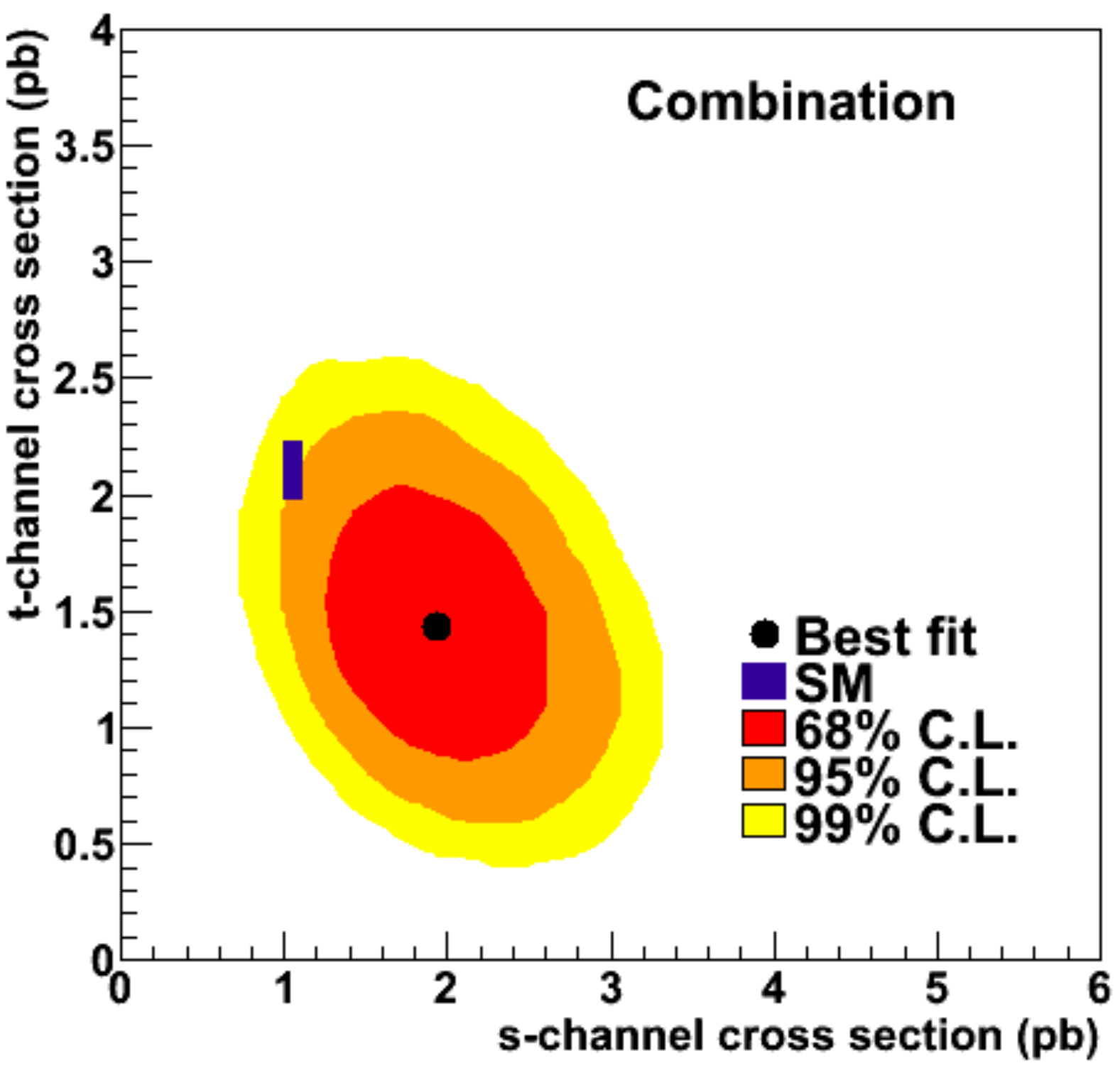}
\hspace{1cm} 
  \includegraphics[width=0.4\textwidth]{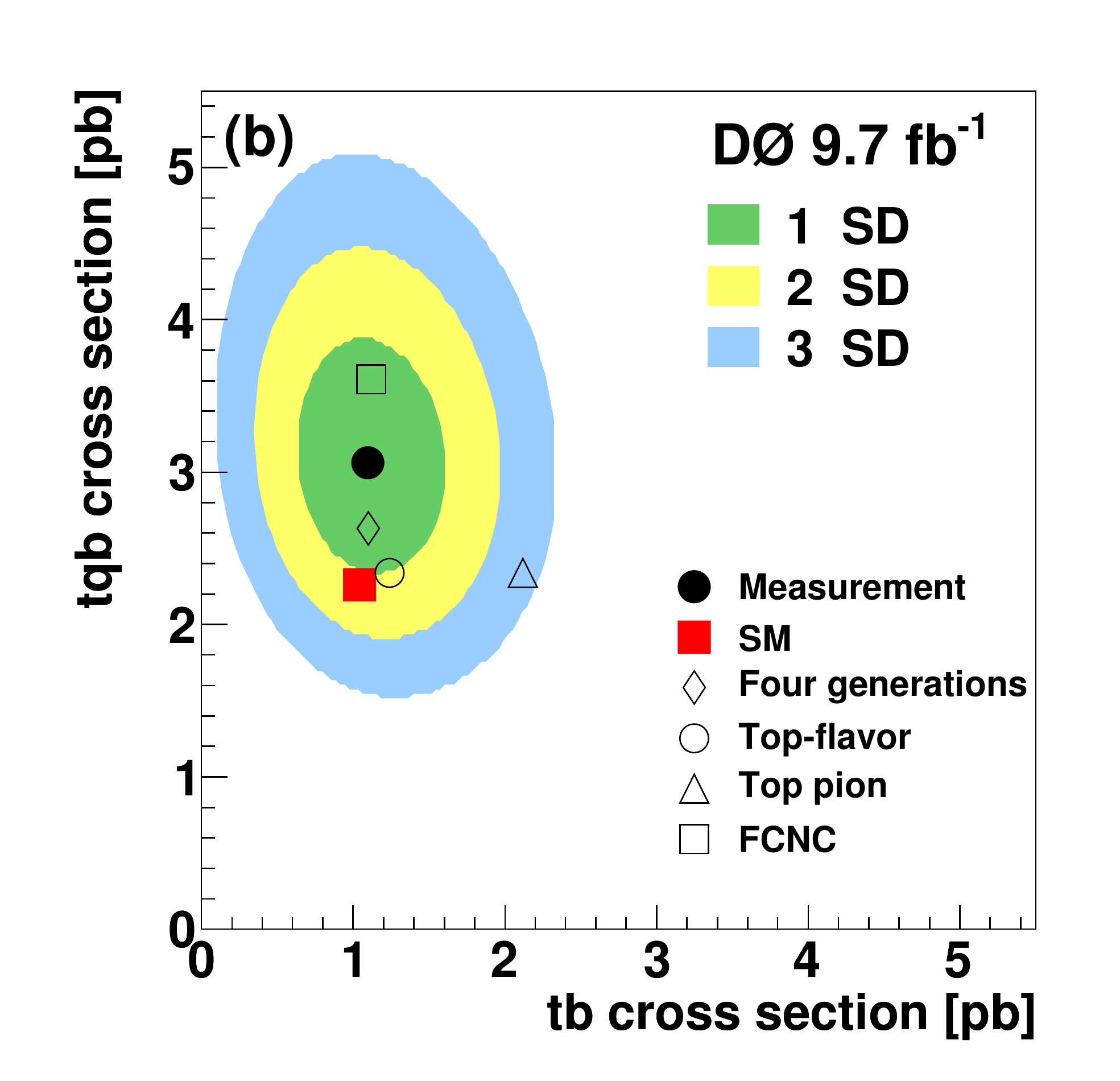}
  \caption{Two-dimensional posterior probability density as a function of the $t$-channel and $s$-channel single top-quark production cross sections for (left) the combined CDF analysis  (from \textcite{Aaltonen:2014mza}) and (right) the D0 analysis (from \textcite{Abazov:2013qka}). Overlaid on the D0 plot are several representative new physics models: FCNC top-gluon interactions~\cite{Tait:2000sh,Abazov:2007ev}, a fourth generation model~\cite{Alwall:2006bx}, a top-flavor model~\cite{Tait:2000sh}, and a top pion~\cite{Hill:1994hp,Tait:2000sh}.}
  \label{fig:PostCDFD0}
\end{figure}

The $l$+jets and MJ discriminants are combined in a likelihood fit that includes all bins of the MVA distributions in all channels of both measurements, with a coherent treatment of the systematic uncertainties and their correlations~\cite{Aaltonen:2014mza}. The resulting two-dimensional posterior probability density as a function of the $t$-channel and $s$-channel cross sections for CDF is shown in Fig.~\ref{fig:PostCDFD0}(left).

\begin{figure}[!htbp]
  \includegraphics[width=0.4\textwidth]{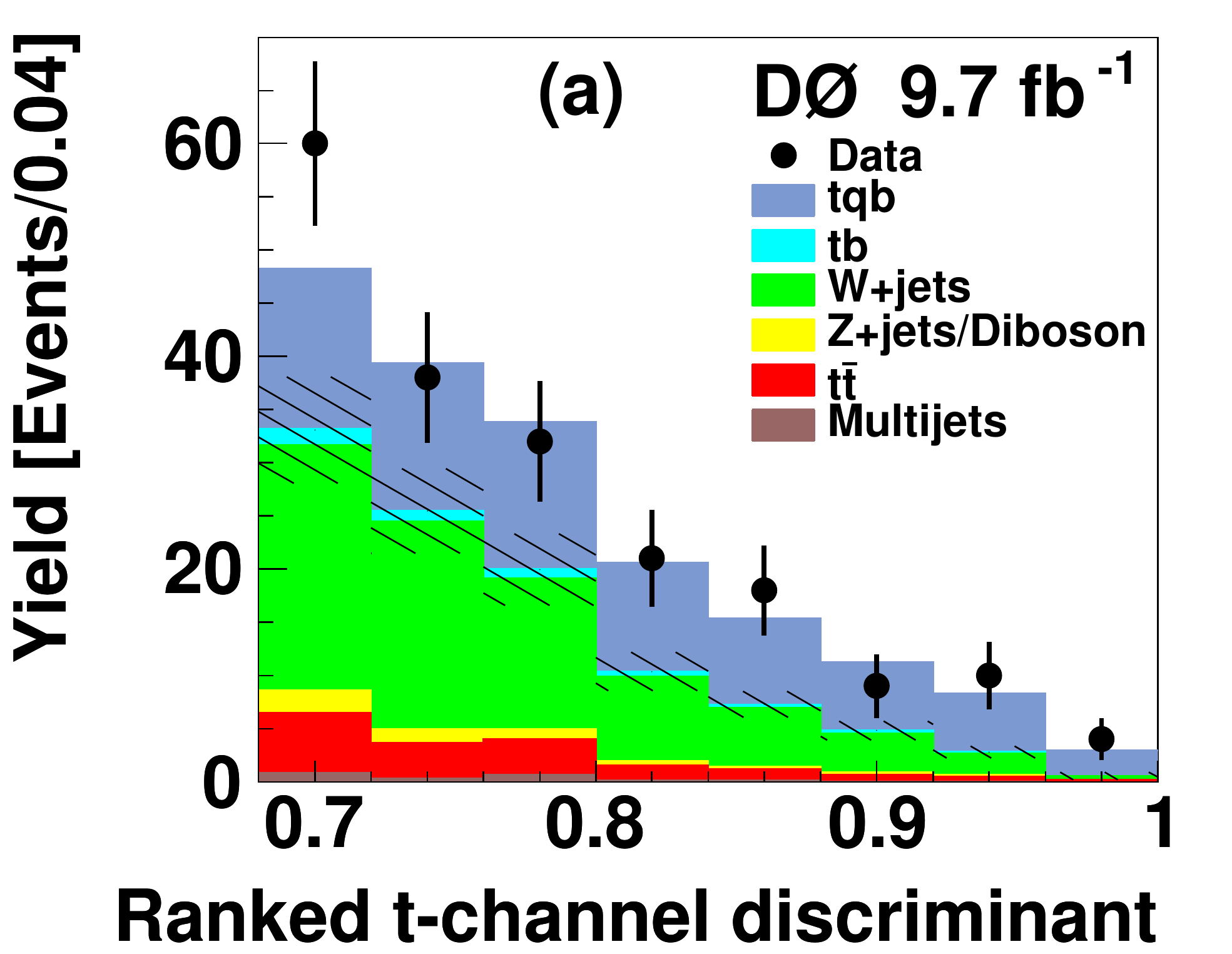}
\hspace{1cm} 
  \includegraphics[width=0.4\textwidth]{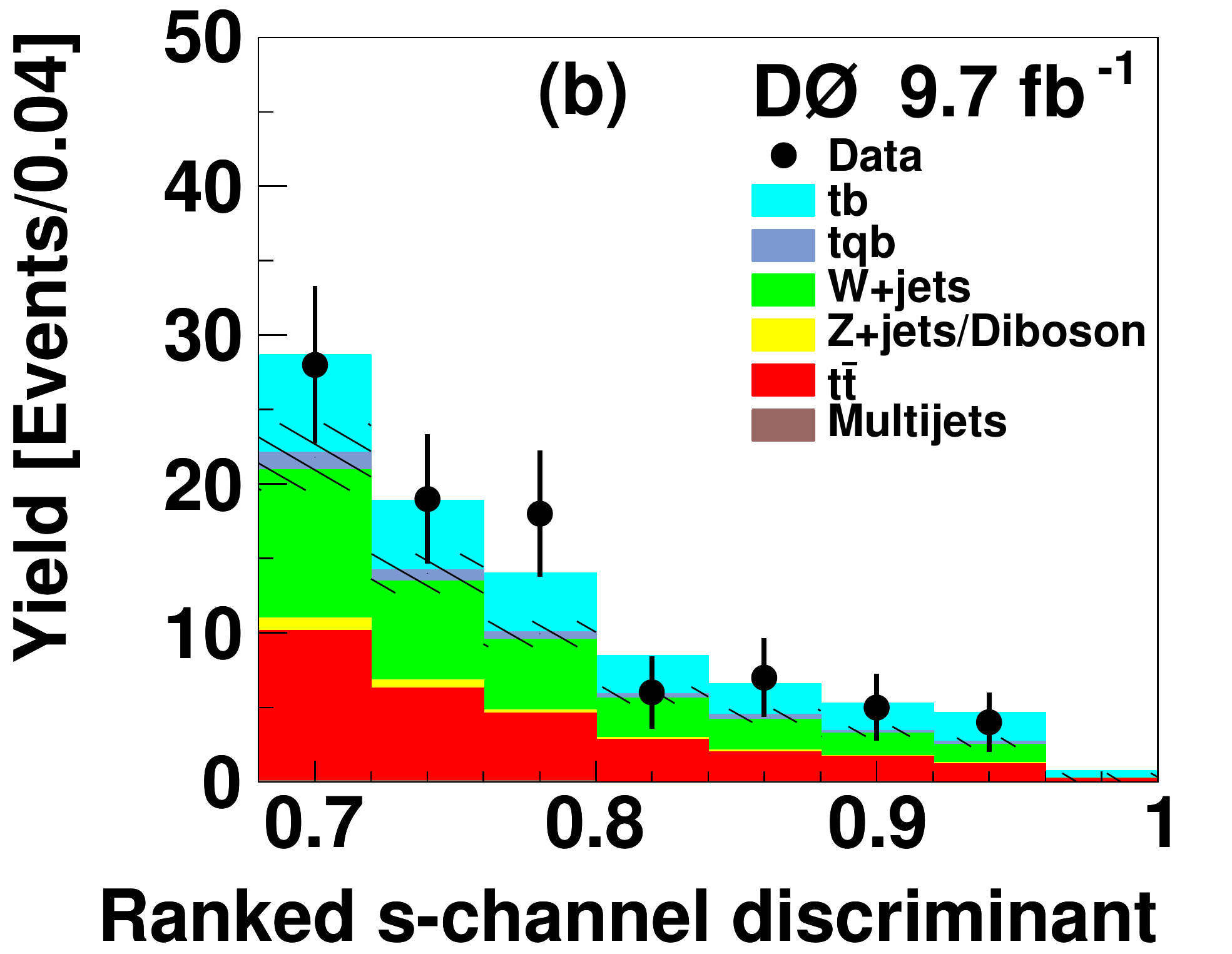}
  \caption{Signal region of the multivariate discriminant (ranked by expected signal-to-background ratio) for the D0 single top-quark analysis for  (a) the $t$-channel discriminant and (b) the $s$-channel discriminant (from \textcite{Abazov:2013qka}).}
  \label{fig:BNND0Run2}
\end{figure}

D0 measures the combined single top-quark cross section using a combination of several MVA techniques~\cite{Abazov:2013qka} using 9.7~\fb of data, selecting events in the $l$+jets channel. Each event is required to have an electron or a muon with $p_T>20$~GeV and two or three jets, at least one of which is required to be $b$-tagged. The leading jet is required to have $p_T>25$~GeV, while all other jets have $p_T>20$~GeV. The missing transverse momentum is required to be $\MET >20$~GeV for 2-jet events and $\MET >25$~GeV for 3-jet events. Events where a hadronic jet is misidentified as a lepton are rejected through additional event topology requirements. In total, 12,000 data events are selected, of which 630 are expected to be from single top-quark production. The $t$-channel and $s$-channel signals are separated from the large background with three MVA discriminants: a Bayesian NN  (BNN), a boosted decision tree (BDT), and a matrix element (ME) discriminant.
The inputs to the BNN and the BDT are kinematic properties of individual analysis objects and whole-event features, and include the output of the $b$-tag algorithm. In the ME method, also known as dynamic likelihood method~\cite{Kondo:1988yd,Kondo:1991dw}, a discriminant is built using probabilities calculated from the squared matrix element for each signal and background process hypothesis based on the corresponding leading-order Feynman diagrams, and thus in principle uses all the kinematic information available for the event. The three individual discriminants are then combined in another BNN to form the final discriminant. The methods are optimized separately for $t$-channel (where $s$-channel is included as part of the background) and  $s$-channel (where $t$-channel is included as part of the background) in each of four regions (2 or 3 jets, 1 or 2 $b$-tags). The signal region for the two discriminants is shown in Fig.~\ref{fig:BNND0Run2}. The cross section is measured in a Bayesian likelihood analysis~\cite{Bertram:2000br}.  The resulting two-dimensional posterior as a function of $t$-channel and $s$-channel single top-quark production cross sections for D0 is shown in Fig.~\ref{fig:PostCDFD0}(right).

\subsubsection{Tevatron combination}
\label{sec:TevComb}

\begin{figure}[!htbp]
  \includegraphics[width=0.6\textwidth]{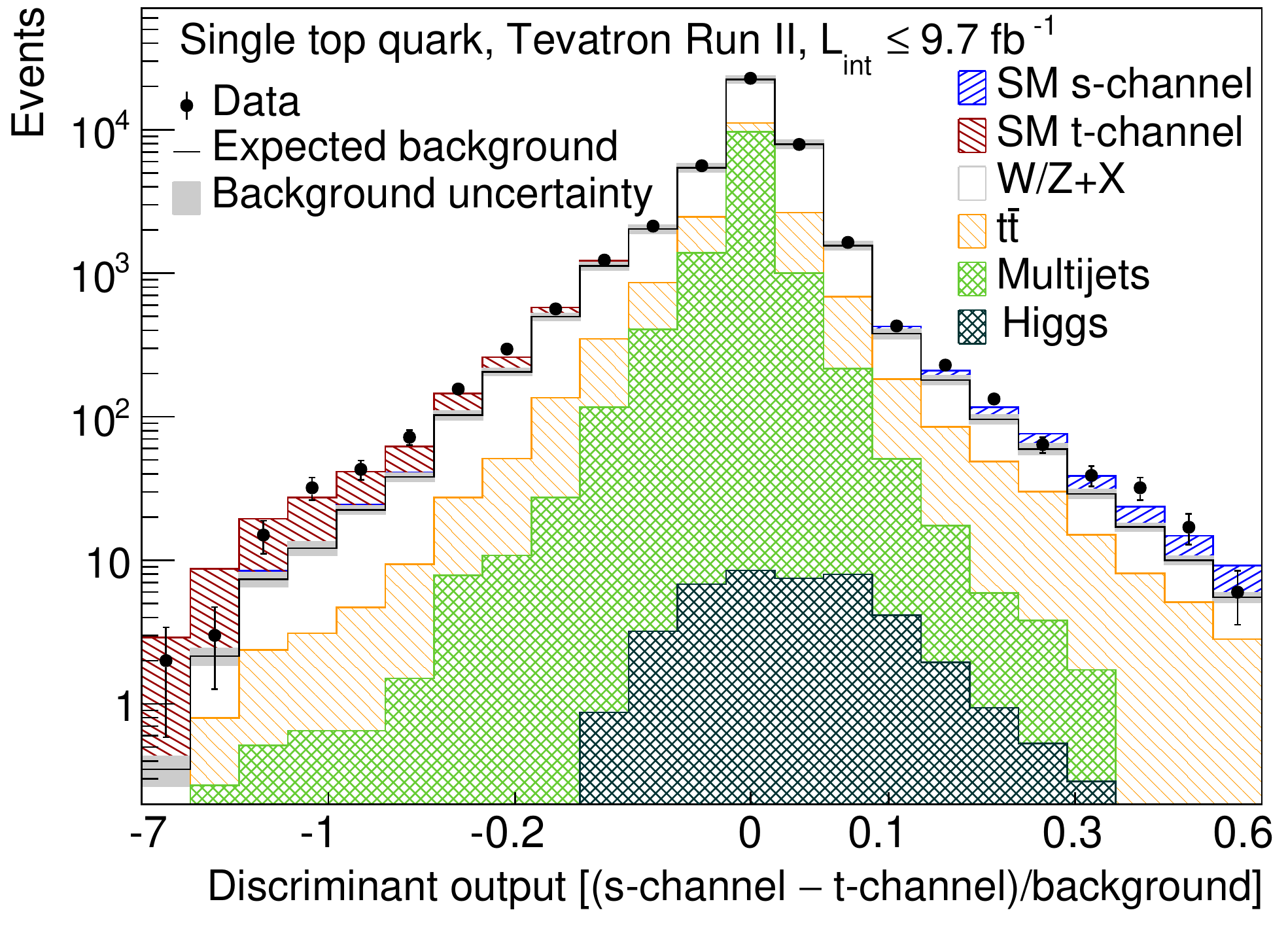}
  \caption{Distribution of the discriminant histograms, summed over bins with similar ratios ($(s-t)$/background) (from \textcite{Aaltonen:2015cra}). A non-linear scale is used on the horizontal axis to better bring out the signal regions of the discriminant.}
  \label{fig:TevDiscr}
\end{figure}

The results from the two experiments are combined starting from the $s$- and $t$-channel discriminants in the two CDF~\cite{Aaltonen:2014ura,Aaltonen:2014mza}. and one D0~\cite{Abazov:2013qka} analyses listed above. The various channels of the different analyses are combined by taking the product of their likelihoods and simultaneously varying the correlated uncertainties and by comparing data to the predictions for each contributing signal and background process. The combined Tevatron cross sections are measured using a Bayesian statistical analysis~\cite{Bertram:2000br}. No assumption is made about the ratio of the $t$-channel and $s$-channel cross sections (unlike for the single top-quark discovery). The several hundred bins of the individual discriminants are sorted by their $t$-channel and $s$-channel signal/background ratios as $s-t$ and rebinned. This discriminant is shown in Fig.~\ref{fig:TevDiscr}. The $t$-channel signal appears on the left, at large negative values. The $s$-channel signal appears on the right, at large positive values. The signal+background distribution shows good agreement with the data over the full discriminant range. The largest background in both the $t$-channel and $s$-channel signal regions is from $W$-boson production in association with jets ($W$+jets), with smaller contributions from \ttbar production and other backgrounds.

\begin{figure}[!htbp]
  \includegraphics[width=0.45\textwidth]{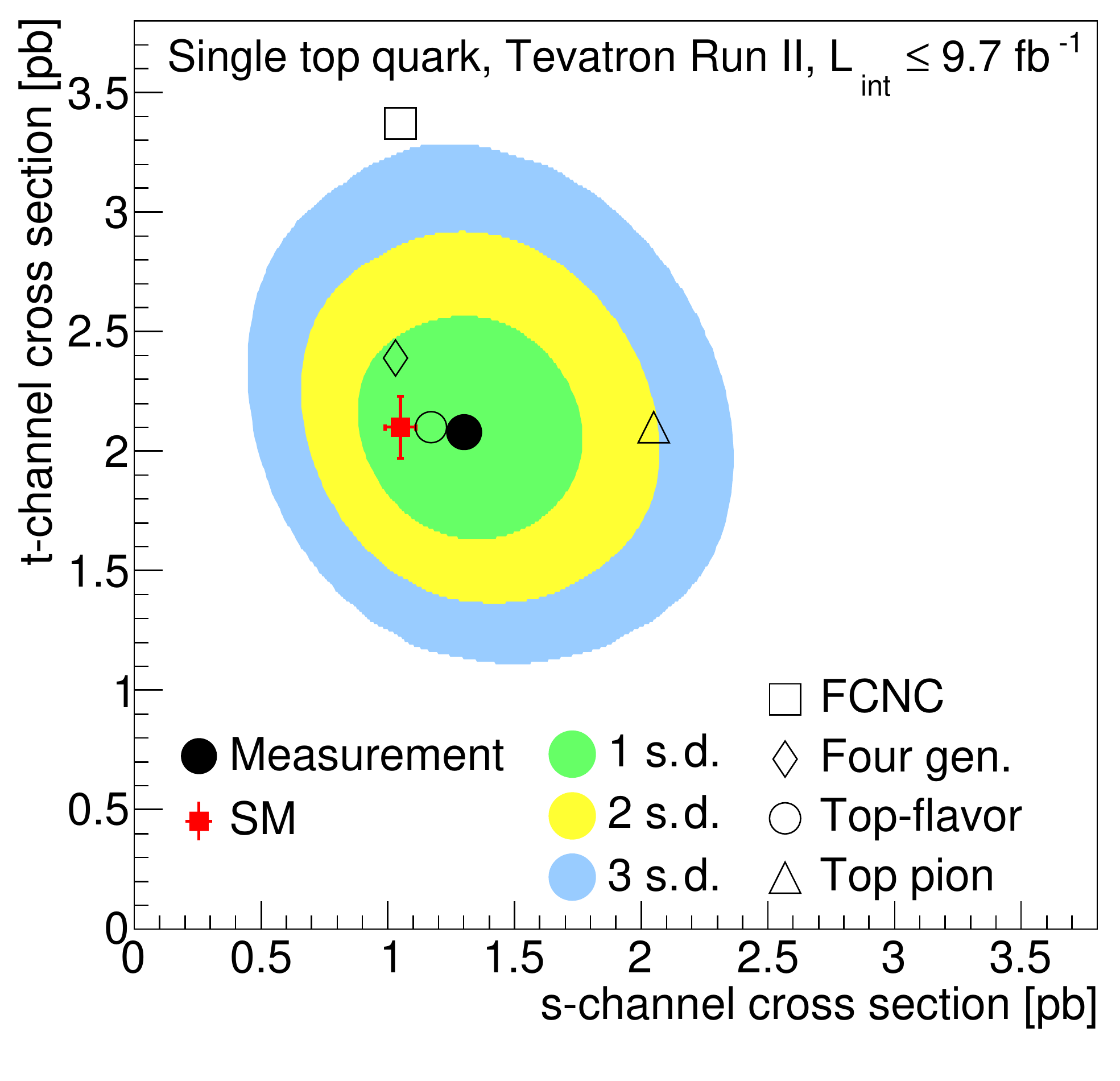}
\hfill
  \includegraphics[width=0.45\textwidth]{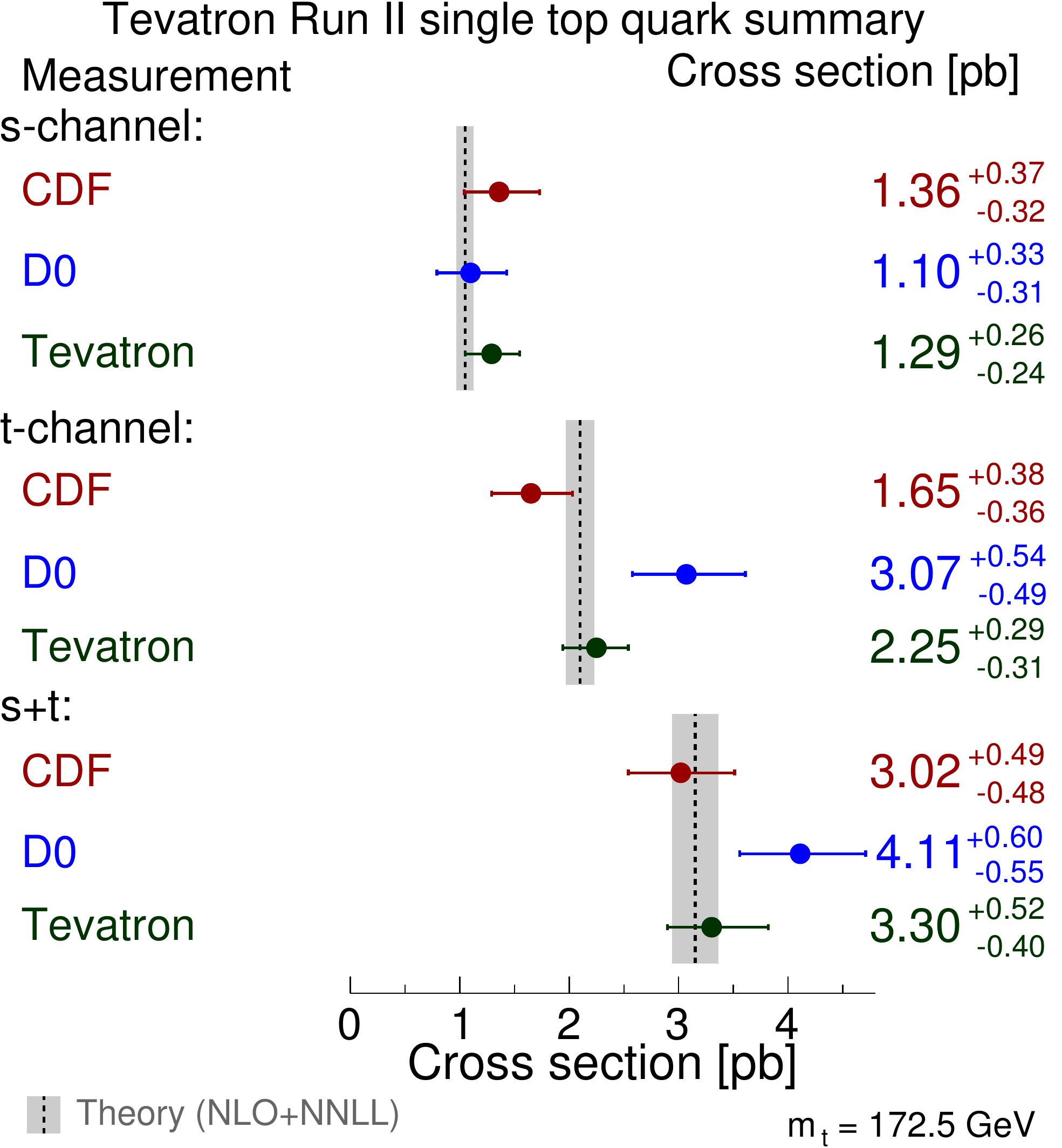}
  \caption{(Left) Posterior probability density as a function of the $t$-channel and $s$-channel cross sections (adapted from \textcite{Aaltonen:2015cra}). Also shown are new physics models: FCNC top-gluon interactions~\cite{Tait:2000sh,Abazov:2007ev}, a four-generation model~\cite{Alwall:2006bx}, a top-flavor model~\cite{Tait:2000sh}, and a top pion~\cite{Hill:1994hp,Tait:2000sh}. (Right) Summary of the Tevatron single top-quark measurements (adapted from \textcite{Aaltonen:2015cra}).}
  \label{fig:Tev2d}
\end{figure}

The two-dimensional Bayesian posterior density as a function of the $t$-channel and $s$-channel cross sections is shown in Fig.~\ref{fig:Tev2d}(left). The measurement agrees with the SM prediction and is also compared to several new physics models for illustration. FCNC couplings of the top quark to the gluon~\cite{Tait:2000sh,Abazov:2007ev} increase the $t$-channel cross section. A possible fourth generation~\cite{Alwall:2006bx} results in an increased top-quark coupling to first- and second-generation quarks and thus reduces the $s$-channel cross section while increasing the $t$-channel cross section. A top-flavor model~\cite{He:1999vp,Tait:2000sh} with an additional boson coupling to the top quark increases the $s$-channel cross section and has no impact on $t$-channel production. A charged ``top pion''~\footnote{The term ``top pion'' refers to hypothetical composite bosons formed by top and bottom quarks and antiquarks, predicted in models with additional strong interactions that only act on third-generation quarks, generally known as ``top-color'' models~\cite{Hill:1991at,Hill:1994hp}. These models seek to explain the largeness of the top-quark mass by a top-quark condensation that plays the role of the Higgs field, in analogy with the phenomenon of superconductivity. Top pions play for such a theory the same role that the SM pions, formed by up and down quarks and antiquarks, play in QCD.} results in a $s$-channel resonance decaying to a top quark and a bottom quark~\cite{Tait:2000sh}.

\subsubsection{$s$-channel}

\begin{figure}[!htbp]
  \includegraphics[width=0.45\textwidth]{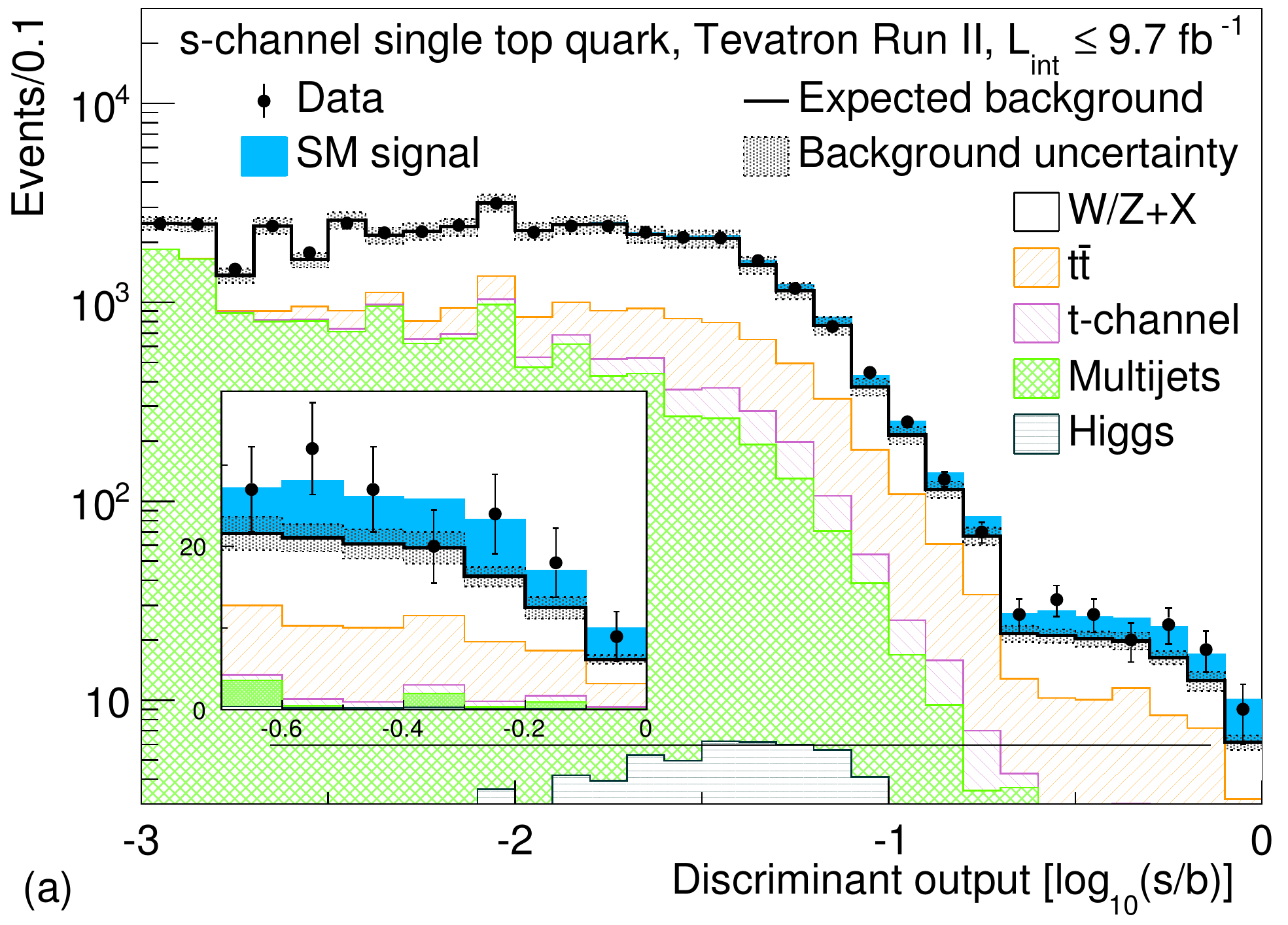}
\hfill
  \includegraphics[width=0.45\textwidth]{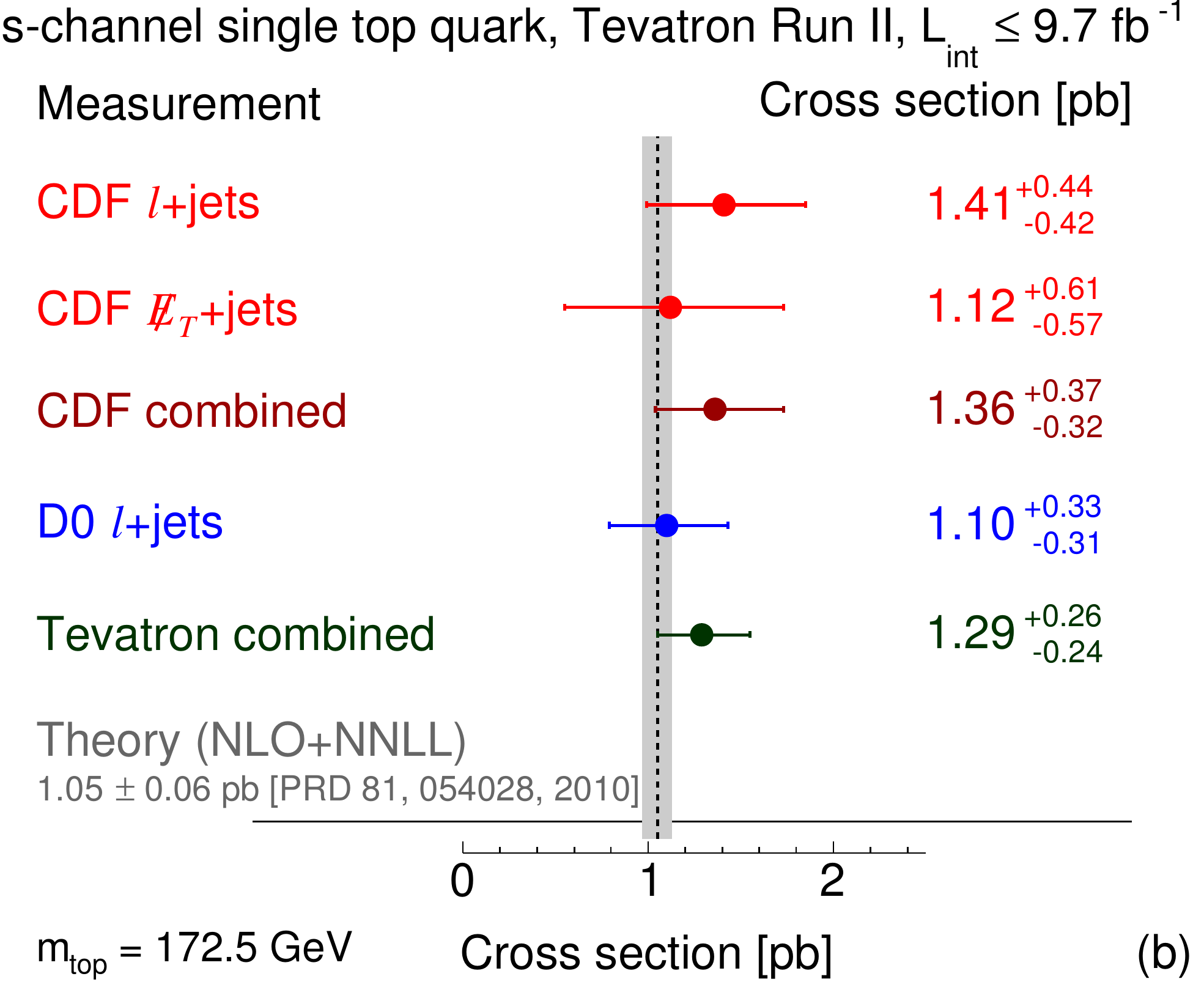}
  \caption{(Left) Tevatron $s$-channel discriminant, with bins sorted by signal/background yields and (right) summary of Tevatron $s$-channel cross section measurements (from \textcite{CDF:2014uma}).}
  \label{fig:Tevschan}
\end{figure}

The existence of $s$-channel production has been established few years ago by the combination of Tevatron measurements~\cite{CDF:2014uma} and it is one of the few ``Tevatron legacies'' that have not been surpassed in precision by the LHC experiments. The input measurements and procedure are the same as described in Section~\ref{sec:TevComb}, but here, the likelihood fit is one-dimensional for the $s$-channel signal, including $t$-channel single top-quark production in the background. The combined discriminant, rebinned to bring out the s-channel signal, is shown in Fig.~\ref{fig:Tevschan}(left). The dominant background in the signal region is from $W$+jets production and top-quark pair production. The $t$-channel contribution in the $s$-channel signal region is negligible.

The cross section is measured to be $1.29^{+0.26}_{-0.24}$~pb, consistent with the SM expectation. The significance of the excess of the data over the background expectation is 6.3 standard deviations. A summary of the Tevatron $s$-channel measurements is shown in  Fig.~\ref{fig:Tevschan}(right).

The Tevatron cross section measurements are summarized in Fig.~\ref{fig:Tev2d}(right) and are compared to the LHC measurements in Fig.~\ref{fig:sqrts}.

\FloatBarrier

\subsection{LHC}
\label{sec:xslhc}

Single top-quark production at the LHC is dominated by the $t$-channel, even more than at the Tevatron. The production cross section for the $t$-channel, shown in Table~\ref{tab:lhctchanpred}, is sufficiently large to produce millions of single top quarks, enough to measure the cross section inclusively and differentially and to measure top-quark properties precisely (see Section~\ref{sec:params}).
The cross section for the production of a top quark in association with a $W$~boson, shown in Table~\ref{tab:lhctWpred}, is second-largest, and is sufficiently high to observe this process at the LHC. The $s$-channel cross section, shown in Table~\ref{tab:lhcschanpred}, is small due to its quark-antiquark initial state and so far only evidence for this process has been reported.

\subsubsection{$t$-channel}
\label{sec:tchannel}

The ATLAS and CMS experiments have recorded proton-proton data at various CM energies. 
The $t$-channel production mode (Fig.~\ref{fig:FG}(a)) has the largest cross section, and is the only single top-quark process whose cross section has been measured at four CM energies so far. Effort has also gone into providing precise theoretical predictions for this mode. 
The $t$-channel cross sections have been calculated at next-to-next-to-leading order (NNLO) in QCD~\cite{Brucherseifer:2014ama,Berger:2017zof,Berger:2016oht} and at NLO with NNLL resummation~\cite{Kidonakis:2011wy}. Automatic calculations as a function of various parameters can be performed with the \HATHOR v2.1 program at NLO~\cite{Aliev:2010zk,Kant:2014oha}, based on MCFM~\cite{Campbell:2004ch}. The dependence of the theory predictions on the flavor-number scheme in the predictions has also been studied by comparing the full NLO calculations in the 4FNS (Fig.~\ref{fig:FG23}(a)) with that in the 5FNS (Fig.~\ref{fig:FG23}(b))~\cite{Frederix:2012dh}. The different predictions are compared in Table~\ref{tab:lhctchanpred}. The NLO+NNLL predictions are slightly larger than the NLO ones, while the NNLO calculations predict a smaller cross section.
The cross sections have also been computed differentially~\cite{Berger:2017zof,Kidonakis:2015wva,Schwienhorst:2010je}.

\begin{table}[!htbp]
\begin{center}
\begin{tabular}{l|c@{\hskip 0.3in}c@{\hskip 0.3in}c}  
$t$-channel & 7 TeV & 8 TeV & 13 TeV \\
cross section in pb & \\ \hline
NNLO & \\
~~$t$                      & - & $54.2^{+0.5}_{-0.2}$ & $134.3^{+1.3}_{-0.7}$ \\
~~$\overline{t}$     & - & $29.7^{+0.3}_{-0.1}$ & $79.3^{+0.8}_{-0.6}$  \\
~~$t+\overline{t}$ & - & $83.9^{+0.8}_{-0.3}$ & $213.6^{+2.1}_{-1.1}$  \\
NLO+NNLL & \\
~~$t$                      & $43.0^{+1.8}_{-0.9}$ & $56.4^{+2.4}_{-1.2}$ & $136^{+4}_{-3}$ \\
~~$\overline{t}$     & $22.9^{+0.9}_{-1.0}$ & $30.7^{+1.5}_{-1.6}$ & $82^{+3}_{-2}$  \\
~~$t+\overline{t}$ & $65.9^{+2.6}_{-1.8}$ & $87.2^{+3.4}_{-2.5}$ & $218^{+5}_{-4}$  \\
NLO & \\
~~$t$                      & $41.8^{+1.8}_{-1.5}$ & $54.9^{+2.3}_{-1.9}$ & $136 \pm 5$ \\
~~$\overline{t}$     & $22.0^{+1.3}_{-1.2}$ & $29.7^{+1.7}_{-1.5}$ & $81 \pm 4$  \\
~~$t+\overline{t}$ & $63.8^{+2.9}_{-2.2}$ & $84.7^{+3.8}_{-3.2}$ & $217^{+9}_{-8}$  \\
\end{tabular}
\caption{Theoretical predictions for the $t$-channel production cross sections at the LHC. 
The NNLO predictions at 8~TeV~\cite{Brucherseifer:2014ama} and 13~TeV~\cite{Berger:2016oht} use a top-quark mass of 172.5~GeV and 173.2~GeV, respectively, and the uncertainties include scale variations.
The NLO+NNLL predictions~\cite{Kidonakis:2011wy,Kidonakis:2013zqa,Kidonakis:2016vlv} have been calculated for a top-quark mass of 173 GeV and the uncertainties include scale and PDF~\cite{MSTW2008NLO} variations.
The NLO predictions have been computed using the \HATHOR v2.1 program~\cite{Aliev:2010zk,Kant:2014oha} based on MCFM~\cite{Campbell:2009ss}. They are obtained at a top-quark mass of 172.5 GeV and the uncertainties include scale, PDF and $\alpha_S$~\cite{Botje:2011sn,MSTW2008NLO,Martin:2009bu,CT10,NNPDF23} variations.}
\label{tab:lhctchanpred}
\end{center}
\end{table}

At the LHC, the inclusive $t$-channel cross sections have been measured at 7~TeV~\cite{Aad:2014fwa,Chatrchyan:2011vp,Chatrchyan:2012ep}, 8 TeV~\cite{Aaboud:2017pdi,Khachatryan:2014iya} and 13~TeV~\cite{Aaboud:2016ymp,Sirunyan:2016cdg} by ATLAS and CMS. All these analyses enhance the $t$-channel signal by selecting events with one isolated electron or muon, significant \MET and/or large invariant mass (\mT) of the lepton plus \MET system~\footnote{Defined as $\mT = \sqrt{ \left(p_T^{l} + \MET \right)^2 - \left( p_x^{l} + \MET{}_{,x} \right)^2 - \left( p_y^{l} + \MET{}_{,y} \right)^2 }$.}, and two or three jets. Exactly one of the jets is required to pass a tight threshold on the $b$-tagging discriminant and is interpreted as coming from the decay of the top quark, while the other (failing the same threshold) as originating from the spectator quark that recoils again the top quark. Main backgrounds to this final state are \ttbar\ and $W$+jets. Orthogonal control regions with different multiplicities of jets and/or $b$-tagged jets are used to measure these backgrounds {\it in situ}, to validate the Monte Carlo models used for their predictions, or to constrain the main experimental uncertainties (e.g., $b$-tag modeling).
 QCD multi-jet events constitute a small but non-negligible background. Given the uncertainties in its modeling, it is necessary to predict the size and properties of this process by data. A reliable model of this background is usually extracted from events that fail the isolation requirement or other elements of the charged-lepton selection, while fulfilling all other selection criteria.

The extraction of the signal cross section is performed by both collaborations by profile-likelihood fits~\cite{Verkerke:2003ir,Cranmer:2012sba,Cowan:2010js}. The fit variable is a multivariate discriminant in the case of ATLAS~\cite{Aad:2014fwa,Aaboud:2017pdi,Aaboud:2016ymp} and of some of the CMS analyses~\cite{Chatrchyan:2011vp,Chatrchyan:2012ep,Sirunyan:2016cdg}. 
ATLAS also measured the cross section at 7~TeV in a simple cut-based approach~\cite{Aad:2012ux}. CMS also demonstrated the feasibility of entirely relying on a simple kinematic observable, \etalj, defined as the pseudorapidity of the jet failing $b$-tag requirement~\cite{Chatrchyan:2012ep,Khachatryan:2014iya}.

Table~\ref{tab:tevlhcacc} compares the acceptances and event yields of the LHC $t$-channel analyses to the Tevatron $s+t$-channel analyses. The kinematic thresholds on leptons, jets and \etmiss are higher at the LHC than at the Tevatron, resulting in an acceptance that is about a factor two lower. However, since the cross section is so much larger, the number of signal events and the signal/background ratio are larger. 
\begin{table}[!htbp]
\begin{center}
\begin{tabular}{l|ccc}  
Experiment &  signal         &  number of               & s/b (\%) \\
                  &  acceptance (\%) &  $t$-channel events &    \\ \hline
1.96 TeV Tevatron\\
~~CDF $s+t$ $\ell$+jets & 2.2 & 550 & 6.4 \\
~~CDF $s+t$ \etmiss+jets & 1.7 & 530 & 2.3 \\
~~D0 $s+t$ $\ell$+jets & 2.0 & 630 & 5.3 \\
7 TeV LHC\\
~~ATLAS $t$-channel, 4.6~\fb & 1.0 & 5,700 & 10 \\
~~CMS $t$-channel, 1.2($\mu$), 1.6($e$)~\fb & 0.8($\mu$), 0.6($e$)  & 950 & 31  \\
8 TeV LHC\\
~~ATLAS $t$-channel, 20.3~\fb & 1.0 & 17,700 & 18  \\
~~CMS $t$-channel, 19.7~\fb & 0.6  & 10,400  & 21 \\
13 TeV LHC\\
~~ATLAS $t$-channel, 3.2~\fb & 1.0 & 6,900 & 11  \\
~~CMS $t$-channel, 2.2~\fb & 0.5  & 2,400 &  11 \\
\end{tabular}
\caption{Comparison of Tevatron and LHC single top-quark acceptances , event yields, and signal/background ratio. The 7~TeV CMS analysis was done separately for electron and muon events and the luminosity and single top-quark acceptances are given separately, while the number of events and the signal/background ratio (s/b) are quoted for electron and muon channels combined.}
\label{tab:tevlhcacc}
\end{center}
\end{table}

\begin{figure}[!htbp]
  \includegraphics[width=0.56\textwidth]{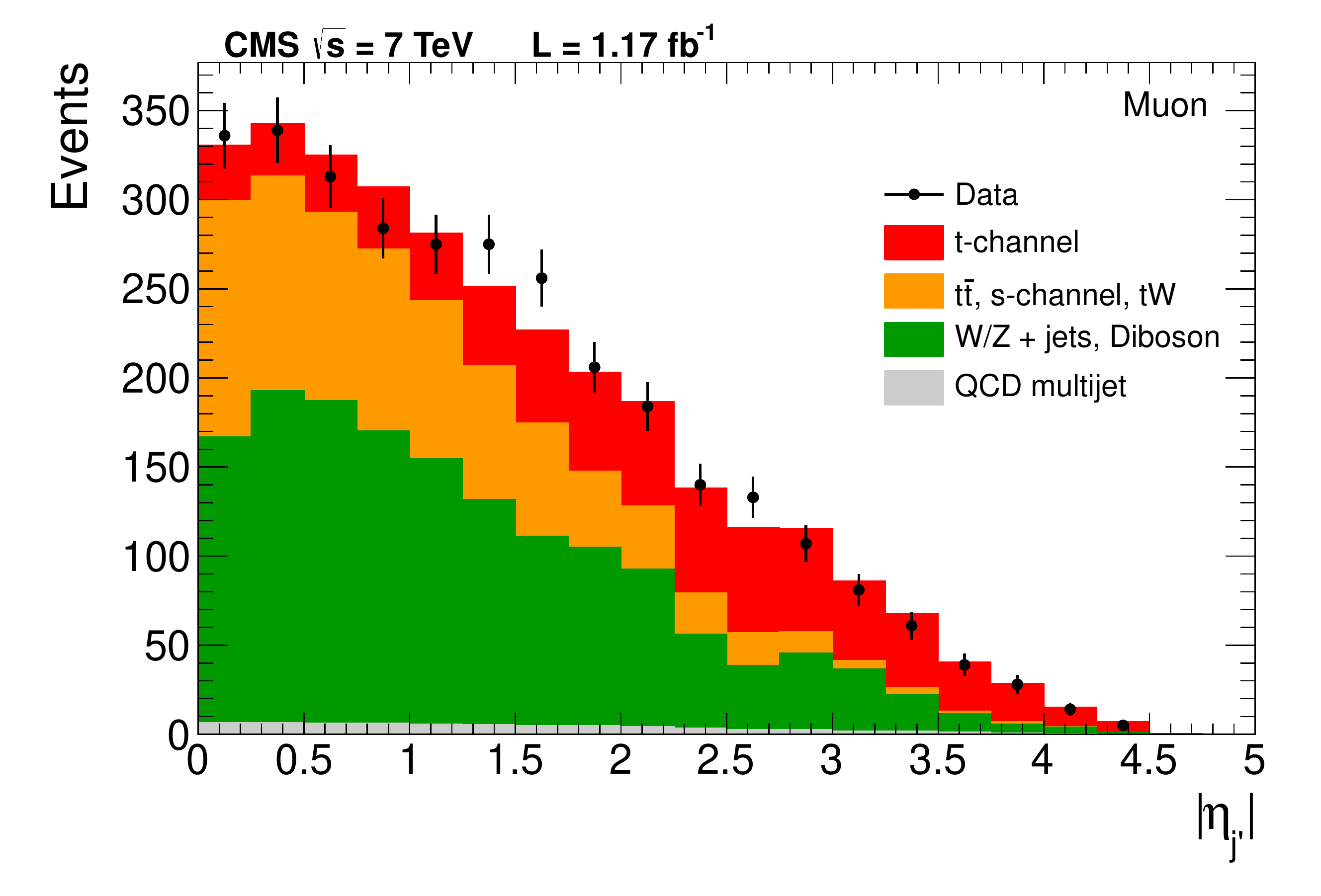}
\hfill
  \includegraphics[width=0.42\textwidth]{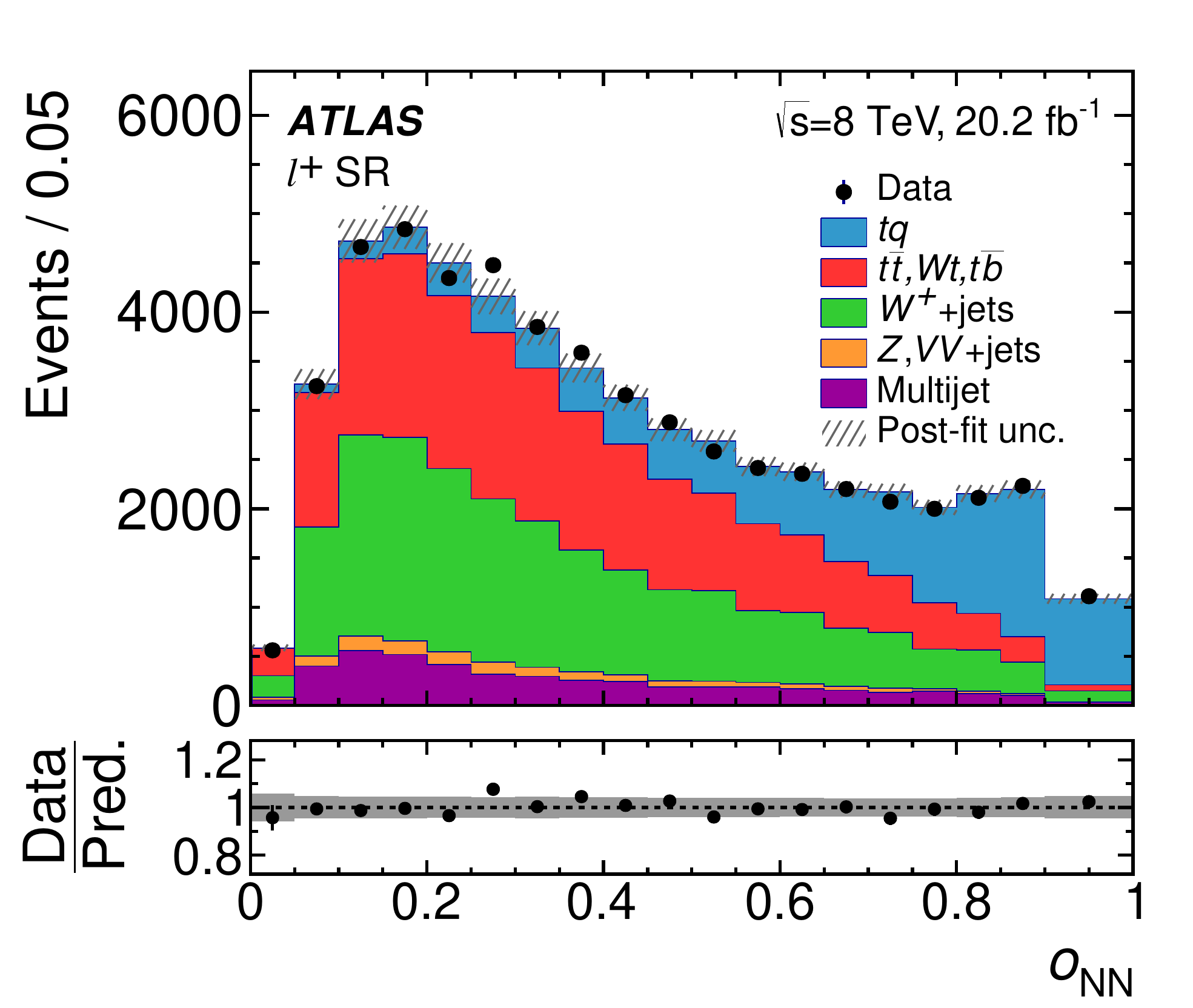}
  \caption{(Left) CMS 7~TeV $t$-channel pseudorapidity distribution of the light-quark jet for muon events (from \textcite{Chatrchyan:2012ep}) and (right) ATLAS 8~TeV $t$-channel NN discriminant distribution (from \textcite{Aad:2014fwa}).}
  \label{fig:LHCtchan1}
\end{figure}

\begin{figure}[!htbp]
  \includegraphics[width=0.42\textwidth]{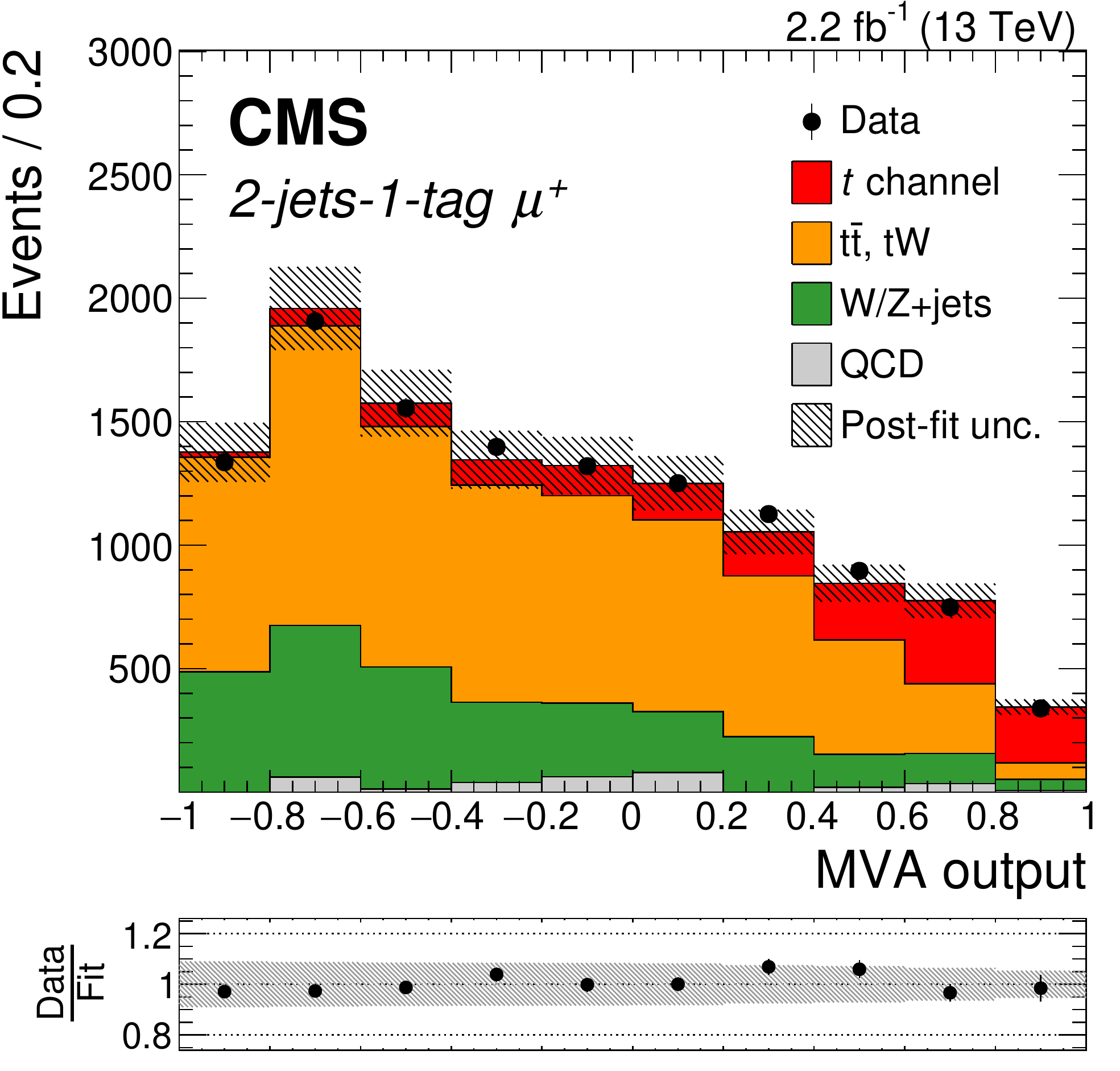}
\hfill
  \includegraphics[width=0.56\textwidth]{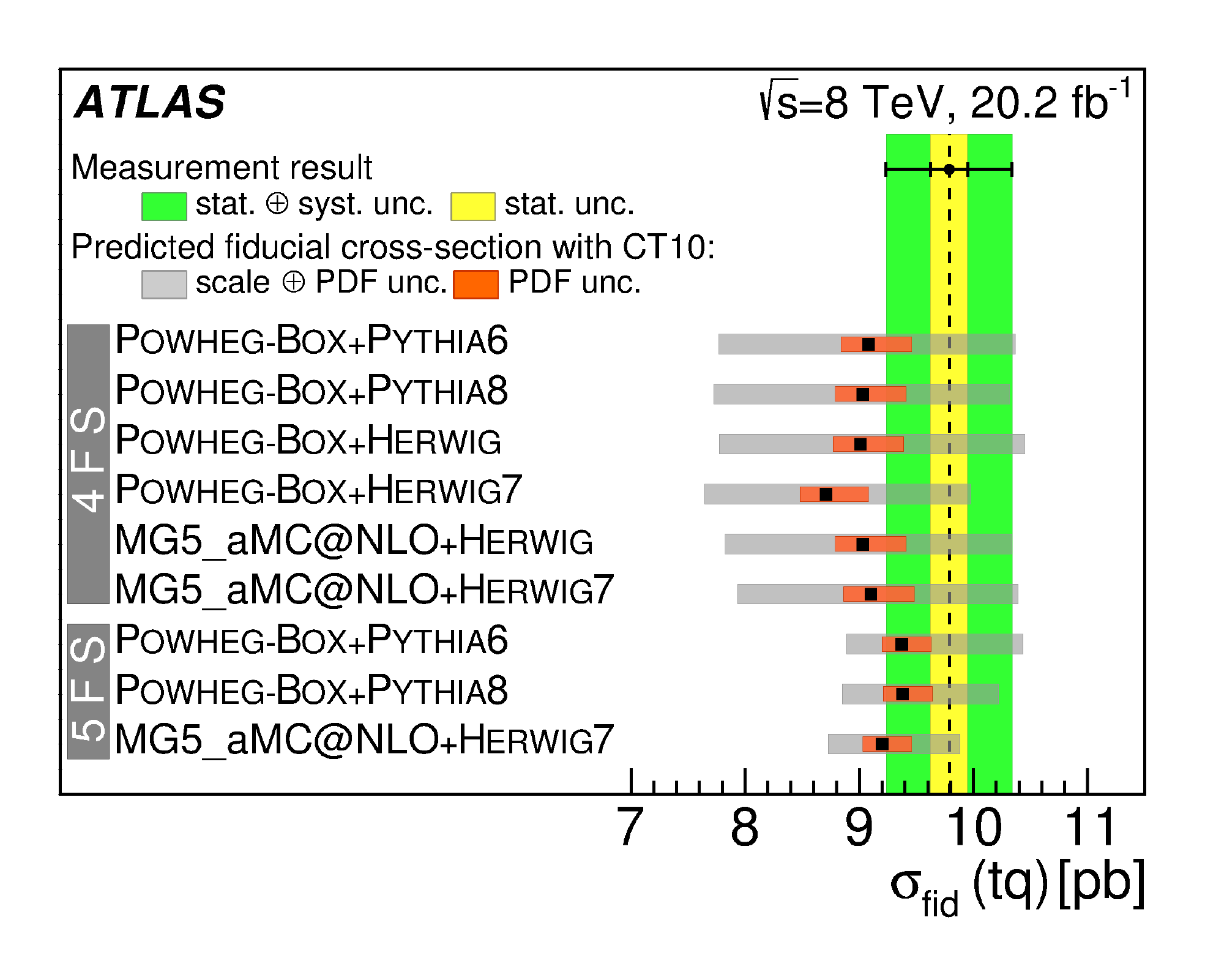}
  \caption{(Left) CMS 13~TeV $t$-channel NN discriminant  (from \textcite{Sirunyan:2016cdg}) and (right) ATLAS 8~TeV $t$-channel fiducial cross-section measurement compared to different signal simulations (from \textcite{Aad:2014fwa}).}
  \label{fig:LHCtchan2}
\end{figure}

Systematic uncertainties are dominant over the statistical uncertainties in these $t$-channel measurements, with the exception of the earliest measurement at 7~TeV using the data collected in 2010~\cite{Chatrchyan:2011vp}. The important detector-related uncertainties are from $b$-tagging and jet energy scale (JES). The theory modeling uncertainties contribute about half of the total systematic uncertainties. These are related to the renormalization and factorization scales in the simulated signal sample, the PDFs, the amount of initial-state and final-state radiation (ISR/FSR), the modeling of the parton shower and the NLO subtraction (treatment of phase-space that is populated by both the NLO corrections in the matrix element and the parton shower). Theory modeling uncertainties are included for both the $t$-channel signal and the background from \ttbar production. The scale and ISR/FSR uncertainties are evaluated by both ATLAS and CMS by varying the relevant parameters in the simulation. The NLO subtraction is evaluated by comparing the \POWHEG method to the \aMCatNLO method~\cite{Frixione:2007vw,Alwall:2014hca,Frederix:2012dh}. For the CMS 8~TeV analysis, this also includes a comparison of events generated in the 4FNS and the 5FNS. The uncertainty due to the description of parton showers is evaluated by comparing Pythia to Herwig, for ATLAS in the entire analysis chain, for CMS only in the JES. The PDF uncertainty is evaluated with the PDF4LHC prescription~\cite{Botje:2011sn}. The background-related uncertainties are dominated by the $t\overline{t}$-modeling and normalization and also have contributions from $W$+jets and fake-lepton background modeling.
Figure~\ref{fig:LHCtchan1} shows the light-quark jet pseudo-rapidity distribution for muon events in the CMS 7~TeV analysis and the NN discriminant for positively charged leptons in the ATLAS 8~TeV analysis. Already with a limited-size sample at 7~TeV, the $t$-channel signal is clearly visible, and at 8~TeV, even bins of the final discriminant where the background is reduced to negligible levels still retain thousands of signal events. Figure~\ref{fig:LHCtchan2}~(left) shows the CMS NN distribution in the 13~TeV $t$-channel analysis. Even with the small data sample analyzed so far in Run~2, the $t$-channel signal can be easily extracted. 
These figures show clearly that in comparison to 7 and 8~TeV, the \ttbar background is now larger than the $W$+jets background, as expected due to the larger increase in the $\ttbar$ cross section.

The cross section is evaluated in a likelihood fit, and some of the uncertainties are constrained by data in the fit, i.e., these nuisance parameters are profiled. For the ATLAS analyses, only the uncertainties on the normalization of the \ttbar and $W$+jets backgrounds (and for the 7~TeV analysis also the $b$-tag scale factor) are profiled, while the other uncertainties are evaluated through pseudo-experiments.
The CMS 7~TeV analysis uses a Bayesian approach to measure the cross section~\cite{jaynes03} and marginalizes the systematic uncertainties, except for the theory modeling uncertainties, which are evaluated in pseudo-experiments.

The cross sections measured by ATLAS and CMS at 7~TeV are $68 \pm 8$~pb and $67.2 \pm 6.1$~pb, respectively. ATLAS also measures the cross section for top-quark production separately from that for top antiquark production, $46 \pm 6$~pb and $23 \pm 4$~pb, respectively.
The CMS measurement is a combination of the electron and muon channels, both of which have a tight event selection that leads to a high s/b ratio, see Table~\ref{tab:tevlhcacc}, resulting in a slightly smaller total uncertainty for CMS than for ATLAS. The cross sections measured by ATLAS and CMS are consistent with each other and with the theory predictions.

At 8~TeV, the inclusive $t$-channel cross section measured by ATLAS is $89.6^{+7.1}_{-6.3}$~pb. 
The cross section has also been measured separately for top quarks and top antiquarks, $56.7^{+4.3}_{-3.8}$~pb for top-quark production and $32.9^{+3.0}_{-2.7}$~pb for top antiquark production. 
At 8~TeV, the inclusive $t$-channel cross section measured by CMS is $83.6 \pm 2.3 \stat \pm 7.4 \syst$~pb, with $53.8 \pm 1.5 \stat \pm 4.4 \syst$~pb for top quarks and $27.6 \pm 1.3 \stat \pm 3.7 \syst$~pb for top antiquarks. The cross sections measured by ATLAS and CMS are again consistent with each other and with the theory predictions, both inclusively and for top quarks and antiquarks separately. The systematic uncertainties are dominant, and the precision of the measurements is comparable.

At 13~TeV, the inclusive cross sections measured by ATLAS and CMS are $247 \pm 46$~pb and $238 \pm 32$~pb, respectively. The largest systematic uncertainty for ATLAS is the parton shower uncertainty  (13\%, when the total uncertainty is 17\%), evaluated by comparing the parton shower models of \PYTHIA and \HERWIG, both applied to events simulated at matrix-element level with \POWHEG. 
ATLAS and CMS also evaluated the cross sections for top quark and antiquark production separately, $156 \pm 28$~pb and $91 \pm 19$~pb, respectively, for ATLAS, and $154 \pm 22$~pb and $85 \pm 16$~pb, respectively, for CMS.
The measured cross sections are consistent with each other and with the theory predictions.

A fiducial $t$-channel cross section has been measured by the ATLAS collaboration using the 8~TeV data set~\cite{Aaboud:2017pdi}.
 The benefit of measuring a production cross section within a fiducial volume is that uncertainties related to event generation can be reduced, as a smaller extrapolation is needed between the reconstruction level and the particle level (unobservable regions of the phase become numerically irrelevant).  Differences between generators, hadronization models or PDFs can be separated into components visible in the measured phase space (similar between particle level and reconstruction level) and in the non-visible phase space (where there would be larger differences between particle level and reconstruction level).
 The fiducial phase space for this analysis is defined close to that of the reconstructed and selected events.
The particle-level objects are constructed from stable particles in the final state, with a very similar definition to the reconstructed objects, in order to minimize the sensitivity of the fiducial cross section to the signal modeling.
 The fiducial measurement is then extrapolated to the full phase space using different Monte Carlo generators, obtaining the spread of results shown in Fig.~\ref{fig:LHCtchan2}(right).

\begin{figure}[!htbp]
  \includegraphics[width=0.48\textwidth]{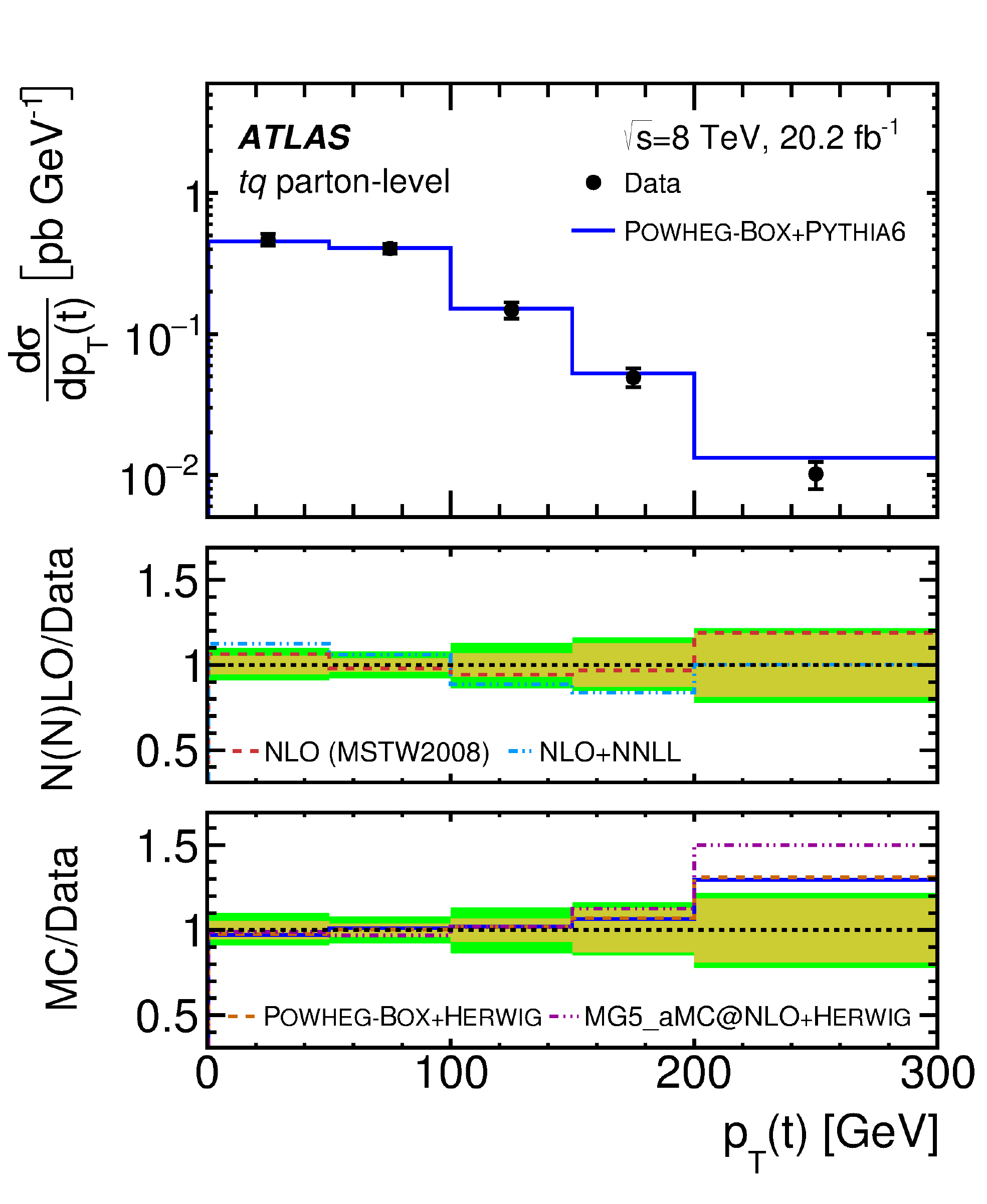}
\hfill
  \includegraphics[width=0.48\textwidth]{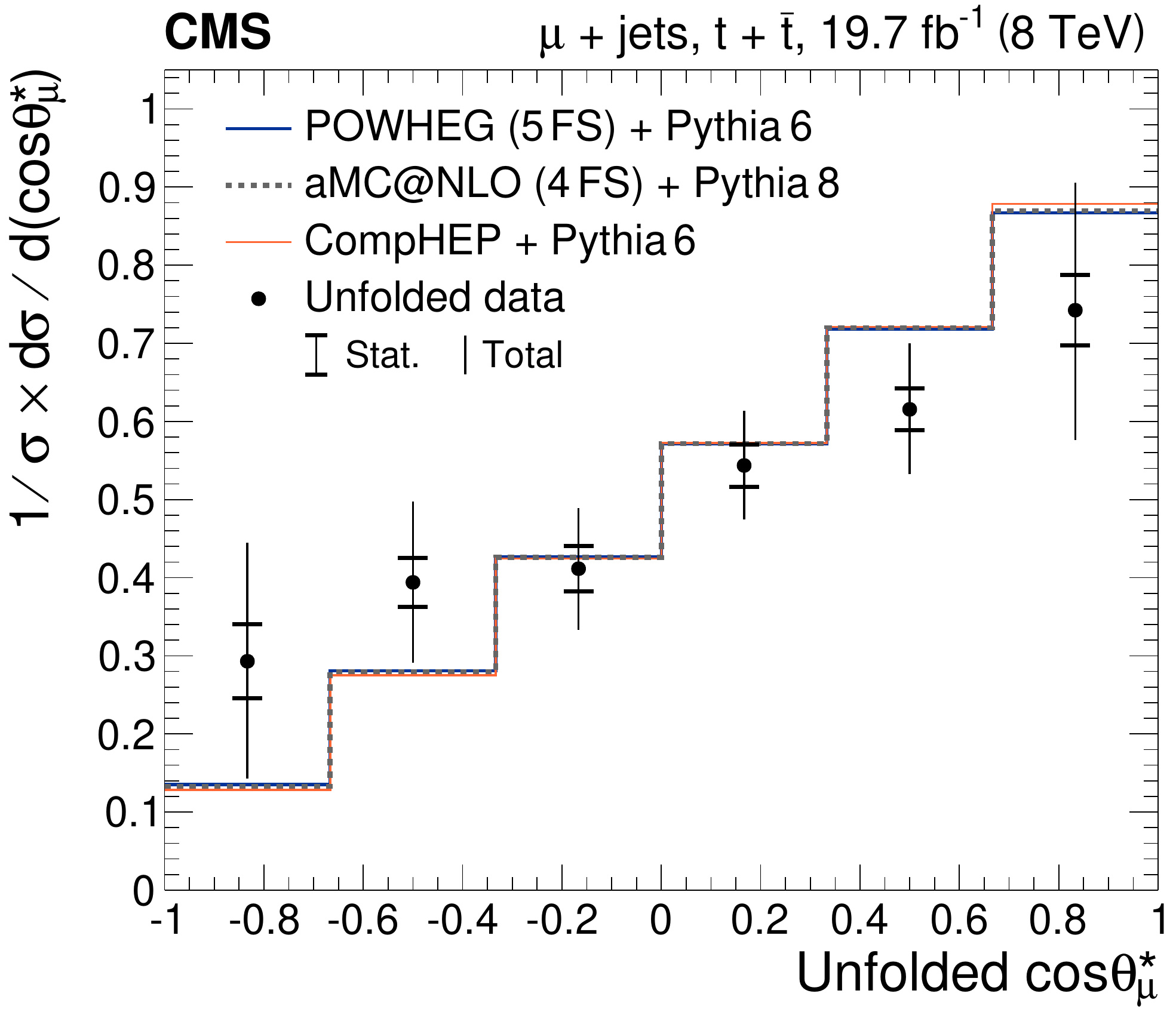}
  \caption{Differential distributions in $t$-channel events unfolded to parton level, (left) of the transverse momentum of the top quark in the ATLAS analysis at 8~TeV (from \textcite{Aaboud:2017pdi}) and (right) of \costheta in the CMS analysis at 8~TeV in the muon channel (from \textcite{Khachatryan:2015dzz}). }
  \label{fig:LHCtdiff}
\end{figure}

Differential cross sections of $t$-channel production as a function of top-quark \pt\ and pseudorapidity have been measured by ATLAS at 7 and 8~TeV~\cite{Aad:2014fwa,Aaboud:2017pdi} at particle and parton level, showing a good agreement with the predictions of various MC generators. Figure~\ref{fig:LHCtdiff}(left) shows the transverse momentum distribution of the top quark (not the antiquark) at parton level. 
 The CMS collaboration reported a relative differential cross-section measurement as a function of \costheta at 8~TeV~\cite{Khachatryan:2015dzz}, where \thetaL is defined at parton level as the angle in the top-quark rest frame between the momentum of the charged lepton from top-quark decay and a polarization axis approximated by the direction of the light quark recoiling against the top quark. This differential measurement, shown in Fig.~\ref{fig:LHCtdiff}(right), is an intermediate step in the extraction of top-quark polarization, see Sec.~\ref{sec:wtb-vertex}, and proves that the observed distribution is linear, as expected in V--A production mechanisms such as the electro-weak force in the SM. The ATLAS collaboration reported a differential measurement in two bins at the parton level in this variable as well as in two additional variables that characterize the angular correlations in top-quark events~\cite{Aaboud:2017aqp}. 

\subsubsection{$W$-associated (\tW)}
\label{sec:wt}

\begin{figure}[!htpb]
  \includegraphics[width=0.23\textwidth]{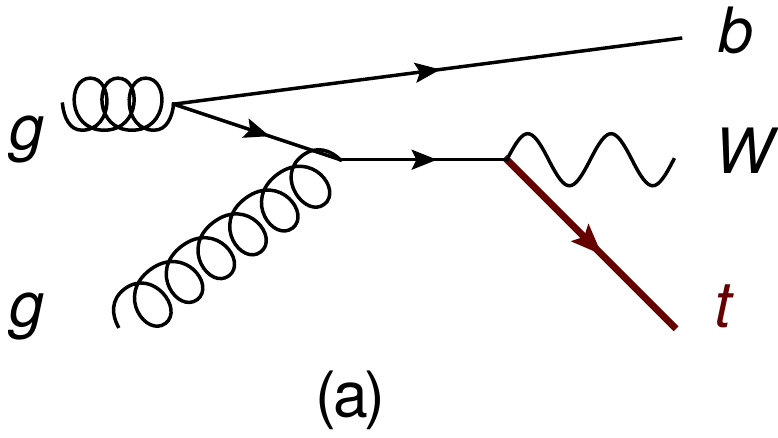}
\hspace{1cm}
  \includegraphics[width=0.18\textwidth]{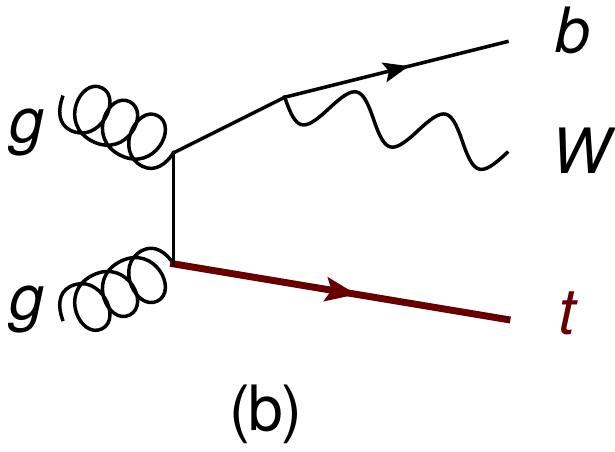}
  \caption{Representative Feynman diagram for $W$-associated single top-quark production (\Wt) from a gluon-gluon initial state, (a) $O(\alpha_s)$ correction that contributes to \Wt and (b) correction with an on-shell top quark that needs to be removed.}
  \label{fig:FGtWb}
\end{figure}

The \tW process, Fig.~\ref{fig:FG}(c), has the second-largest cross section. The theoretical prediction for $\tW$ production has been calculated at NLO with NNLL corrections~\cite{Kidonakis:2010ux} and at NLO~\cite{Aliev:2010zk,Kant:2014oha,Campbell:2004ch}.  This process is of particular interest because it overlaps experimentally and interferes by quantum principles with top-quark pair production.
The \Wt process is well-defined only at Born level. When higher-order QCD diagrams are taken into account, such as the production of $\tW$ with an associated $b$-quark as shown in Fig.~\ref{fig:FGtWb}, quantum interference induces a mixing with \ttbar as exemplified in Fig.~\ref{fig:FGtWb}(b). Some proposals have been made to define the two processes in an unambiguous way~\cite{Belyaev:2000me,Campbell:2005bb,Frixione:2008yi}.
The NLO event generators \MCatNLO~\cite{Frixione:2002ik} and \POWHEG~\cite{Frixione:2007vw}
 allow to choose between the so called ``Diagram Removal'' (DR) and ``Diagram Subtraction'' (DS) approaches~\cite{Frixione:2008yi,White:2009yt,Re:2010bp}. The DR approach removes all diagrams where the associated $W$~boson and the associated $b$-quark that are shown in Fig.~\ref{fig:FGtWb}(b) form an on-shell top quark.  The DS approach makes use of a subtraction term designed to locally cancel the \ttbar contributions. While the latter approach is designed to be gauge-invariant, the former breaks gauge invariance explicitly, but this is demonstrated to have little practical effect in most of the phase space.
This difference has a larger impact in extreme regions of phase space, such as those sampled by supersymmetry searches (see, for example, \textcite{Aad:2014qaa,Khachatryan:2016pup}). The ATLAS and CMS \tW cross-section measurements are tailored for the Born-level description of this process and thus not very sensitive to the difference between the DR and DS approaches, nevertheless a systematic uncertainty is assigned to account for the difference.

The \tW cross section has been calculated at NLO+NNLL (also called approximate N$^3$LO)~\cite{Kidonakis:2016sjf} and at NLO with \HATHOR~\cite{Aliev:2010zk,Kant:2014oha}, based on MCFM~\cite{Campbell:2005bb}. 
The NLO+NNLL calculation is based on a NLO \tW calculation~\cite{Zhu:2001hw} that removes the interference terms at the cross-section level. The MCFM calculation introduces a cut-off on the transverse momentum of the $b$-quark from gluon splitting, and the cross section is somewhat sensitive to this threshold. Table~\ref{tab:lhctWpred} compares the two predictions to each other. The NLO+NNLL prediction is quite a bit higher than the NLO calculation due to the $b$-quark cut-off in the latter.

\begin{table}[!htbp]
\begin{center}
\begin{tabular}{l|c@{\hskip 0.3in}c@{\hskip 0.3in}c}  
\tW & 7 TeV & 8 TeV & 13 TeV \\
cross section in pb & \\ \hline
NLO+NNLL & $17.0 \pm 0.7$ & $24.0 \pm 1.0$ & $76.2 \pm 2.5$  \\
NLO &  $13.2 \pm 1.4$ & $18.9 \pm 1.9$ & $60 \pm 6$  \\
\end{tabular}
\caption{Theoretical predictions for the \tW production cross sections at the LHC. 
The NLO+NNLL predictions~\cite{Kidonakis:2016sjf} have been calculated for a top-quark mass of 172.5 GeV and the uncertainties include scale and PDF~\cite{MMHT14} variations.
The NLO predictions have been prepared using the \HATHOR v2.1 program~\cite{Aliev:2010zk,Kant:2014oha} based on MCFM~\cite{Campbell:2009ss,Campbell:2005bb}. They are obtained at a top-quark mass of 172.5 GeV and the uncertainties include scale, PDF and $\alpha_S$~\cite{Botje:2011sn,MSTW2008NLO,Martin:2009bu,CT10,NNPDF23} variations. The cutoff threshold for the $b$-quark $p_T$ from gluon-splitting is set to 60~GeV.}
\label{tab:lhctWpred}
\end{center}
\end{table}

 The first evidence of \Wt production has been reported by the ATLAS and CMS collaborations using 7~TeV data~\cite{Aad:2012xca,Chatrchyan:2012zca}. The conventional $5\sigma$ threshold has been crossed with 8~TeV data~\cite{Chatrchyan:2014tua,Aad:2015eto}. 
More recently, the ATLAS collaboration measured the \Wt inclusive cross section at 13~TeV using 3~\fb of data collected in 2015~\cite{Aaboud:2016lpj}, and CMS reported a precision measurement of the \tW cross section at the same CM energy with 36~\fb of 2016 data~\cite{Sirunyan:2018lcp}. The cross section measurements at all three CM energies are in agreement with the SM calculation at NLO in QCD with NNLL corrections~\cite{Kidonakis:2013zqa} shown in Table~\ref{tab:lhctWpred}.

\begin{figure}[!htbp]
  \includegraphics[width=0.52\textwidth]{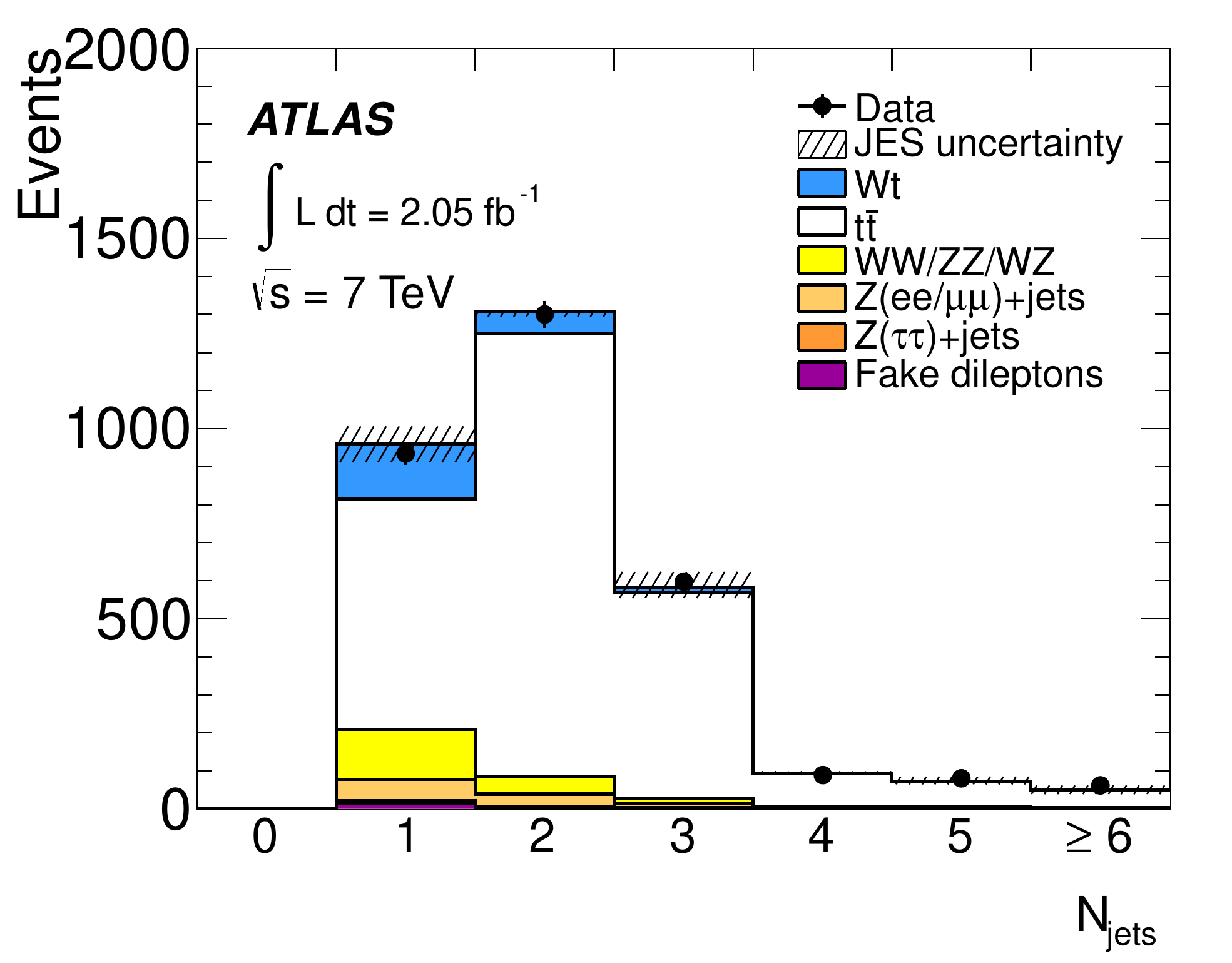}
\hfill
  \includegraphics[width=0.46\textwidth]{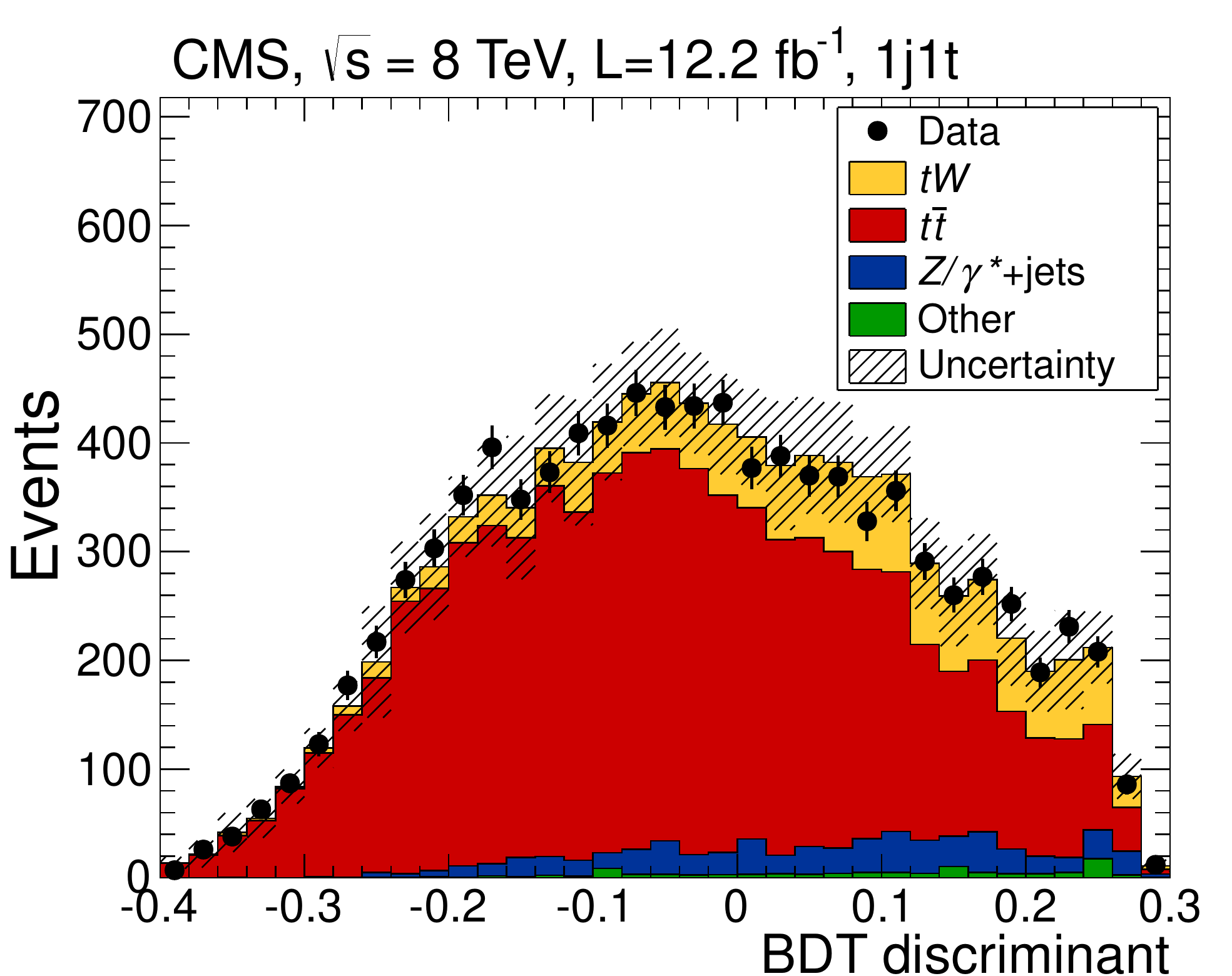}
  \caption{(Left) Distribution of the number of reconstructed jets in the ATLAS 7~TeV $tW$ analysis (from \textcite{Aad:2012xca}) and (right) BDT discriminant for 1-jet events in the CMS 8~TeV $tW$ analysis (from \textcite{Chatrchyan:2014tua}).}
  \label{fig:LHCtW}
\end{figure}

All these analyses are performed in the dilepton final state, exploiting the presence of two real $W$ bosons (the associated one, and the one from top-quark decay), by selecting events with two charged leptons (electrons or muons). The distribution of the number of reconstructed jets in the ATLAS 7~TeV analysis, shown in Fig.~\ref{fig:LHCtW}, shows that even in the signal region with one jet, the $tW$ signal is overwhelmed by a larger background from \ttbar production where one of the two $b$-quark jets is not reconstructed.
Measurements of this process in the $l+$jets final state, i.e., with one $W$~boson decaying leptonically and one hadronically, suffer from the combinatorial problem of quark-parton association and from the difficulty of discriminating the signal from the overwhelming \ttbar background~\cite{Ball:2007zza,Giorgi2016}. A measurement in the $l+$jets channel, however, would have the added value that the top quark/antiquark ratio would become accessible~\footnote{A top-quark-mass constraint allows to assign the charged lepton to either the top quark or the associated $W$ boson. Therefore, the charge of this lepton would provide discrimination between $t W^-$ and $\bar t W^+$ production. This is much more difficult, and so far unfeasible, in the dilepton final state, because of the presence of two neutrinos and an insufficient number of mass constraints to determine all the degrees of freedom.} and could be used as a handle to constrain \vtd, as an initial-state $d$-quark parton makes this ratio deviate from unity~\cite{Alvarez:2017ybk}.

The distributions of multivariate discriminants are used in a likelihood fit to extract the signal cross section. The fit utilizes multiple regions: Not only 1-jet, 1 $b$-tag events that have the largest fraction of \tW signal, see Fig.~\ref{fig:LHCtW}(left), but also 2-jet events with 1 or 2 $b$-tags, which are used to constrain the dominant background from \ttbar production and the large systematic uncertainties. In particular the \ttbar modeling uncertainties would otherwise swamp the precision of the signal measurement. The BDT distribution for the CMS 8~TeV analysis is shown in Fig.~\ref{fig:LHCtW}(right). The \tW signal appears at high discriminant values, with a s/b ratio approaching 1/1.

\begin{figure}[!htbp]
  \includegraphics[width=0.44\textwidth]{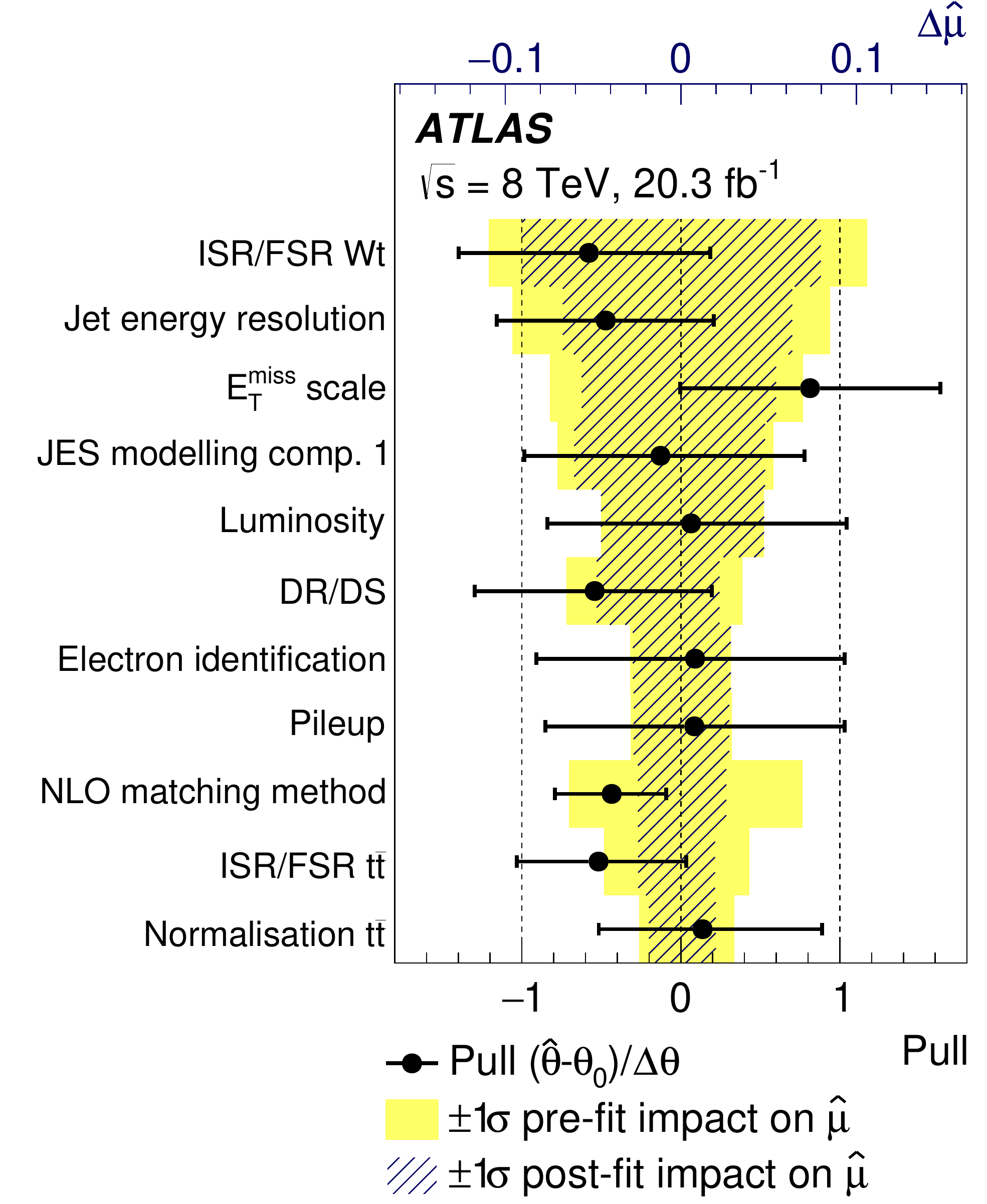}
\hfill \includegraphics[width=0.46\textwidth]{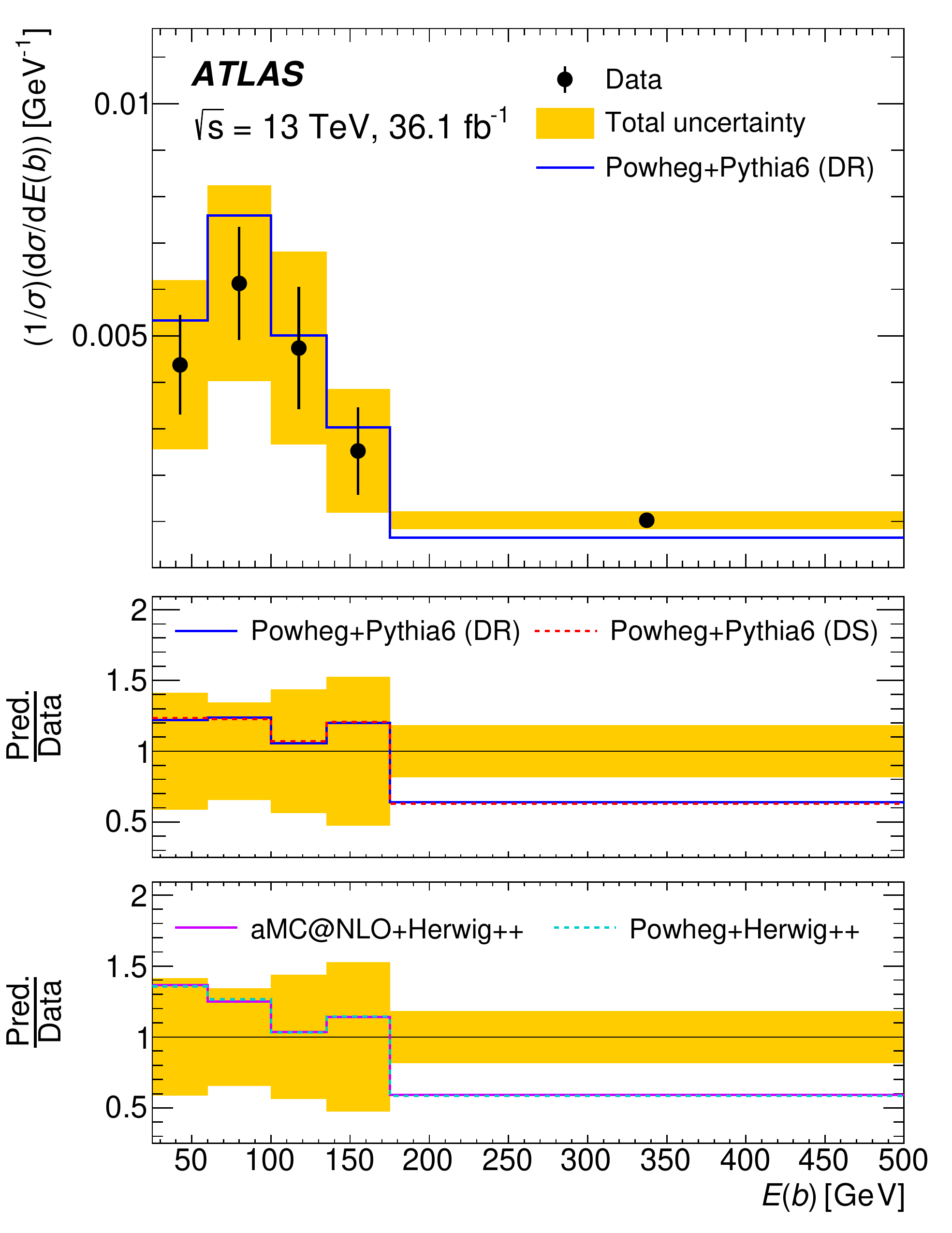}
 \caption{(Left) Constraints on the systematic uncertainties (pull, for which the nominal value is $0 \pm 1 \sigma$) and impact of those uncertainties on the \tW cross section measurement in the ATLAS 8~TeV $tW$ analysis (from \textcite{Aad:2015eto}).  The shaded and hashed areas refer to the top axis: the shaded bands show the initial impact of that source of uncertainty on the precision of the signal strength $\Delta \hat{\mu}$; the hatched areas show the impact on the measurement of that source of uncertainty, after the profile likelihood fit, at the $\pm 1 \sigma$ level. The points and associated error bars show the pull of the nuisance parameters and their uncertainties and refer to the bottom axis. A mean of zero and a width of 1 would imply no constraint due to the profile likelihood fit. 
(Right) Differential \tW cross section as a function of the energy of the $b$~quark measured by ATLAS at 13~TeV (from \textcite{Aaboud:2017qyi}).
}
  \label{fig:LHCtWsys}
\end{figure}

The largest systematic uncertainties in the \tW measurements arise from the modeling of \ttbar as mentioned above and the modeling of the \tW signal. Detector-modeling uncertainties from $b$-tag modeling, JES, and \etmiss modeling are also important. The systematic uncertainties affect not only the signal and background acceptance and the shape of the MVA distributions, but also result in migration between the different analysis regions. The sensitivity to this migration provides constraints on \ttbar uncertainties in the likelihood fit. 
This also has the consequence that the precision with which the signal can be measured is determined in part by the assumptions about correlations of modeling uncertainties between \ttbar and \tW, i.e., how much a strong constraint on \ttbar also applies to \tW. This includes the parton shower and ISR/FSR and other generator modeling uncertainties. 
The DR/DS uncertainty is not constrained in the fit but is also not a large uncertainty contribution. Figure~\ref{fig:LHCtWsys}(left) shows the impact of the systematic uncertainties on the ATLAS 8~TeV \tW measurement and how much each uncertainty is constrained in the fit. The detector-related uncertainties that have the largest impact are only moderately constrained and are shifted somewhat away from their nominal (0) value. The largest constraint is on the NLO matching method, which is obtained by comparing \tW and \ttbar samples generated with \POWHEG~\cite{Frixione:2007vw}  with those generated with \MCatNLO~\cite{Frixione:2002ik}, both interfaced to \HERWIG. This uncertainty, as well as that from ISR/FSR \ttbar, is pulled to a central value below zero and constrained because it shifts events between different jet multiplicities. Care needs to be taken when interpreting this pull. It implies that neither \MCatNLO nor \POWHEG  is able to model the kinematic properties of the \tW event selection. While \MCatNLO is more disfavored in the fit, both need improving. The modeling can be improved with the help of fiducial measurements at particle-level, see Section~\ref{sec:top-lhcb}.

At 7~TeV, ATLAS measures a \tW cross section of $16.8 \pm 5.7$~pb, while CMS measures $16^{+5}_{-4}$~pb.
At 8~TeV, ATLAS measures a \tW cross section of $23.0 \pm 3.8$~pb, while CMS measures $23.4 \pm 5.4$~pb.
At 13~TeV, ATLAS measures a \tW cross section of $94 \pm 28$~pb, while CMS measures $63.6 \pm 6.1$~pb. The cross sections measured by ATLAS and CMS are consistent with each other, and are quite close to each other at 7 and 8~TeV. At 13~TeV, the cross section measured by CMS is based on a dataset about ten times larger than the ATLAS one and about one standard deviation below the measurement by ATLAS (hence the smaller CMS uncertainty). All measurements are consistent with the theoretical predictions.

Differential measurements of the $tW$ cross section have also been reported as a function of the energy and invariant mass of different combinations of final-state objects by ATLAS at 13~TeV~\cite{Aaboud:2017qyi}. The kinematic distributions are unfolded to the particle level (defined by the presence of one lepton and one $b$-quark jet) and are compared to different MC simulations. This first differential measurement shows some conflict with the different MC generators, which all have about the same level of agreement with the data, as can be seen in the distribution of the energy of the $b$~quark from the top-quark decay in Fig.~\ref{fig:LHCtWsys}(right).

\subsubsection{\Wt plus \ttbar in fiducial regions}
\label{sec:top-lhcb}

\begin{figure}[!htbp]
  \includegraphics[width=0.52\textwidth]{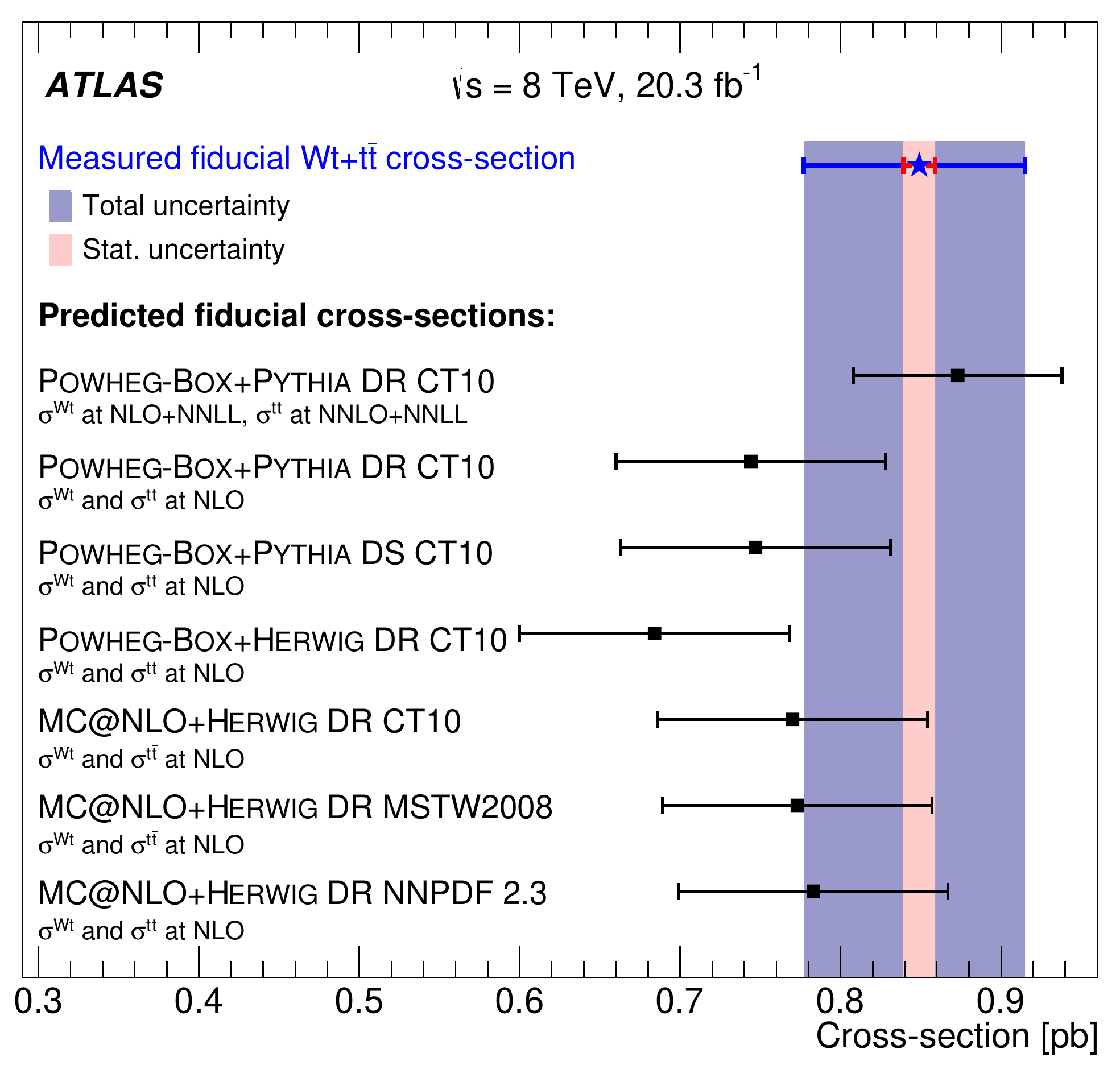}
\hfill
  \caption{Fiducial cross-section measurement in the ATLAS 8~TeV \tW analysis compared to theoretical predictions (from \textcite{Aad:2015eto}).}
  \label{fig:LHCtW2}
\end{figure}

 To reduce the dependence on the theory assumptions, the ATLAS collaboration reports a cross section in a fiducial detector acceptance defined by the presence of two charged leptons and exactly one $b$ jet at particle level~\cite{Aad:2015eto}. This signal definition encompasses not only \Wt production but also \ttbar production where one of the final-state $b$ quarks is outside of the acceptance. The result is shown in Fig.~\ref{fig:LHCtW2} and is found to be in agreement with the predictions from two different NLO matrix-element generators (\POWHEG and \MCatNLO) matched to two different parton-shower generators (\PYTHIA~6~\cite{Sjostrand:2006za} and \HERWIG~6~\cite{Corcella:2000bw}), the DR and DS approaches, and a variety of PDF sets. In this comparison, where the relative normalization of \tW and \ttbar is important, the measurement has the best compatibility with the simulation when \tW is normalized to the NLO+NNLL calculation and \ttbar is normalized to the NNLO+NNLL calculation. In particular the \ttbar normalization plays an important role. While no conclusion about individual generators can be drawn given the size of the uncertainties, it is clear that in the fiducial measurement, \POWHEG predicts a lower cross section than \MCatNLO, when both are interfaced to \HERWIG.

\begin{figure}[!htbp]
  \includegraphics[width=0.52\textwidth]{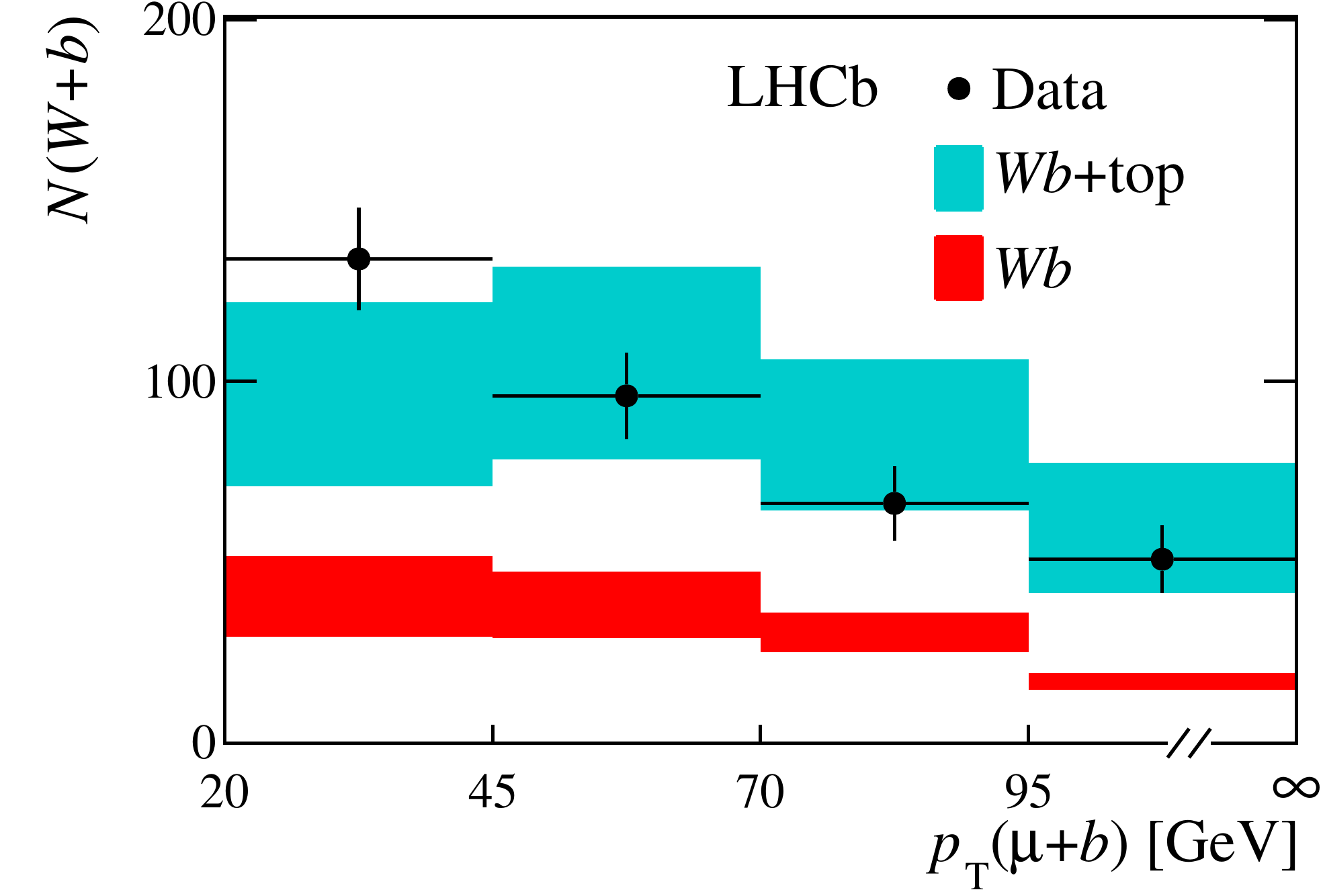}
  \caption{Number of events with a $W$-boson and a $b$ quark observed by LHCb as a function of $p_T(\mu + b)$, compared to expectations with and without top-quark signal ($t\bar t + tW$) at NLO accuracy (from \textcite{Aaij:2015mwa}).}
  \label{fig:LHCb-top}
\end{figure}

 Although top-quark physics was not among the design goals of the LHCb experiment, it has been remarked that, by accessing a kinematical region beyond the reach of ATLAS and CMS, studies of top-quark production with the LHCb data may have a strong impact on constraining parton distribution functions (PDF)~\cite{Gauld:2013aja}, or indirectly probe anomalous top-quark couplings in single and pair production in a complementary way with respect to multi-purpose experiments, in particular in BSM scenarios where top-quark production proceeds via $t$-channel exchange of a new low-mass particle~\cite{Kagan:2011yx}.
 Using samples of 1.0 and 2.0~fb$^{-1}$ collected at CM energies of 7 and 8~TeV in 2011 and 2012 respectively, the \textcite{Aaij:2015mwa} achieved the first observation of top-quark production in the forward region defined by its acceptance to muons ($2.0 < \eta < 4.5$) and to $b$ jets ($2.2 < \eta < 4.2$), see Fig.~\ref{fig:LHCb-top}. Inclusive top-quark production cross sections were measured in a fiducial particle-level region that includes contributions mainly from \ttbar and also from \Wt and presented together with differential yields and charge asymmetries. Results are in agreement with SM predictions at NLO accuracy.

\subsubsection{$s$-channel}
\label{sec:schannel}

The $s$-channel process, Fig.~\ref{fig:FG}(b), poses particular challenges at the LHC because of the very small cross section in comparison with backgrounds with very similar final state, a situation comparatively worse than at Tevatron. The theoretical prediction for $s$-channel production has been calculated at NLO with NNLL corrections~\cite{Kidonakis:2010tc} and at NLO~\cite{Aliev:2010zk,Kant:2014oha,Campbell:2004ch,Heim:2009ku}. Table~\ref{tab:lhcschanpred} compares the two predictions to each other. The cross section rises by only a factor 2 from 8 to 13~TeV, making this process even harder to observe in Run~2 than in Run~1 at the LHC.

\begin{table}[!htbp]
\begin{center}
\begin{tabular}{l|c@{\hskip 0.3in}c@{\hskip 0.3in}c}  
$s$-channel  & 7 TeV & 8 TeV & 13 TeV \\
cross section in pb & \\ \hline
NLO+NNLL & \\
~~$t$                     & $  3.1 \pm 0.1 $ & $ 3.8 \pm 0.1 $ & $ 7.1 \pm 0.2 $  \\
~~$\overline{t}$     & $ 1.4 \pm 0.1 $ & $ 1.8 \pm 0.1 $ & $ 4.1 \pm 0.2 $  \\
~~$t+\overline{t}$ & $ 4.6 \pm 0.2 $ & $ 5.6 \pm 0.2 $ & $ 11.2 \pm 0.4 $  \\
NLO & \\
~~$t$                     & $ 2.8 \pm 0.1 $ & $ 3.3 \pm 0.1 $ & $ 6.3 \pm 0.4 $  \\
~~$\overline{t}$     & $ 1.5 \pm 0.1 $ & $ 1.9 \pm 0.1 $ & $ 4.0 \pm 0.2 $  \\
~~$t+\overline{t}$ & $ 4.3 \pm 0.2 $ & $ 5.2 \pm 0.2 $ & $ 10.3 \pm 0.2 $  \\
\end{tabular}
\caption{Theoretical predictions for the $s$-channel production cross sections at the LHC. 
The NLO+NNLL predictions~\cite{Kidonakis:2010tc} have been calculated for a top-quark mass of 173 GeV and the uncertainties include scale and PDF~\cite{MSTW2008NLO} variations.
The NLO predictions have been prepared using the \HATHOR v2.1 program~\cite{Aliev:2010zk,Kant:2014oha} based on MCFM~\cite{Campbell:2004ch}. They are obtained at a top-quark mass of 172.5 GeV and the uncertainties include scale, PDF and $\alpha_S$~\cite{Botje:2011sn,MSTW2008NLO,Martin:2009bu,CT10,NNPDF23} variations. }
\label{tab:lhcschanpred}
\end{center}
\end{table}

The ATLAS and CMS $s$-channel analyses select events with one isolated electron or muon, significant \MET and/or large \mT, and two jets, both b-tagged. Main backgrounds are \ttbar, $W$+jets, QCD multi-jet production, and the other single top-quark processes. Several orthogonal control regions with different multiplicities of jets and/or $b$-tagged jets are used to measure these backgrounds {\it in situ} or to validate the Monte Carlo models used for their predictions,  or to constrain the main experimental systematics (e.g., $b$-tagging efficiency).

\begin{figure}[!htbp]
  \includegraphics[width=0.48\textwidth]{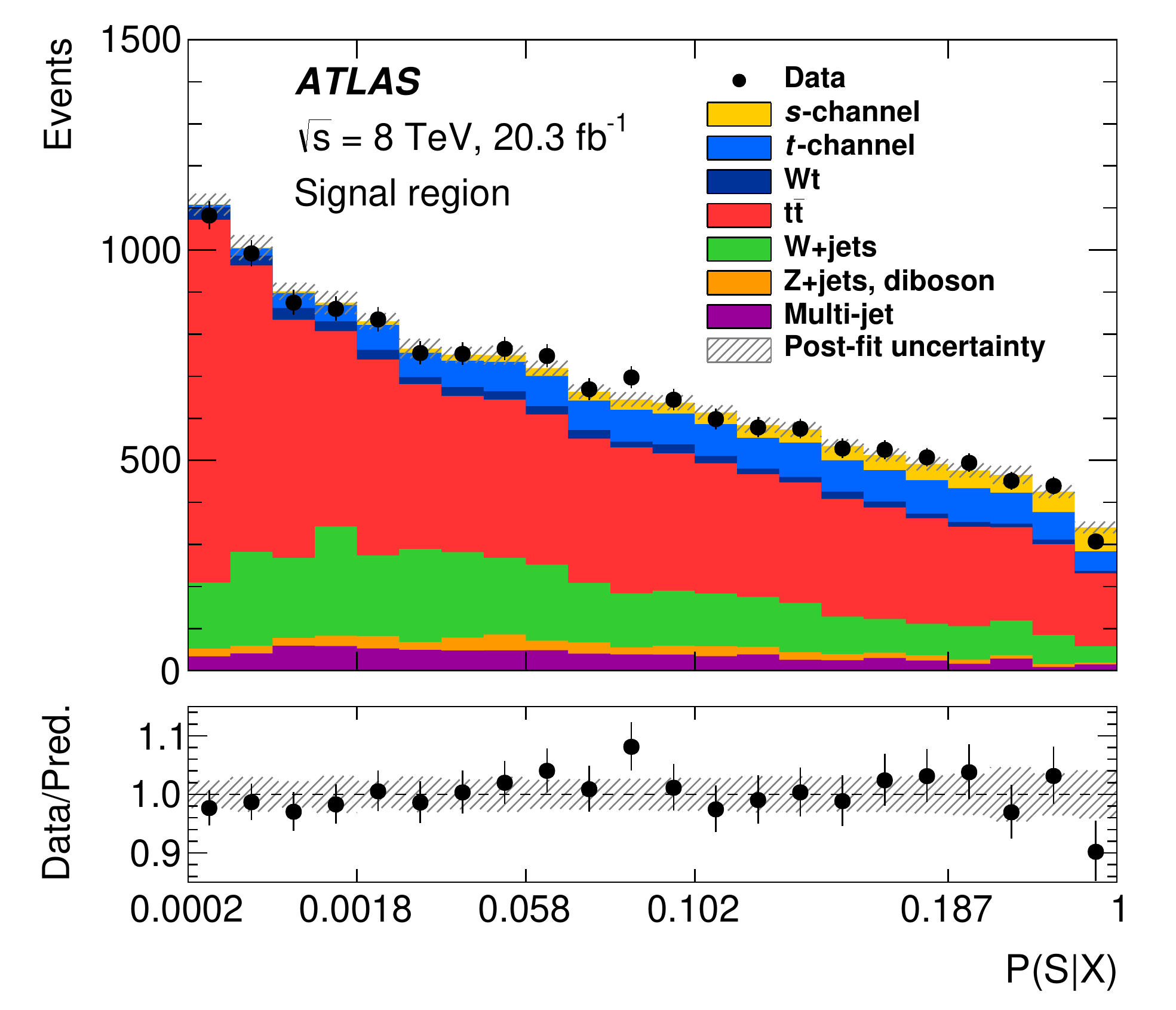}
\hfill
 \includegraphics[width=0.48\textwidth]{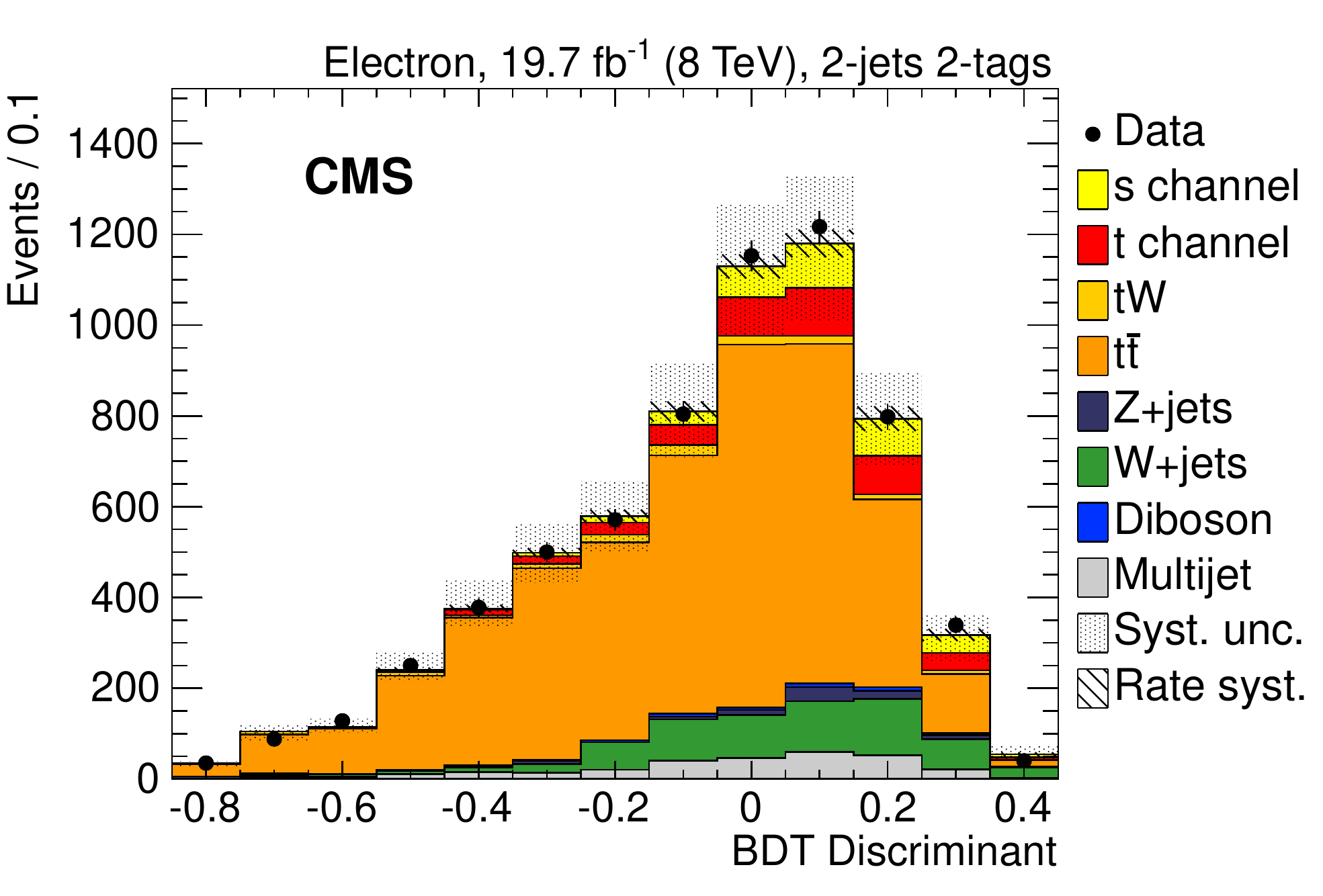}
  \caption{(Left) Matrix element discriminant, expressed as the probability for an observed event $X$ to be a signal event ($S$), $P(S|X)$, in the ATLAS 8~TeV $s$-channel analysis (from \textcite{Aad:2015upn}) and (right) BDT discriminant in the CMS 8~TeV $s$-channel analysis (from \textcite{Khachatryan:2016ewo}).}
  \label{fig:LHCschan}
\end{figure}

With the 7~TeV dataset, ATLAS and CMS were not able to observe the $s$-channel process and only set upper limits on its production cross section~\cite{schanAT7,Khachatryan:2016ewo}. With the 8~TeV dataset, ATLAS first published a search~\cite{Aad:2014aia}, and then improved the sensitivity of the analysis to report evidence for $s$-channel single top-quark production~\cite{Aad:2015upn}. The latter analysis employs a matrix element (ME) method (see Section~\ref{sec:tev-legacy}) to optimize the sensitivity to the $s$-channel signal. Here, the likelihood for each event to originate from the signal or one of the backgrounds is computed based on the four-vectors of the particles in the corresponding LO Feynman diagrams. Un-observed four-vector components and detector resolution effects are integrated over, resulting in large computing-time requirements.  The final ME discriminant for the ATLAS $s$-channel analysis is shown in Fig.~\ref{fig:LHCschan}(left). The background is subtracted from the data in this figure, making the otherwise small signal visible.
CMS  measured the cross section simultaneously at 7 and 8~TeV~\cite{Khachatryan:2016ewo}, taking advantage of the correlations between the different CM energies to constrain backgrounds and systematic uncertainties. The signal is separated from the large backgrounds using a BDT discriminant, which is shown in Fig.~\ref{fig:LHCschan}(right), with the small $s$-channel signal visible on the right-hand side of the distribution.

The $s$-channel analyses are limited by large backgrounds in the signal region, in particular from \ttbar as Fig.~\ref{fig:LHCschan} shows. The bins with the largest signal fraction correspond to unusual phase-space regions for the largest backgrounds, thus very large amounts of simulated events are necessary for the analysis. The MC statistics uncertainty is the largest of all systematic uncertainties. For both the ATLAS and CMS analyses, large detector-related uncertainties arise from JES and b-tag modeling, and the theory modeling uncertainties are dominated by $t$-channel and \ttbar modeling uncertainties.

At 7~TeV, the limit set by ATLAS on the $s$-channel cross section is 26.5~pb (20.5~pb expected). The limit set by CMS is 31.4~pb (20.2~pb expected). 
At 8~TeV, ATLAS reported evidence with an observed (expected) significance of 3.2 (3.9) standard deviations. The measured cross section is $4.8 \pm 1.8$~pb. The CMS limit at 8~TeV is 28.8~pb (15.6~pb expected). The combined CMS 7+8~TeV analysis, which assumes the SM ratio between the cross sections at the two CM energies, has an observed (expected) significance of 2.5 (1.1) standard deviations. The measured cross section value for CMS at 8~TeV is $13.4 \pm 7.3$~pb. The limits and measurements are all consistent with each other and with the theory predictions. The two analyses have similar selections and amounts of signal and background, but the Matrix-element based discriminant in use by ATLAS is able to better separate the single top-quark signal from the large backgrounds.
The $s$-channel measurements will improve with the large Run~2 dataset and better understanding of the theory modeling for \ttbar and $t$-channel single top-quark production.

\subsubsection{$Z$-associated (\tZ)}
\label{sec:zt}

\begin{figure}[!htpb]
  \includegraphics[width=0.19\textwidth]{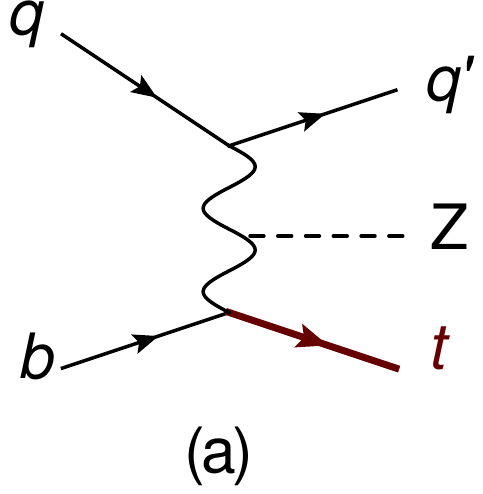}
\hspace{1cm} 
\includegraphics[width=0.2\textwidth]{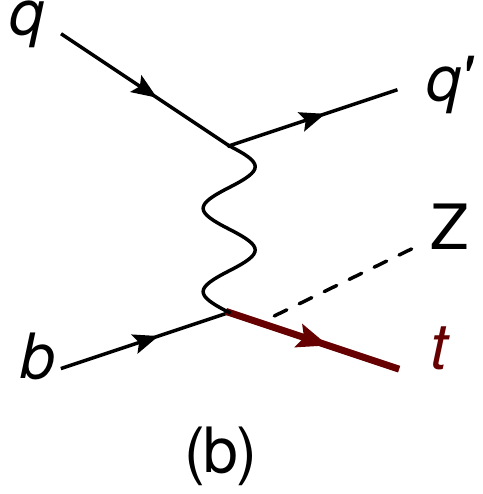}
  \caption{Representative Feynman diagrams for electroweak single top-quark production in association with a $Z$ boson (\tZ), (a) with the $Z$ boson coupling to the exchanged $W$~boson and (b) the $Z$~boson coupling to the top quark.}
  \label{fig:FGtZ}
\end{figure}

The cross section for single top-quark production at the LHC is sufficiently large, in particular in the $t$-channel mode, that it is possible to observe the coupling to additional particles in single top-quark events. Figure~\ref{fig:FGtZ} shows an example of this where single top quarks in the $t$-channel mode are produced in association with a $Z$~boson. This process probes both the $WZ$ coupling and the top-$Z$ coupling. 
The production cross section for this process has been calculated at NLO~\cite{Campbell:2013yla}. At 8~TeV, the cross section is $236 \pm 15$~fb, while at 13~TeV it is $800 \pm 60$~fb.
 
The signature of \tZ production is that of $t$-channel single top-quark production, plus a $Z$~boson. Thus, the description of the process, background estimates, kinematic properties described in Section~\ref{sec:tchannel} all apply here, except that a $Z$~boson is added to each. The experimental signature consists of a leptonically decaying top-quark, with a central high-$p_T$ $b$~quark, and a forward light quark, plus a leptonically decaying $Z$~boson. The main backgrounds are $WZ$+jets (instead of $W$+jets), $Z$+jets with a jet mis-identified as an isolated lepton (instead of multi-jets with a mis-identified lepton), and $ttZ$ (instead of \ttbar). The requirement of the presence of the $Z$~boson reduces the event rates for all of these processes by three orders of magnitude compared to Section~\ref{sec:tchannel}. In addition, the requirement of a leptonically decaying $Z$~boson reduces the rate by about another order of magnitude. Selecting events in a narrow region around the $Z$~boson mass peak is important to effectively reject non-$Z$ backgrounds, and this is not viable for hadronically decaying $Z$~bosons, for which there is an overwhelmingly large QCD background. Final states with hadronically decaying top quarks and leptonically decaying $Z$~bosons is similarly challenging, analogous to $t$-channel production, where hadronic top quark decays are also overwhelmed by a large QCD background.

Using the full data set at 8~TeV, the CMS collaboration presented a search for the \tZ production mechanism~\cite{Sirunyan:2017kkr}, exploiting the very clean signature of three charged leptons (electrons or muons), two of them consistent with originating from the decay of a $Z$ boson, accompanied by a $b$ quark, a forward jet, and significant \MET. About 16 signal events are expected with basic selection requirements, compared to the 17,700 events selected in the 8~TeV $t$-channel analysis (see Table~\ref{tab:tevlhcacc}). The signal is separated from the background using a BDT discriminant, and the cross section is measured in a fit to the BDT output and to the $W$ transverse mass in a control region to control the systematic uncertainties and backgrounds. The observed significance is 2.4 standard deviations (1.8 standard deviations expected), and the measured cross section is $10^{+8}_{-7}$~fb. The 95\% CL limit on the \tZ signal is 21~fb, consistent with the theory expectation.

\begin{figure}[!htbp]
  \includegraphics[width=0.56\textwidth]{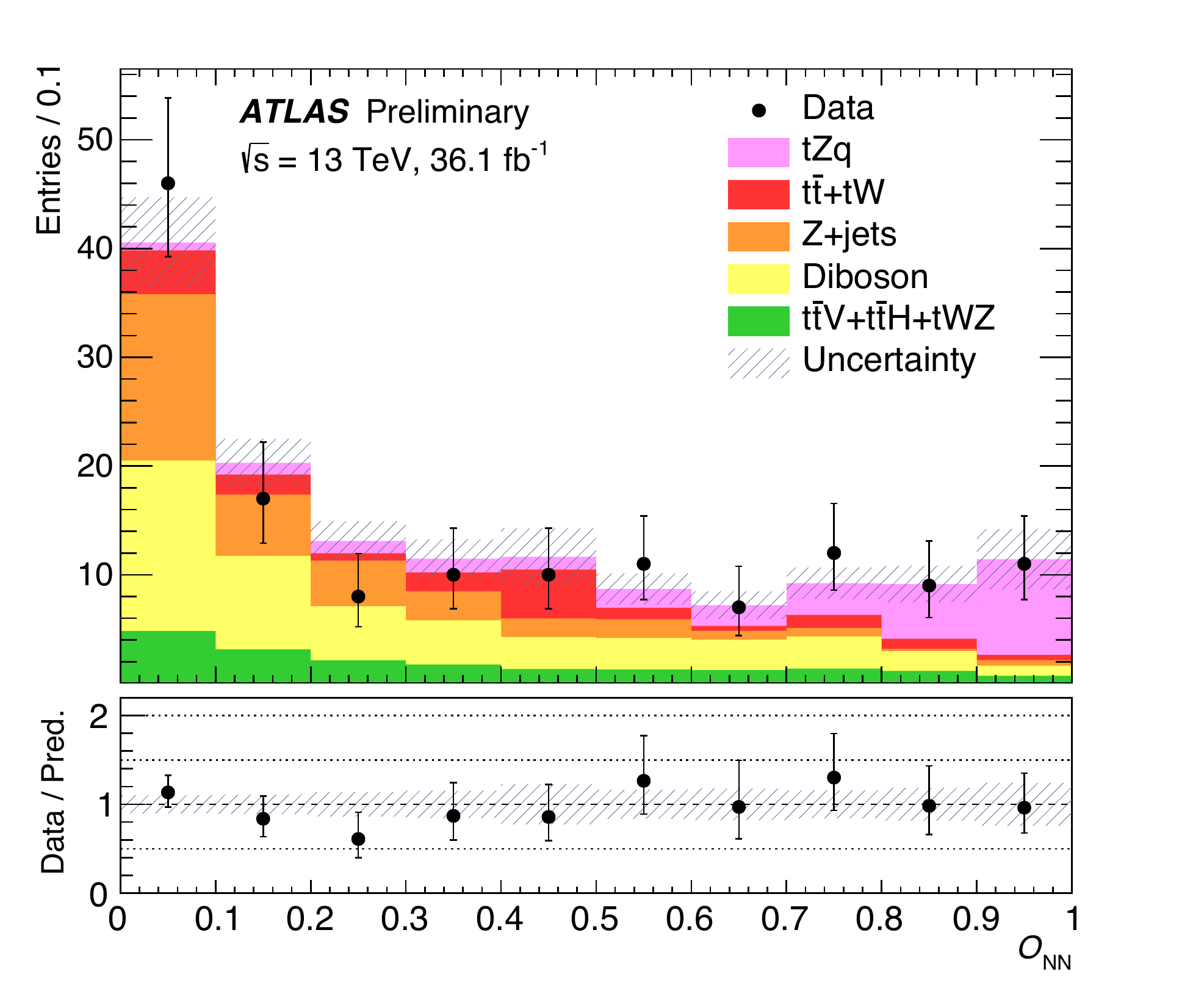}
  \caption{Post-fit neural network discriminant distribution in the ATLAS search for the \tZ process in 13~TeV data (from \textcite{Aaboud:2017ylb}).}
  \label{fig:LHCtZ1}
\end{figure}

ATLAS reported evidence for \tZ production with 13~TeV data~\cite{Aaboud:2017ylb}, also relying on the three-lepton final state. Exactly two jets are required, one $b$-tagged jet and one light-quark jet. This selects 143 events in data with 35 signal events expected from a LO simulation in the 4FNS rescaled to NLO. A neural network is utilized to separate the \tZ signal from the background, and the signal is extracted from a profile likelihood fit to the NN discriminant in the signal region. The post-fit NN distribution is shown in Fig.~\ref{fig:LHCtZ1}. The observed (expected) significance is 4.2 (5.4) standard deviations. The measured cross section is $600 \pm 170 \stat \pm 140 \syst$~fb.

\begin{figure}[!htbp]
  \includegraphics[width=0.9\textwidth]{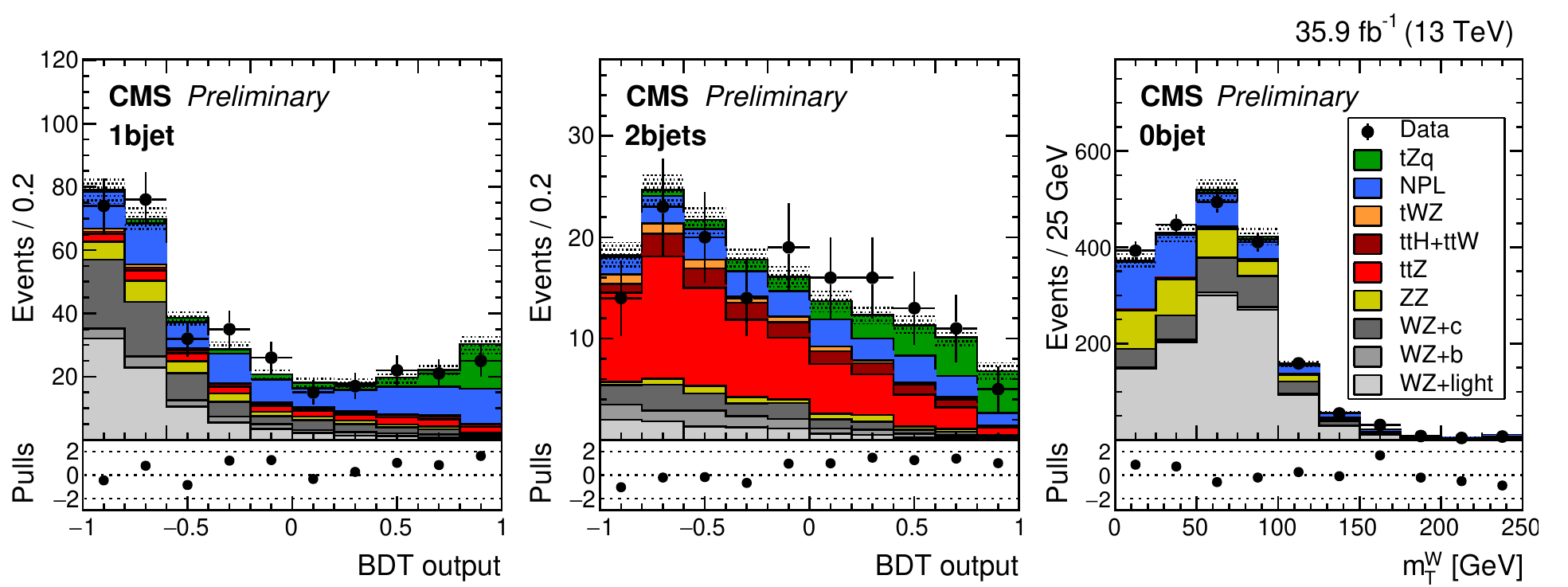}
\hfill
  \caption{Post-fit discriminant distribution in the CMS \tZ analysis at 13~TeV: (left) BDT for 1-$b$-jet events, (middle) BDT for 2-$b$-jet events and (right) $W$ transverse mass distribution for 0-$b$-jet events. (from \textcite{Sirunyan:2017nbr}).}
  \label{fig:LHCtZ2}
\end{figure}

CMS also reported evidence for \tZ production with 13~TeV data~\cite{Sirunyan:2017nbr}. Three-lepton events are selected separately for each lepton combination, and two or three jets are required, with 1-$b$-jet events defining the signal region and 2-$b$-jet and 0-$b$-jet events defining two control regions that are also included in the final likelihood fit to constrain uncertainties. The signal region has 343 data events, 25 of which are expected to come from the \tZ signal according to a NLO simulation of the signal in the 5FNS. The discriminant used in each of the three regions is shown in Fig.~\ref{fig:LHCtZ2}. 
The observed (expected) significance is 3.7 (3.1) standard deviations. The measured cross section, including only leptonic $Z$~boson decays, is $123~^{+33}_{-31} \stat~{}^{+29}_{-23} \syst$~fb. This corresponds to an inclusive cross section of $1040 \pm 370$~fb. The ATLAS and CMS measurements are consistent with each other within about one standard deviation. ATLAS observes a small deficit compared to the theory prediction, while CMS observes an excess. The expected signal event yield in the highest bin of the MVA distribution is comparable for the two experiments, while the background is larger for CMS, in part due to the better $b$-tag performance in the ATLAS analysis thanks to their upgrade of the pixel detector at the beginning of Run~2, see Section~\ref{sec:atlas} (the corresponding upgrade was made by CMS at the beginning of 2017).

The approaches followed by the two experiments differ under a few aspects, each exemplifying a particular issue in single top analyses in general. The most important differences are the inclusion of three signal regions in the CMS analysis compared to just one for ATLAS, the treatment of the non-prompt lepton (NPL) background, and the signal simulation. 
\begin{itemize}
\item The background in the highest signal bins is larger for CMS than for ATLAS, thus CMS benefits from profiling background normalizations and systematic uncertainties that affect the background estimate, which would have less of an impact on the ATLAS analysis.
\item It can be seen, by comparing the ATLAS (Fig.~\ref{fig:LHCtZ1}) and CMS signal regions (Fig.~\ref{fig:LHCtZ2}, left), that the NPL background is larger in the high-discriminant region for CMS than for ATLAS. This corresponds to \ttbar dilepton and $Z$+jets events where an additional jet is mis-identified as an isolated lepton. The ATLAS approach is to estimate separately the $\ttbar$ (real top quark, misidentified $Z$~boson) and $Z$+jets (misidentified top quark, real $Z$~boson) backgrounds, both from simulation samples normalized to and checked in control regions in data. Both samples are included in the MVA training. CMS groups these sources together and focuses instead on the origin of the NPL separately for each lepton flavor. This results in a smaller NPL uncertainty, but the background is larger in the high-discriminant region. 
\item The signal simulations of the two experiments also differ, affecting the MVA training. Although both normalize the event yields to NLO predictions, the simulation samples generated by ATLAS are at LO in the 4FNS, while those simulated by CMS are at NLO in the 5FNS. Generating events at LO avoids negative event weights and the associated MC statistics issues, making it easier to obtain optimal MVA training. Generating events at NLO gives improved modeling of the kinematic properties of the signal and smaller signal-modeling uncertainties. However, a large fraction of simulated events in the signal region that have negative weights results in a non-optimal MVA. 
\item A significant fraction of events have three jets in the final state, the two from the Feynman diagram shown in Fig.~\ref{fig:FGtZ}, plus the forward $b$ jet shown in Fig.~\ref{fig:FG23}(b) or a gluon. This migration to 3-jet events is more pronounced at NLO in the 5FNS. This motivates the inclusion of 3-jet events in the CMS analysis, which recovers signal events, but also adds more $\ttbar Z$ background, similar to 3-jet events in the $t$-channel analysis. 
\end{itemize}

It should be stressed that the modeling differences affect the expectations, and indirectly the selection strategy, but do not bias the cross-section measurement itself. 


\subsection{Summary of the inclusive cross-section measurements}

\begin{figure}[!ht]
 \begin{center}
  \includegraphics[width=0.9\textwidth]{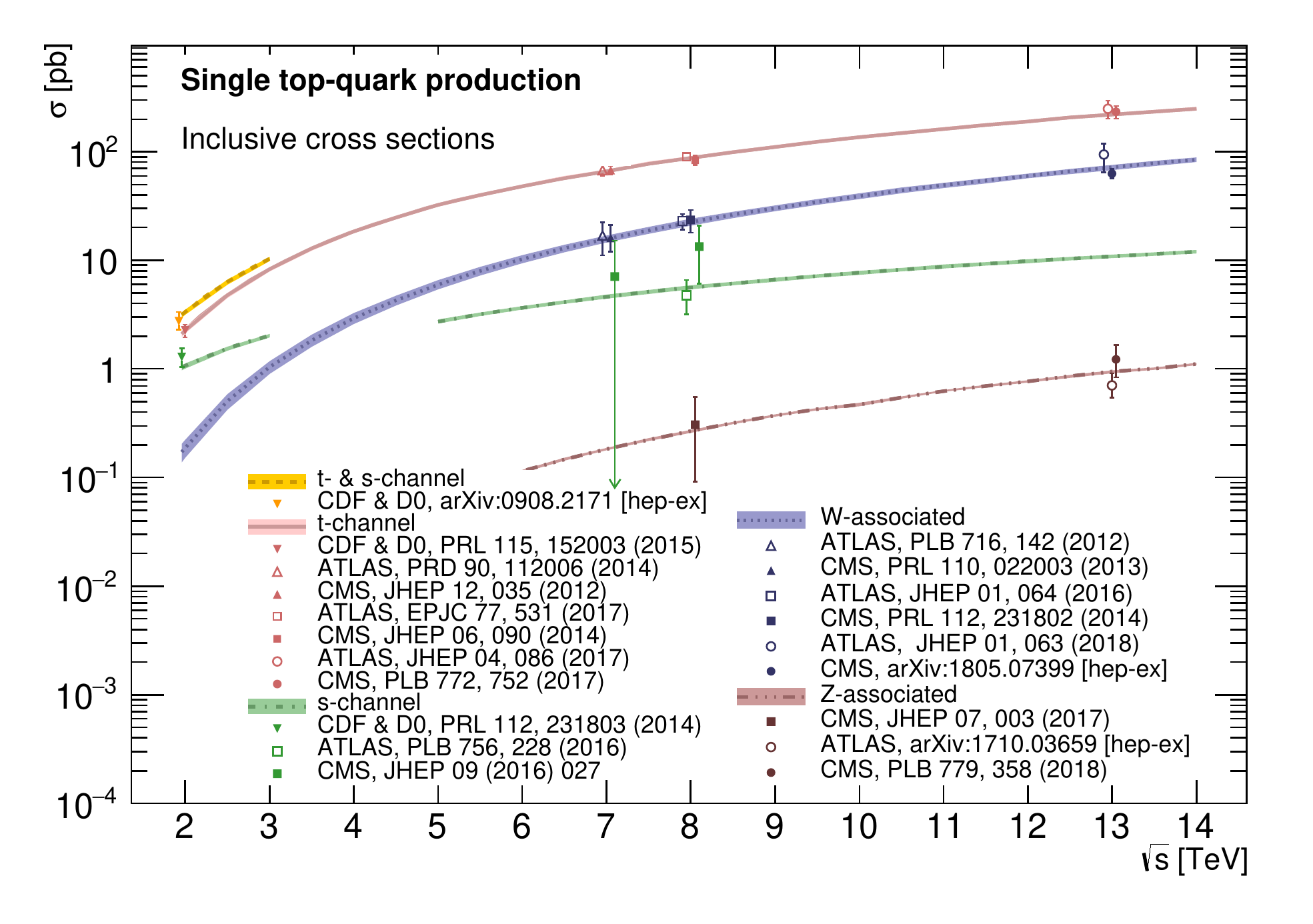}
  \caption{Summary of Tevatron and LHC measurements of the inclusive single top-quark production cross sections in $t$-channel, $s$-channel, \tW and \tZ production. The measurements are compared to theoretical calculations based on NLO QCD complemented with NNLL resummation. The full theory curves as functions of the CM energy are calculated as in Refs.~\cite{Kidonakis:2011wy,Kidonakis:2010ux,Kidonakis:2010tc} for $t$-channel, $s$-channel, and \tW, and are calculated with \aMCatNLO (v.254)~\cite{Alwall:2014hca} for \tZ. The curves for $s$-channel and the sum of $s$- and $t$-channel are calculated for \ppbar collisions up to 3~TeV and for pp collisions beyond; for $t$-channel, \tW and \tZ the curves for pp and \ppbar coincide at the considered accuracy.}
\label{fig:sqrts}
 \end{center}
\end{figure}

Figure~\ref{fig:sqrts} summarizes all of the experimental measurements of the inclusive cross sections for single top-quark production at the Tevatron and at the LHC. The measurements are compared to the NLO+NNLL predictions for $t$-channel, \tW and $s$-channel, and to a NLO calculation with \MCatNLO for \tZ, using the NNPDF3.0 PDF set~\cite{NNPDF30}.

\begin{figure}[!ht]
 \begin{center}
  \includegraphics[width=0.6\textwidth]{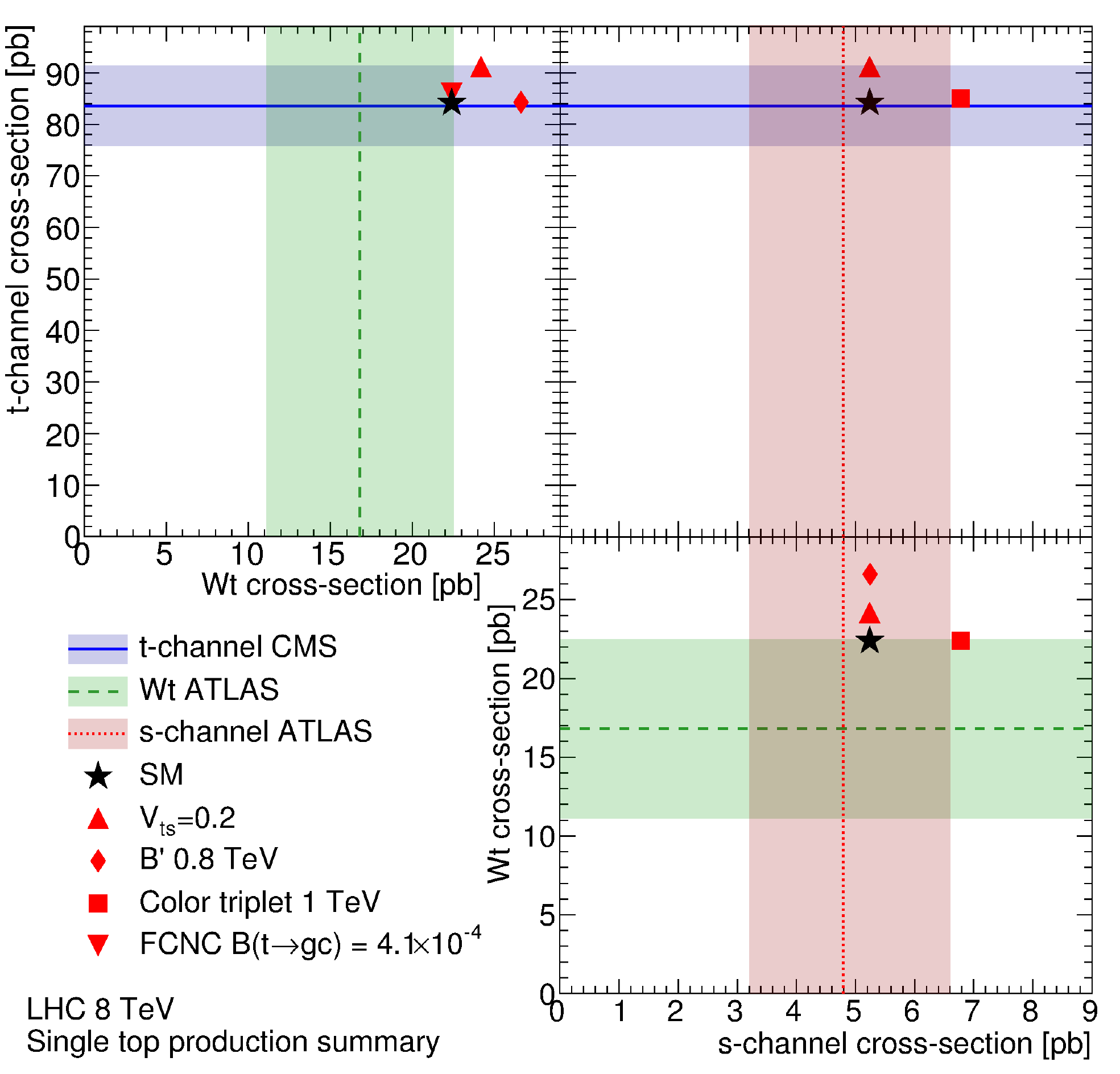}
  \caption{Inclusive single top-quark cross sections measured at 8 TeV at the LHC, $t$-channel vs \tW and $s$-channel and \tW vs $s$-channel. The SM theory predictions are calculated as in Refs.~\cite{Kidonakis:2011wy,Kidonakis:2010ux,Kidonakis:2010tc}.
Also shown are example BSM scenarios: A model with CKM element $V_{ts}=0.2$~\cite{Alwall:2006bx}, a vector-like fourth generation quark with chromo-magnetic couplings~\cite{Nutter:2012an}, a color triplet~\cite{Drueke:2014pla}, and flavor-changing neutral current interactions of the top quark with the gluon and the charm quark~\cite{AguilarSaavedra:2008zc}.}
\label{fig:xs2d}
 \end{center}
\end{figure}

Figure~\ref{fig:xs2d} visualizes the most precise single top-quark cross section measurements at 8 TeV at the LHC for the three dominant channels, displayed versus each other. For each channel only one result from either ATLAS or CMS is shown, thus the correlations between individual measurements can be assumed to be small. The measurements are compared to examples of new physics models that lead to deviations in one or more of the cross sections. If the CKM matrix is not unitary, then deviations from 1 are possible for $V_{tb}$, and in turn, large non-zero values are possible for $V_{td}$ and $V_{ts}$~\cite{Alwall:2006bx}. Here, we calculate the corrections to the single top-quark cross sections for a value of $V_{ts}=0.2$, keeping $V_{td}=0$ and thus setting $V_{tb}=0.98$. Thus, the impact of this model on the top-quark decay is not detectable given the uncertainty of the branching ratio of $t\to Wb$ (see Section~\ref{sec:vtb}), and only the production cross sections for $t$-channel and \tW are increased. As another example, a vector-like fourth-generation quark $B'$ with a mass of 0.8~TeV and chromo-magnetic couplings~\cite{Nutter:2012an} modifies the \tW production cross section but only has a negligible impact on $t$-channel and $s$-channel production. A color triplet with a mass of 1~TeV decays to $tb$ and thus enhances the $s$-channel cross section but has no effect on $t$-channel or \tW. And finally, a small FCNC interaction corresponding to a branching ratio of $4.1\times 10^{-4}$ for $t\rightarrow gc$~\cite{AguilarSaavedra:2008zc} increases the $t$-channel cross section but has no impact on \tW or $s$-channel. It should be noted that for all of these examples, a proper evaluation of the sensitivity includes not just the modification of the cross section but also of the experimental acceptance. In particular, since the experimental analyses use MVA techniques, the sensitivity is mainly to SM-like production mechanisms. Dedicated searches, such as those presented in the next sections, are generally more sensitive for each possible BSM scenario.

\FloatBarrier

%% file: tex/extraction.tex
\section{SM parameter extraction and searches for new physics leading to anomalous couplings}
\label{sec:params}

Since the mass of the top quark is of the order of the electroweak symmetry-breaking scale ($|\yt|\approx 1$, where $y_t$ is the top-quark Yukawa coupling), several new-physics models assign a special role to the top quark, with the consequence of typically predicting larger anomalies in the top-quark sector than for other quarks. 
Examples include top-flavor models with a seesaw mechanism~\cite{He:1999vp}, top-color seesaw models~\cite{Dobrescu:1997nm}, models with vector-like quarks~\cite{Okada:2012gy}, and others. 

The large data sets accumulated so far allow the use of single top-quark events as tools to constrain the parameters of the SM and to search for evidences of new physics, directly and indirectly. Beyond measuring the cross section, which provides access to the CKM matrix element \vtb, single top-quark events are now also used to measure asymmetries and angular correlations with increasing complexity. 
The $t$-channel production mode has the largest production cross section and the smallest background and is thus the only channel where these measurements have been made so far. 
These measurements provide indirect limits on effective field theory couplings of the top quark to the $W$~boson and other bosons~\cite{AguilarSaavedra:2018nen}.

\subsection{Constraints on \vtb and other CKM matrix elements}
\label{sec:vtb}

The moduli of the elements of the CKM matrix that connect the top quark with the down-type quarks, \vtd, \vts, and \vtb, are precisely determined from measurements of $B$-meson oscillations and loop-mediated rare $K$ and $B$ decays~\cite{Charles:2004jd}. 
From these data, and with some model assumptions such as the existence of only three generations of quarks and the absence of non-SM particles in the loops~\cite{Alwall:2006bx}, the value of \vtb is derived with a precision of order $10^{-5}$: $\vtb = 0.999097\pm 0.000024$~\cite{PDG2016}. 
The strong reliance of this derivation on the aforementioned assumptions motivates alternative inferences based on different sets of hypotheses. 
There is interest, for example, in exploring the possibility that a hypothetical heavier quark-like particle, such as a fourth-generation up-type quark or a heavy vector-like quark~\cite{Aguilar-Saavedra:2013qpa} (both named \tprime in the following) mixes with the top quark, yielding a lower value of \vtb than expected from $3\times 3$ unitarity. 
 Mixing may happen not only with sequential replicas of the known quarks, easily accommodated in the SM framework but severely constrained by the Higgs cross section measurements~\cite{Lenz:2013iha}), but in general with any hypothetical quark-like particle with the appropriate quantum numbers. Differently from the new-generations case, the effective mixing matrix may be rectangular, as in the case of vector-like quarks~\cite{Okada:2012gy,Aguilar-Saavedra:2013qpa}. 
 While the sum $\vtd^2+\vts^2+\vtb^2+\vtbprime^2$ and, a fortiori, the sum $\vtd^2+\vts^2+\vtb^2$ is bound to be $\le 1$ also in the extended matrix, the constraints on \vtd and \vts derived from precision physics~\cite{PDG2016} do not hold when their underlying assumptions (e.g., no non-SM particles in the loops) are relaxed~\cite{Alwall:2006bx}.

\textcite{Swain:1997mx} made a first attempt to extract \vtb without relying on $3\times 3$ unitarity, using electroweak loop corrections, in particular from the $Z\to b\bar b$ branching ratio, and combining several electroweak data from LEP, SLC, the Tevatron, and neutrino experiments, to obtain $\vtb = 0.77^{+0.18}_{-0.24}$. 
\textcite{Alwall:2006bx} applied the same principle to derive a lower limit on the mixing angle between the top quark and a \tprime from the branching fraction of the $Z$ boson into $b$ quarks measured at LEP and SLD.

Another complementary approach links \vtb with measurements of the ratio 
 $\Rb\equiv \RbDefOne$ in \ttbar events~\cite{Abazov:2011zk,Aaltonen:2013luz,Aaltonen:2014yua,Khachatryan:2014nda}, where $q=d,s,b$. 
The SM with three fermion families imposes the $3\times 3$ unitarity condition $\vtd^2+\vts^2+\vtb^2 = 1$, implying that this quantity can be written as $\Rb = \RbDefTwo$ and can thus be used to infer \vtb\ directly. 
 The most precise measurement of this ratio, $\Rb = 1.014 \pm 0.032$~\cite{Khachatryan:2014nda}, yields a $1.6\%$ precision on \vtb if no unitarity assumption is made ($\vtb = 1.007\pm 0.016$), and a lower limit $\vtb > 0.975$ at 95\% confidence level is obtained with the Feldman-Cousins frequentist approach~\cite{Feldman:1997qc} if $3\times 3$ unitarity is imposed to the CKM matrix. 

The ratio \Rb can be combined with the $t$-channel cross-section measurement in order to extract an indirect measurement of the top-quark width, which is directly proportional to the $t$-channel cross section as long as $\vtb \simeq 1$. Using this approach, the width measured by D0 is $\Gamma_t=2.0^{+0.47}_{-0.43}$~GeV~\cite{Abazov:2012vd}, which is significantly improved upon in the measurement by CMS of $\Gamma_t=1.36^{+0.14}_{-0.11}$~GeV~\cite{Khachatryan:2014nda}. These measurements assume that the initial-state $W$~boson is on-shell in the $t$-channel exchange, which of course is not generally valid.
The width of the top quark will be measurable directly, in a theoretically well defined approach, by exploiting a selection targeting $t$-channel single top quarks, and distinguishing between resonant and non-resonant $Wb$ production ($t \to W^+ b$ and $\bar t \to W^- \bar b$, versus $W^- b$ and $W^+ \bar b$ production)~\cite{Giardino:2017hva}.

The single top-quark production cross sections in $t$- and $s$-channel and $W$-associated mode can be written, in the SM, as the sum of three contributions:
\begin{equation}
\sigma_{tot} = \vtd^2 \sigma_d + \vts^2 \sigma_s + \vtb^2 \sigma_b\, ,
\end{equation}
where $\sigma_d$, $\sigma_s$, and $\sigma_b$ represent the cross sections expected for the sub-processes where, respectively, a down, strange, and bottom quark are connected to a top quark, see Fig.~\ref{fig:FG}. 
Therefore, these production modes are potentially sensitive to all three elements of the third row of the CKM matrix. 
The single top-quark cross sections in $t$-channel and \Wt production modes in particular have an enhanced sensitivity to \vtd and \vts due to the large parton densities of $d$ and $s$ quarks in the proton~\cite{Tait:2000sh,Alwall:2006bx,Lacker:2012ek}, differently from the $s$-channel mode.

\begin{figure}[!ht]
 \begin{center}
  \includegraphics[width=0.9\textwidth]{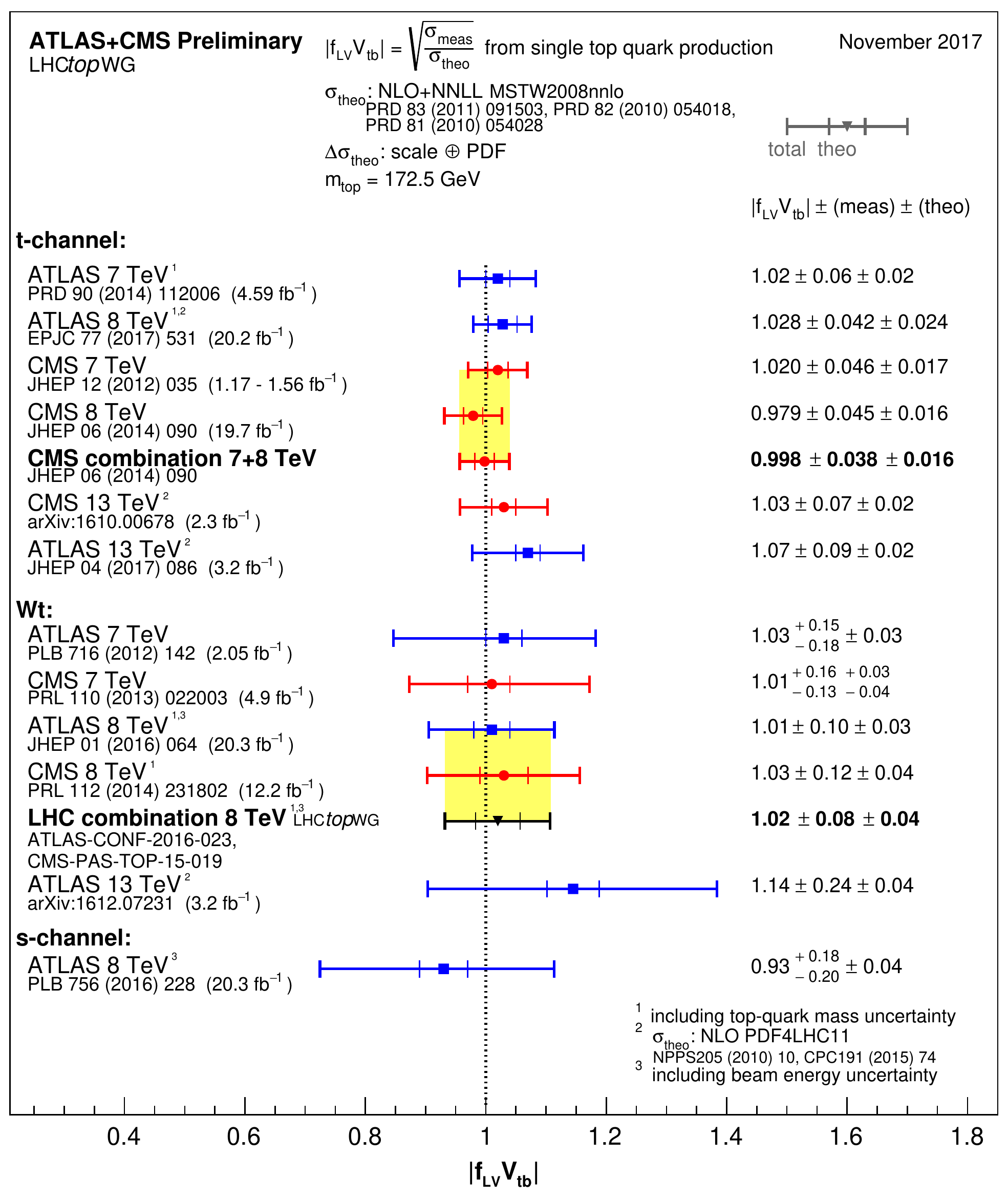}
  \caption{\label{fig:vtb}{} Summary of ATLAS and CMS extractions of $\fLVVtb$ from the single top-quark cross section measurements, using NLO+NNLL theoretical predictions. From \textcite{toplhcwg}, including some preliminary results.}
 \end{center}
\end{figure}

Single top-quark cross section measurements can be used to derive \vtb without the need to rely on the $3\times 3$ unitarity condition, under the simplifying assumption that, whatever the values, the relationships $\vtb \gg \vtd$ and $\vtb\gg \vts$ hold true, which makes the cross section of the processes in Fig.~\ref{fig:FG} directly proportional to $\vtb^2$. 
Under these conditions, the product \fLVVtb is extracted by dividing the measured cross section for each channel by the corresponding theory prediction and then taking the square root. The factor $\fLV$ is the form factor for the purely left-handed vector \tbW coupling, see Eq.~\ref{eq:tbW}. It is unity in the SM but could be larger than unity if anomalous couplings due to new physics are present.
It is customary to also quote the 95\% confidence level interval obtained by setting $\fLV=1$, i.e. with the additional unitarity constraint $0\le \vtb\le 1$. 
The procedure outlined so far ignores the possibility that the \tbW coupling may receive contributions from right-handed or non-vectorial operators that are instead usually considered in studies such as those reported in Section~\ref{sec:wtb-vertex}.
Figure~\ref{fig:vtb} shows the \vtb\ values times \fLV extracted by the LHC experiments from single top-quark cross section measurements under these assumptions~\cite{toplhcwg}. 
At the Tevatron, the CKM matrix element \vtb is extracted from the $s+t$ cross section measurement, obtaining $\fLVVtb = 1.02^{+0.06}_{-0.05}$, corresponding to a lower limit at the 95\% confidence level of \vtb$ > 0.92$~\cite{Group:2009qk}.

\textcite{Alwall:2006bx,AguilarSaavedra:2010wf,Lacker:2012ek} illustrated how to derive less model-dependent limits on all three \vtq matrix elements by re-examining the measurements of single top-quark cross sections and \Rb published at the time.
 Not having direct access to the data requires several approximations in the analysis.
 A particularly tricky case for the reinterpretation is that single top-quark analyses are  based on multivariate techniques. The MVA input variables are related to the kinematic properties of the reconstructed top quark and the event, which would be modified in production through \vts or \vtd, thus modifying the acceptance. Moreover,  the jet coming from the top-quark decay is assumed to be a $b$-jet, thus $\vtb^2 \gg \vtd^2+\vts^2$ is assumed.

\cite{AguilarSaavedra:2010wf} propose to use the rapidity of the single top quark and antiquark in $t$-channel and \Wt production modes to set direct limits on \vtd. Similarly, in \textcite{Alvarez:2017ybk}, it is proposed to use the integrated charge asymmetry in \Wt to extract \vtd. Both methods rely on the consideration that $b$-quark-initiated \Wt production, Fig.~\ref{fig:FG}, has exactly the same kinematic properties and rate whether the initiator quark is a $b$ or $\bar b$, while $d$-quark-initiated processes feature different rate, spectra and angular distributions, depending on the initiator being a $d$ or $\bar d$, due to the different \xB spectrum of quark and antiquark.

\subsection{Cross section ratios as inputs for PDF extraction}
\label{sec:rt}

 A feature of SM single top-quark production at the LHC, absent in \ppbar collisions and therefore unmeasurable in Tevatron data, is the difference in production rate (integrated charge asymmetry, $\Rt\equiv \RtDef$) between top quark and antiquark production in the $t$- and $s$-channel modes.
 The magnitude of these ratios is primarily driven by the relative importance of the up- and down-quark densities and is therefore potentially helpful to constrain those densities, making single top-quark production a useful input to global PDF fits. 
 This sub-section focuses on the integrated charge asymmetry in $t$-channel production, as no measurement of this quantity has been performed yet for the other single top-quark production modes. The interest of charge asymmetry in \tW is discussed in Section~\ref{sec:vtb}. 

The \Rt expectations depend on the CM energy: predictions at 13~TeV are, in general, significantly smaller than those at 8~TeV, which are in turn smaller than at 7~TeV, as intuitively understandable from the consideration that ``sea'' quarks contribute more than ``valence'' quarks at large \xB. 
The \Rt measurements are complementary to $W$-boson cross-section ratios (that are similarly sensitive to up- and down-quark densities) by probing larger \xB values. The ABMP16 PDF set~\cite{Alekhin:2017kpj} already includes this information in the fit, and the relative importance of $R_t$ in PDF extractions is expected to grow with more integrated luminosity available to the LHC experiments in Run~2.

\begin{figure}[!ht]
 \begin{center}
  \includegraphics[width=0.45\textwidth]{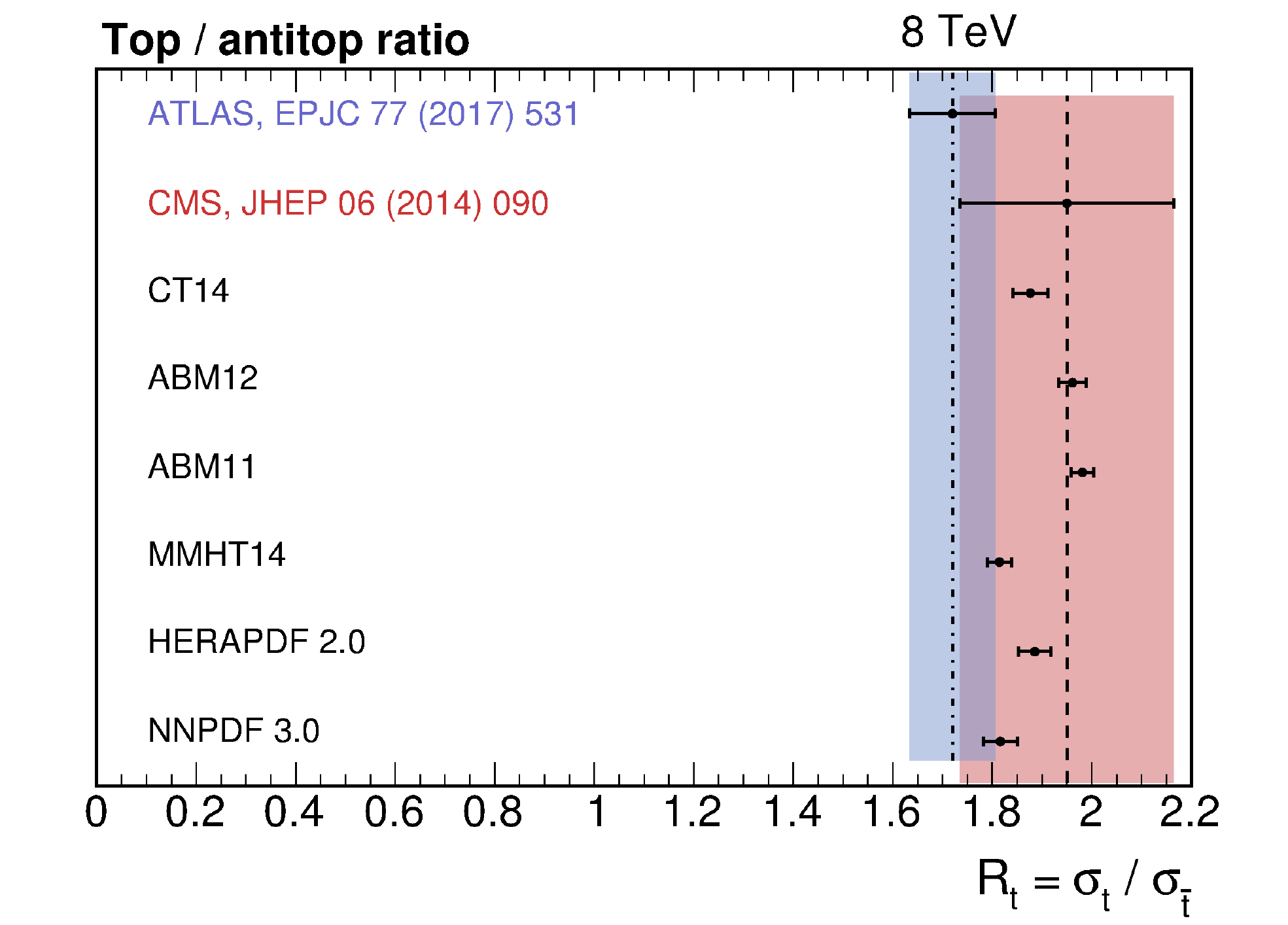}
  \includegraphics[width=0.45\textwidth]{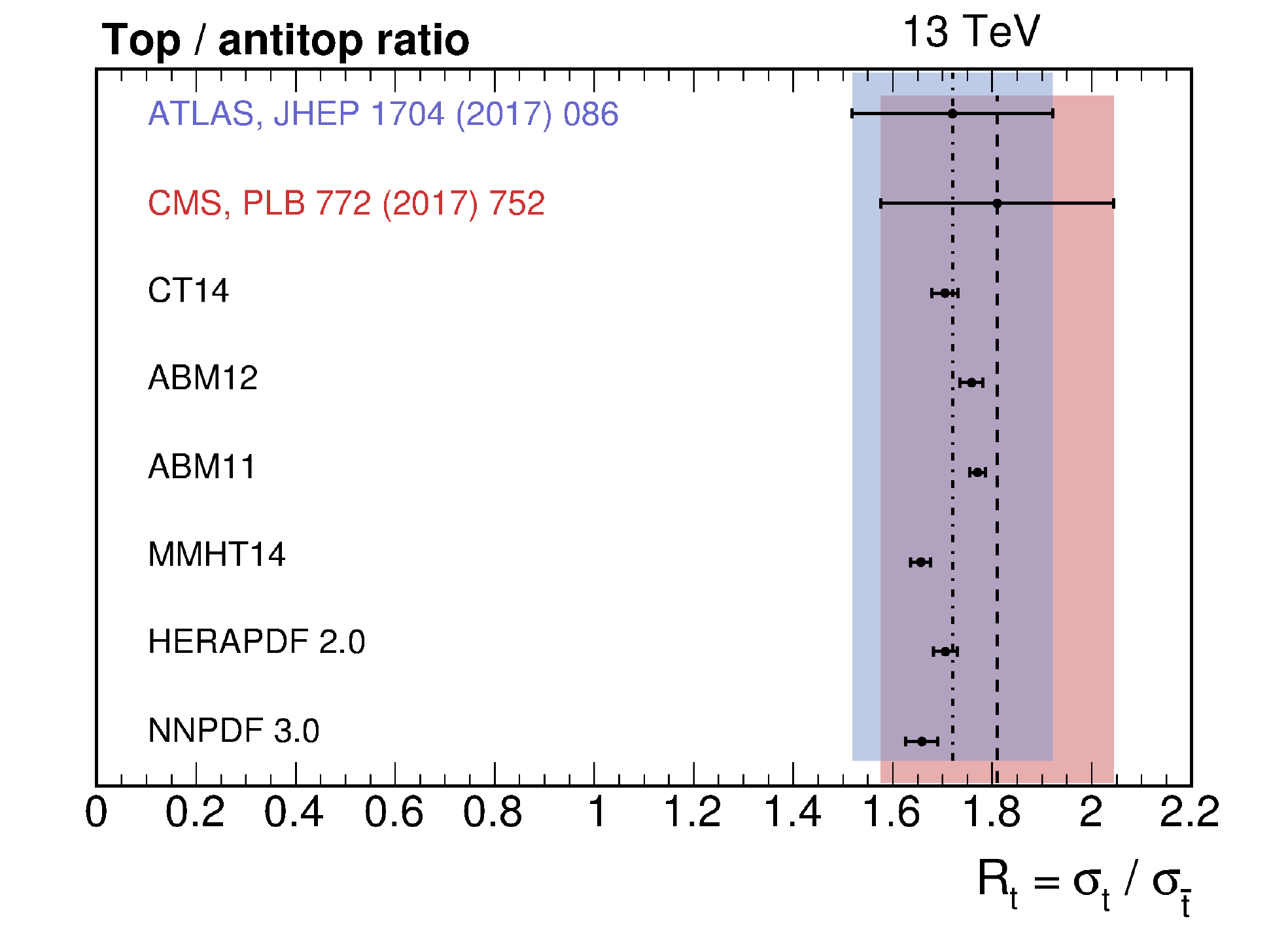}
  \caption{\label{fig:Rt}{} Summary of ATLAS and CMS measurements of $\Rt\equiv \RtDef$ at (left) 8~TeV~\cite{Aaboud:2017pdi,Khachatryan:2014iya} and (right) 13~TeV~\cite{Aaboud:2016ymp,Sirunyan:2016cdg}, compared with theoretical expectations at NLO obtained with \HATHOR~\cite{Aliev:2010zk,Kant:2014oha} and a variety of PDF sets~\cite{CT14,ABM12,ABM11,MMHT14,HERAPDF,NNPDF30}. Error bars for the different PDF sets represent the quadratic sum of the following uncertainty components: the 68\% confidence level interval of the predictions of the eigenvectors in the set, the statistical uncertainty due to the finite number of iterations employed for the calculation, the uncertainty in the factorisation and renormalisation scales, derived varying both of them by a factor 1/2 and 2, and the uncertainty in the top-quark mass. }
 \end{center}
\end{figure}

 The values of \Rt measured by the ATLAS collaboration at 7, 8 and 13~TeV~\cite{Aad:2014fwa,Aaboud:2017pdi,Aaboud:2016ymp} and the CMS collaboration at 8 and 13~TeV~\cite{Khachatryan:2014iya,Sirunyan:2016cdg} have been compared to the predictions for a variety of PDF sets.
 Figure~\ref{fig:Rt} compares the \Rt measurements at 8 and 13~TeV between the two experiments and with predictions for several PDF sets: HERAPDF 2.0 NLO~\cite{HERAPDF}, ABM11 NLO~\cite{ABM11}, ABM12 NNLO~\cite{ABM12}, MMHT14 NLO~\cite{MMHT14}, CT14 NLO~\cite{CT14}, NNPDF 3.0 NLO~\cite{NNPDF30}.
 The perturbative part of these calculations is performed at NLO with the \HATHOR program~\cite{Aliev:2010zk,Kant:2014oha} and has been cross-checked with the \POWHEG generator~\cite{Alioli:2009je, Re:2010bp}. The scale and top-quark mass uncertainty components on the predictions are numerically small in comparison with the PDF and number of iterations components.
\HATHOR and \POWHEG are found to yield compatible predictions within the statistical uncertainty. The ratio computed from the NNLO predictions shown in Table~\ref{tab:lhctchanpred} are 1.82 at 8~TeV and 1.69 at 13~TeV, computed with MSTW2008, though no PDF uncertainty is available. This NNLO ratio is slightly higher than the MMHT-based calculation at 8~TeV and consistent with it at 13~TeV.

\textcite{Alekhin:2015cza} (Fig.~13 of that paper) showed that the ATLAS measurement of \Rt at 7~TeV and the one by CMS at 8~TeV give consistent pictures, with the CT10~\cite{CT10}, CT14, MMHT14, NNPDF 3.0 sets slightly disfavored, while ABM12 and ABM15~\cite{Alekhin:2015cza} are favored. The latter includes $W$-boson charge ratios in the fit, while the single top-quark charge ratio in the $t$-channel is used as a ``standard candle'' to validate the predictions of their PDF set~\footnote{The individual cross section measurements of single top quark and antiquark production at the LHC, not yet including the 8~TeV ATLAS measurement, have been used to extract the ABMP16 set~\cite{Alekhin:2017kpj}.}.
However, this picture became inconsistent with the later publication of the most precise \Rt result in the literature, which is the ATLAS measurement at 8~TeV: this yields smaller values than most PDF sets, and is in tension with most of the PDF set predictions for this observable, as shown in Fig.~\ref{fig:Rt}, while the aforementioned ATLAS and CMS measurements at 7 and 8~TeV both yield larger values than most PDF sets. The small uncertainty of the ATLAS measurement highlights the value of time in hadron collider analyses. The ATLAS analysis was published almost three years after the CMS analysis, and that time was used to improve the detector understanding and theory modeling, and to devise an optimal analysis strategy. Rather than obtaining \Rt from the ratio of measured cross sections, ATLAS extracts \Rt in one simultaneous fit to the top quark and antiquark cross sections. This directly accounts for all correlations, including those between the two analysis regions and those between different systematic uncertainties that are induced in the fit.

The currently available \Rt measurements at 13~TeV, based on the data collected in 2015, are limited by their statistical uncertainty and do not shed light on this inconsistency yet.
However, future measurements of \Rt based on the full Run~2 data set may be expected to surpass the best Run~1 measurements in precision, and, in conjunction with them, may provide strong constraints on future global PDF fits. 
Moreover, with more data, differential distributions of \Rt as a function of the rapidity and transverse momentum of the top quark will provide significant additional discriminating power~\cite{Berger:2016oht}.

Another useful input for constraining PDFs is the measurement of the ratios of single top-quark cross sections between different CM energies, as done by the CMS collaboration in the $t$-channel case. The ratio of the cross sections of the \etalj-based analysis at 7 and 8~TeV~\cite{Khachatryan:2014iya} is ($R_{\rm 8~TeV/7~TeV}=1.24\pm 0.08 \stat \pm 0.12 \syst$. Measurements of the ratios $R_{\rm X~TeV/Y~TeV}$ profit from cancellations of several important systematic uncertainties and are sensitive to the evolution of the partonic distributions in the proton.
 Given the larger jump in energy, it will be instructive to see the results of the same exercise using the 13~TeV results, as well as the double-ratio obtained by taking the ratio of \Rt between different CM energies. 
Unfortunately, these measurements have not been reported by the LHC experiments yet.

\subsection{Top-quark mass}
\label{sec:mass}

Similarly to \ttbar, single top-quark events can be exploited for the measurement of the top-quark mass, \mtop, either directly by kinematic reconstruction of a top-quark candidate, or indirectly through the dependence of the cross section on the mass.

The ~\textcite{Sirunyan:2017huu} has performed a direct measurement of the top-quark mass with $t$-channel single top-quark events using the 8~TeV dataset. 
Top-quark candidates are reconstructed in the $t$-channel topology from their decay to a $W$ boson and a $b$ quark, with the $W$ boson decaying leptonically to a muon and a neutrino. 
At variance with respect to \ttbar events, there is typically only one central $b$ jet in the $t$-channel single top-quark process. 
Top-quark pair events constitute a relatively large fraction of the events even in a single top-quark optimized signal region, but in the context of this measurement they are treated as a component of the signal, as they carry information on the parameter of interest. However, care is taken in making the selection orthogonal to the \ttbar-based measurements of the same quantity in the single- and di-lepton final states, in order to facilitate future combinations~\cite{Khachatryan:2015hba}. 
The interest of performing this measurement in a single top-quark topology lies in the complementarity with \ttbar, with which the systematic uncertainties are partially uncorrelated as the color flow is very different (there is no color flux between the two quark lines in $t$-channel production), and the statistical uncertainty is uncorrelated.

The event selection and the procedure to reconstruct the top-quark candidates follow closely the $t$-channel cross section measurement in the same dataset~\cite{Khachatryan:2014iya}, with two additional conditions imposed in order 
to enhance the purity of the sample: the absolute value of \etalj, defined as in Section~\ref{sec:tchannel}, is required to be larger than 2.5; and in order to exploit the large charge asymmetry of the $t$-channel production mode, the main result is restricted to events with positive muons, hence with top quarks, while those with negative muons (top antiquarks) are only used to cross-check the result on an independent dataset.
A fit to the invariant mass distribution of reconstructed top-quark candidates`\footnote{The fit assumes, of course, the same top-quark mass in single top-quark and \ttbar events; therefore, the latter are effectively treated as a component of the signal.} yields a value of the top-quark mass of $172.95 \pm 0.77 \stat {}^{+0.97}_{-0.93} \syst$~GeV, in agreement with the results from \ttbar~\cite{TevatronElectroweakWorkingGroup:2016lid,Aad:2015nba,Khachatryan:2015hba}.
Several systematic uncertainties are larger than in the standard analyses in the $l$+jets \ttbar topology, where the invariant mass of the jets failing $b$-tagging is expected to peak at the mass of the $W$ boson, allowing to calibrate the jet energy scale {\it in situ} and also reducing several modeling uncertainties related to soft QCD effects. Moreover, in comparison with \ttbar-optimized selections, the $t$-channel signal region is more contaminated by $W/Z$+jets backgrounds, whose modeling parameters are relatively poorly constrained, due to its lower multiplicity of jets and $b$~jets.

Similarly to the \ttbar case~\cite{Aad:2014kva,Khachatryan:2016mqs,Abazov:2016ekt}, the inclusive single top-quark cross sections can be used to extract the top-quark pole mass thanks to the strong dependence of the theoretical predictions on this parameter~\cite{Kant:2014oha}. The strongest dependence is found for $s$-channel production ($\frac{\Delta\sigma_s}{\sigma_s} = -3.9\frac{\Delta \mtop}{\mtop}$ at $\sqrt{s}=8$~TeV), 
followed by \tW ($\frac{\Delta\sigma_{\tW}}{\sigma_{\tW}} = -3.1\frac{\Delta \mtop}{\mtop}$ at $\sqrt{s}=8$~TeV), 
while the $t$-channel shows a weaker dependence ($\frac{\Delta\sigma_t}{\sigma_t} = -1.6\frac{\Delta \mtop}{\mtop}$ at $\sqrt{s}=8$~TeV). 
However, for a practical use of this method, particular care should be taken to minimize the dependence of the experimental measurement of the cross section on \mtop~\cite{Schuh:2016xfi}. The 8~and 13~TeV ATLAS $t$-channel analyses~\cite{Aaboud:2017pdi,Aaboud:2016ymp} measure a cross section that decreases with the assumed top-quark mass. 
This is the same  behavior as in the theoretical prediction, and this imposes an additional limitation on the precision of the extraction of the top-quark mass.

\FloatBarrier

%% file: tex/searches.tex
\subsection{\tWb vertex structure}
\label{sec:wtb-vertex}

All single top-quark production processes are sensitive to anomalous couplings in the \tWb vertex and provide sensitivity beyond \ttbar because the \tWb vertex appears both in the production of the top quark and in its decay. In particular, since the top-quark lifetime is shorter than the timescale of spin decoherence induced by QCD, its decay products retain memory of its polarization imprinted by the production mechanism.  This provides additional powerful tools in the search for BSM physics in single top-quark studies: 
 in  single top-quark production via the $t$-channel, the SM predicts that top quarks are produced almost fully polarized through the V--A coupling along the direction of the momentum of the quark
 that recoils against the top quark~\cite{Mahlon:1999gz,Jezabek:1994zv}, while new physics models may lead to a depolarization in production or decay by altering the coupling structure~\cite{AguilarSaavedra:2010nx, AguilarSaavedra:2008gt,AguilarSaavedra:2008zc,Bach:2012fb}.

The most general Lagrangian term that one can write for the \tbW coupling up to dimension-six gauge invariant operators~\cite{AguilarSaavedra:2008zc}, under the approximation $\vtb = 1$, is:
\begin{equation}
\label{eq:tbW}
\mathcal{L}_{tWb} = -\frac{g}{\sqrt{2}}\bar b\left[ \gamma^{\mu}(\fLV P_L + \fLR P_R) + \frac{i\sigma^{\mu\nu}q_{\nu}}{M_W} (g_L P_L + g_R P_R) \right] t W^-_{\mu} + h.c. \, ,
\end{equation}
where the form factors \fLV and \fLR denote the strength of the left- and right-handed vector-like couplings, and $g_L$ and $g_R$ denote the left- and right-handed tensor-like couplings. Slightly different notations are used in the figures in this review, $f_L=f_{LV}=f^L_V=V_L$. Similarly, $g_R=f^R_T$.
 The SM predicts $\fLV = 1$, $\fLR = g_L = g_R = 0$ at tree level. In single top-quark production, the production and the decay of the top quark are both sensitive to anomalous couplings. When considering one form factor at a time, the cross section is proportional to the form factor squared. When considering two or more simultaneously, interference effects may also come into play. For consistency, the Tevatron limits are given in terms of absolute value of couplings squared.

\begin{figure}[!htbp]
 \includegraphics[width=0.32\textwidth]{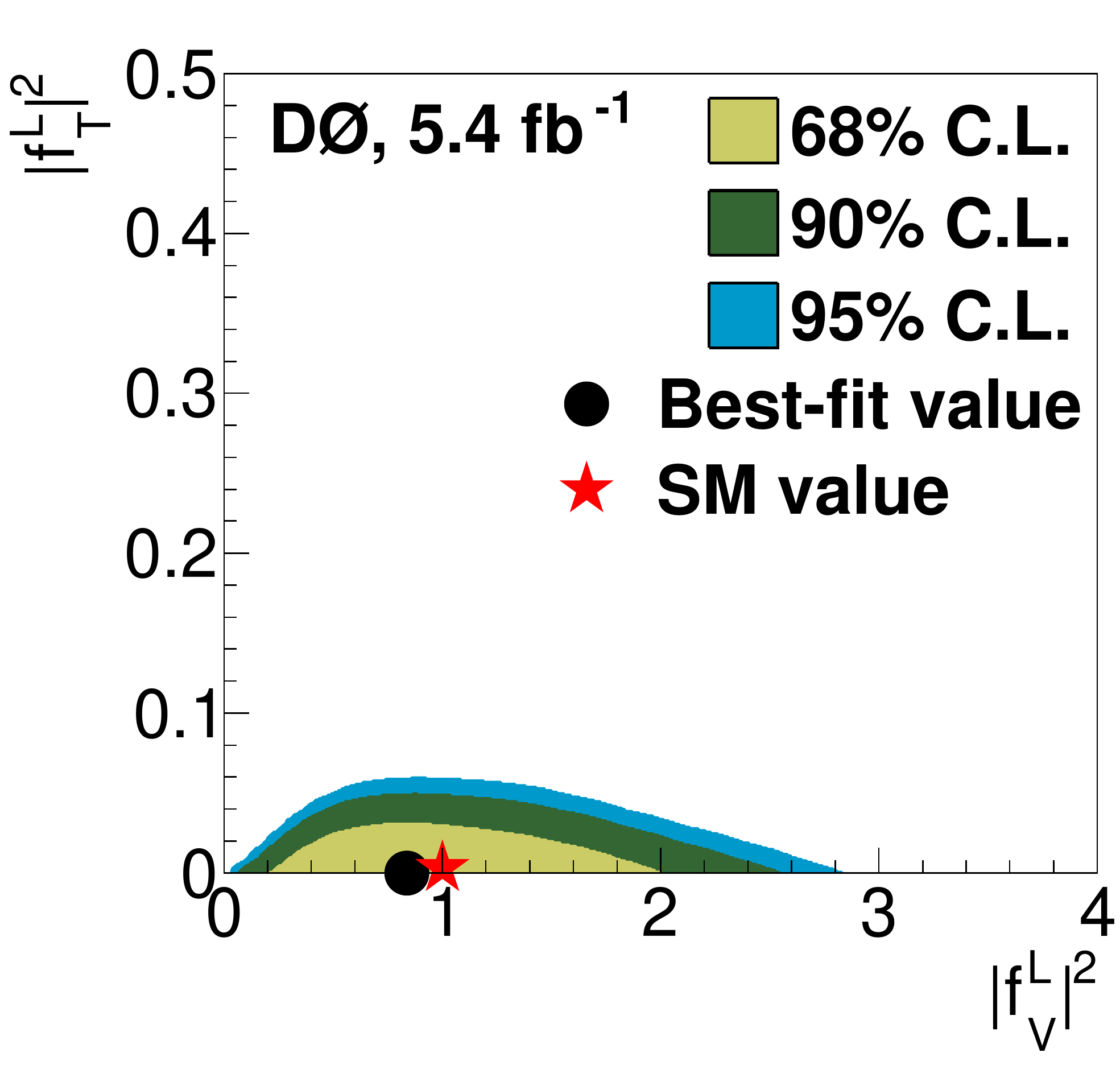}
  \caption{Limits on pairs of anomalous couplings squared from the D0 combination of single top and \ttbar anomalous couplings searches:  left-handed tensor coupling vs left-handed vector coupling (from \textcite{Abazov:2012uga}).}
  \label{fig:d0anom}
\end{figure}

At the Tevatron, anomalous coupling searches have focused on the magnitude of the four form factors.
D0 optimized the single top-quark anomalous couplings search in the two-dimensional plane of one anomalous coupling and the SM-like left-handed vector coupling \fLV~\cite{Abazov:2011pm}. The D0 single top-quark anomalous couplings search uses an MVA, which is trained on samples with either purely left-handed or purely right-handed vector couplings, in both production and decay. The single top-quark search was also combined with a $W$-boson helicity measurement in \ttbar to set stringent limits on pairs of form factors~\cite{Abazov:2012uga}. Figure~\ref{fig:d0anom} shows the two-dimensional Bayesian posterior density for one such pair of anomalous couplings. 
Note that the limit is set as a function of the coupling squared since the cross section is proportional to that. For comparison with the LHC experiments below, one should take the square root.

\begin{figure}[!htbp]
\includegraphics[width=0.52\textwidth]{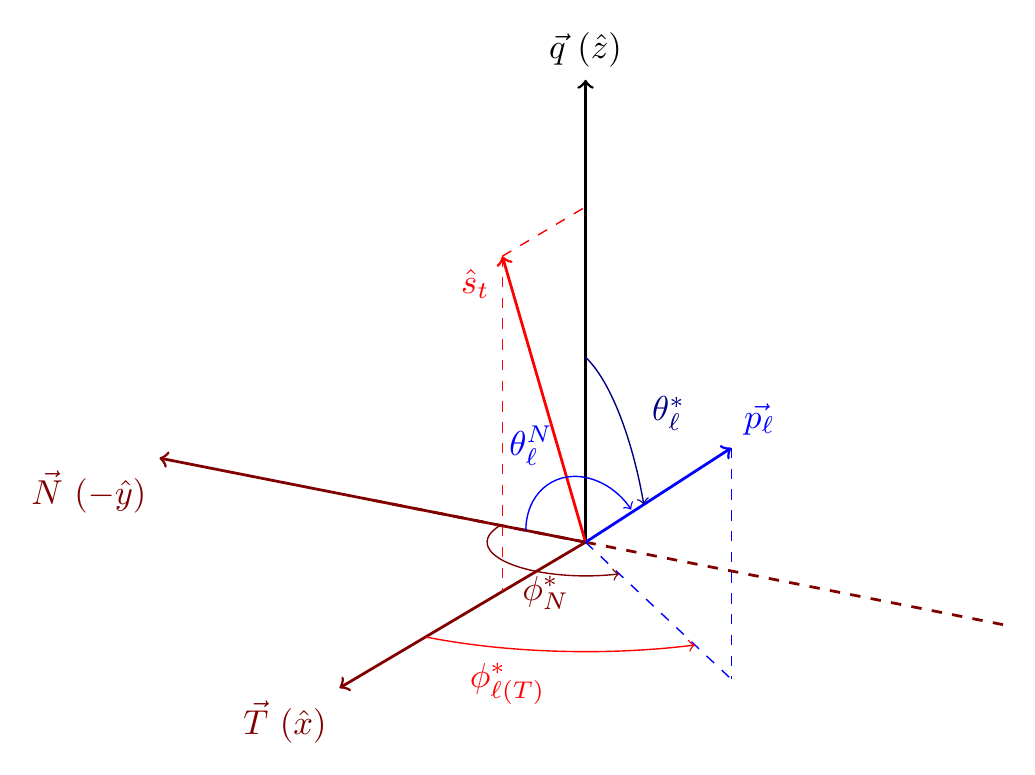}
\hfill
\includegraphics[width=0.45\textwidth]{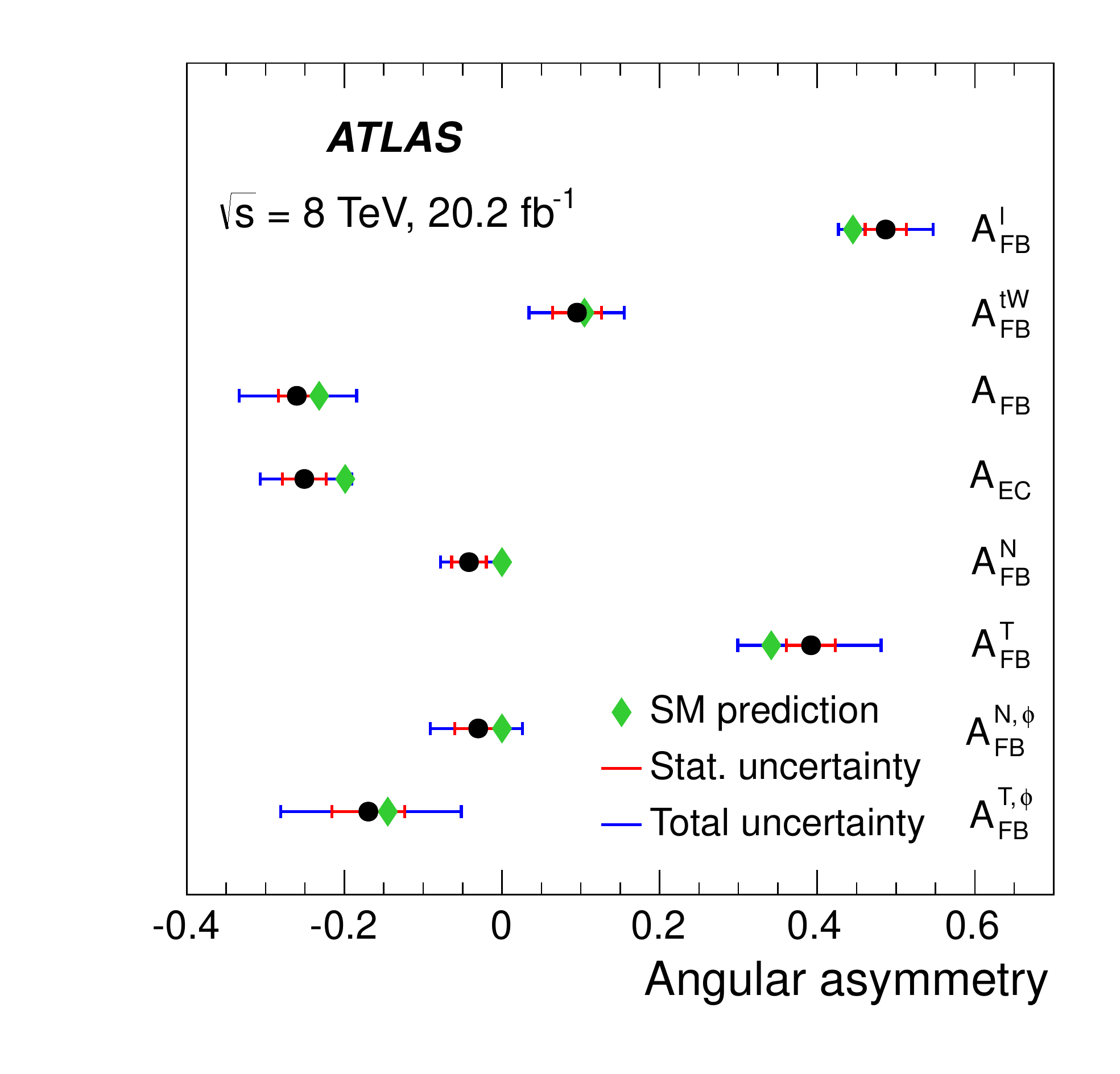}
\caption{(Left) Illustration of the definition of the polarization angles in $t$-channel single top-quark production, and (right) predicted and observed angular asymmetries (from \textcite{Aaboud:2017aqp}).}
  \label{fig:PolAngles}
\end{figure}

At the LHC, the approach followed by ATLAS and CMS has been to consider the relationship between top-quark production and decay. At 8~TeV, ATLAS relied on the definition of eight polarization variables, together with the magnitude of the polarization. The angular distributions of the decay products of the top quark are given by 
$$
\label{eq:pol}
\frac{1}{\Gamma} \frac{d\Gamma}{d\cos \theta} = \frac{1}{2} \left( 1+ \alpha P \cos \theta \right) \, ,
$$
where $\theta$ is the angle between the direction of flight of the decay product and a properly chosen spin quantization axis, $P$ is the top-quark degree of polarization along this quantization axis, and $\alpha$ is the spin analyzing power for this decay product, which takes a value of $\pm 0.998$ at NLO for charged leptons in the SM~\cite{Jezabek:1994zv,Brand:2002,Aaboud:2017aqp}. 
The relevant angles $\theta$ are illustrated in Fig.~\ref{fig:PolAngles}. The $z$~axis is given by the direction of the $W$~boson in the top-quark rest frame, the $x$-axis is given by the top-quark spin component that is orthogonal to $z$, and the $y$ axis is orthogonal to these two, defining a right-handed coordinate system. With these definitions, three angles are defined: $\theta_\ell$  is the angle between the $z$~axis and the lepton momentum in the top-quark rest frame, the $\phi_\ell(T)$  is the angle between the projection of the lepton momentum in the top-quark rest frame onto the $x-y$ plane and the $x$~axis and $\theta_\ell^N$ is the angle between the lepton momentum in the top-quark rest frame and the $y$ axis. Quantifying the degree of polarization along the direction of the spectator quark gives 0.91 for top quarks and -0.86 for top antiquarks~\cite{Schwienhorst:2010je}. 

The ATLAS and CMS experiments select single top-quark events in the $t$-channel final state consisting of a charged lepton from the decay of the $W$~boson from the top-quark decay, large \etmiss, and two jets, one of which is $b$-tagged and the other one is in the forward detector region.
In the ATLAS analysis~\cite{Aaboud:2017aqp}, using 8~TeV data, the signal region contains about 9000 events, half of which are expected to come from $t$-channel production. The angular observables are unfolded to the parton level in two bins, one for positive cosine of the relevant angle (i.e., forward-going direction of the decay product with respect to the corresponding spin quantization axis) and one for negative cosine (backward-going with respect to the same axis). 
Based on these angular observables as well as for the \costheta variable, forward-backward asymmetries are defined.
The measured asymmetries and the corresponding theory predictions are shown in Fig.~\ref{fig:PolAngles}(right). From the asymmetries, a limit on the imaginary part of $g_R$ is also derived. The limit interval at the 95\% confidence level is $[-0.18, 0.06]$.

CMS measured the single top-quark polarization with 8~TeV data~\cite{Khachatryan:2015dzz}. A model-independent selection targets $t$-channel production, then the observed \costheta distribution (Fig.~\ref{fig:LHCtdiff}) is used to infer the differential cross section as a function of the parton-level \costheta (see Section~\ref{sec:tchannel}). This is found to be compatible with the linear expectation of Eq.~(\ref{eq:pol}), and a linear fit yields $P\times\alpha_{\ell} = 0.52 \pm 0.06 \stat \pm 0.20 \syst$, compatible with the SM expectation within two standard deviations. 

With the same data set, CMS also used a different selection, targeting $t$-channel events but tolerating a larger contamination from \ttbar with respect to typical analyses in the same final state, to extract the $W$-boson helicity amplitudes with 8~TeV data~\cite{Khachatryan:2014vma}. The sensitivity to those parameters comes mostly from the decay vertex of the top quark rather than from the production vertex, exploiting the helicity angle \thetaW defined as the angle between the $W$-boson momentum in the top-quark rest frame and the momentum of the down-type fermion from the $W$-boson decay, in the rest frame of the mother particle. 
A fit to the distribution of \thetaW discriminates the components of the signal originating from the right-handed ($F_R$), left-handed ($F_L$) and longitudinal ($F_0$) helicity fractions of the W boson. 
Similarly to the top-quark mass case described in Section~\ref{sec:mass}, the interest of an analysis in this final state lies in the complementarity with the measurements traditionally performed with selections targeting \ttbar production. 
In this measurement, \ttbar events, that constitute the majority of the population in the signal region, are treated as a component of the signal as they carry information on the parameters of interest. 
The measured helicity fractions are $F_L = 0.298 \pm 0.028 \stat \pm 0.032 \syst$, $F_0 = 0.720 \pm 0.039 \stat \pm 0.037 \syst$, and $F_R = -0.018 \pm 0.019 \stat \pm 0.011 \syst$. These results are used to set limits on the real part of the tWb anomalous couplings, $g_L$ and $g_R$, assuming no CP violation (hence no imaginary components for those couplings). 

\begin{figure}[!htbp]
\includegraphics[width=0.52\textwidth]{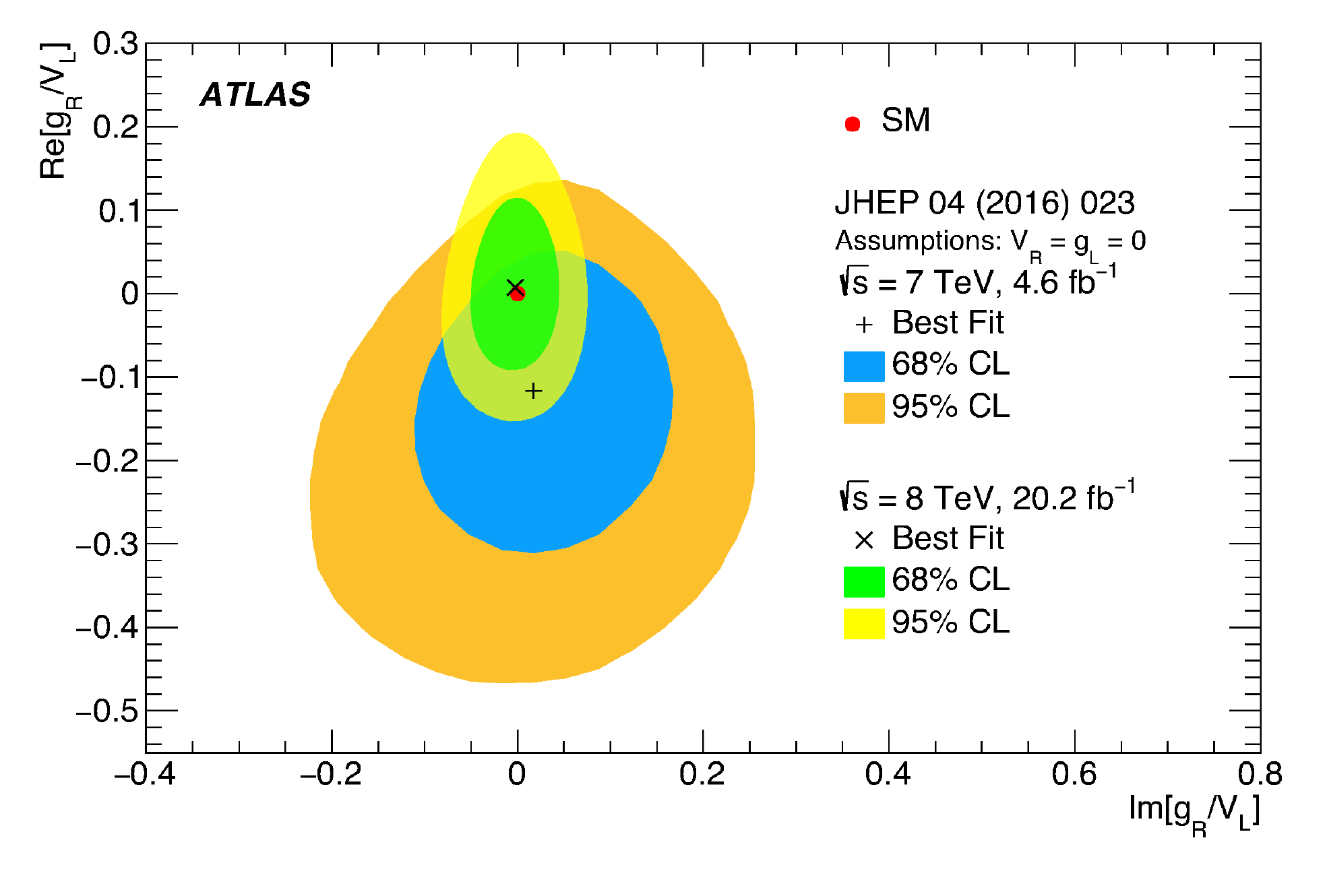}
\hfill
\includegraphics[width=0.45\textwidth]{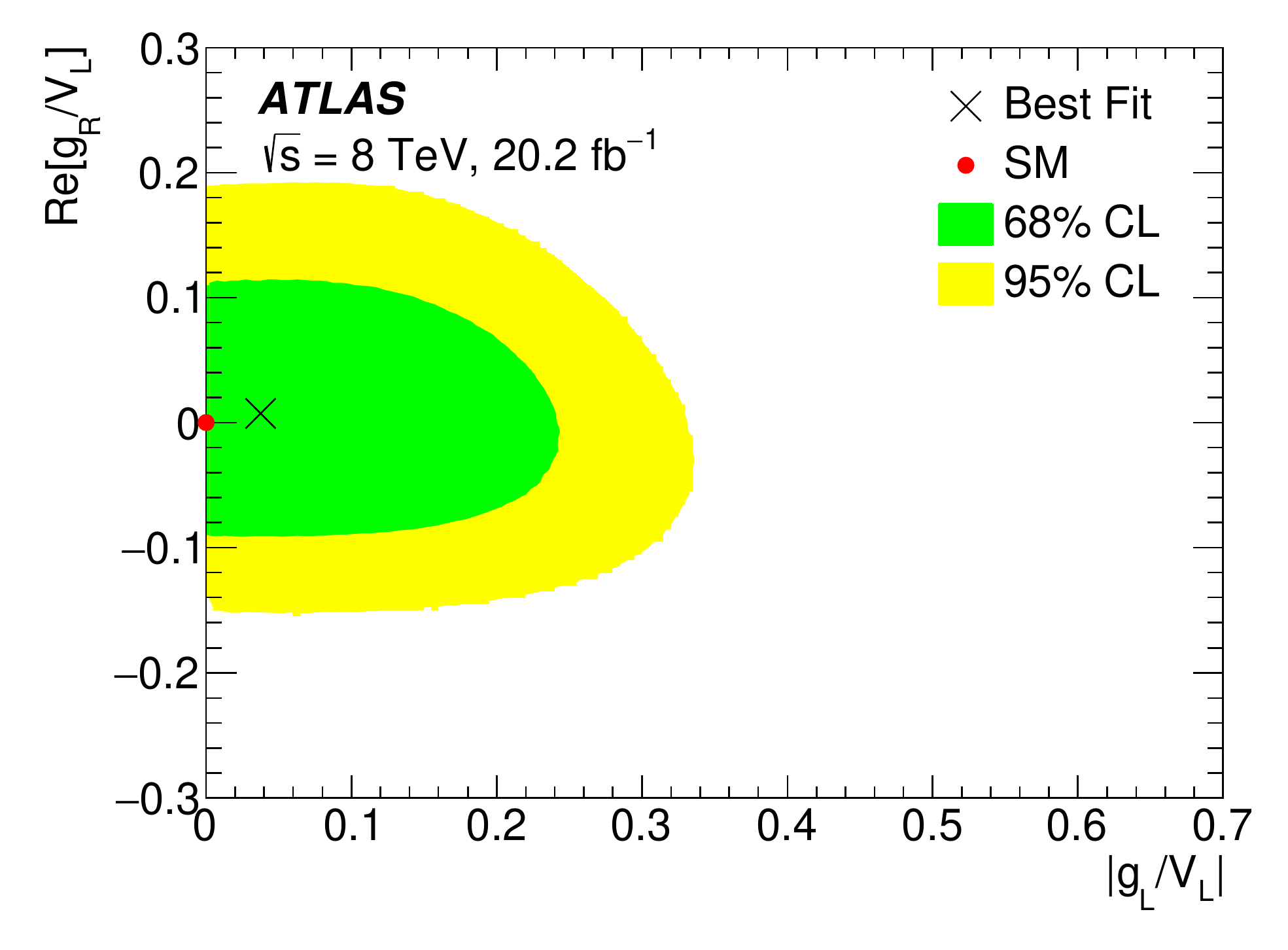}
  \caption{Limits on anomalous couplings from the ATLAS two- (left) and three-angle (right) analyses (from \textcite{Aaboud:2017yqf}).}
  \label{fig:Atl23ang}
\end{figure}

ATLAS also measured double-differential angular correlations in 7~TeV data~\cite{Aad:2015yem} and triple-differential angular correlations in 8~TeV data~\cite{Aaboud:2017yqf}. The angular observables are expressed in terms of spherical harmonics in the 7~TeV analysis and in terms of orthonormal functions that are the products of spherical harmonics~\cite{Boudreau:2013yna,Boudreau:2016pdi}. 
Figure~\ref{fig:Atl23ang} summarizes the results at both CM energies, shown as a function of the ratio of the anomalous coupling over the SM-like left-handed vector coupling, including  both the real and imaginary parts for the right-handed tensor coupling ($g_R$). The measurements are consistent with the SM prediction, and the 8~TeV measurement is a significant improvement over the 7~TeV one.

\begin{figure}[!htbp]
\includegraphics[width=0.48\textwidth]{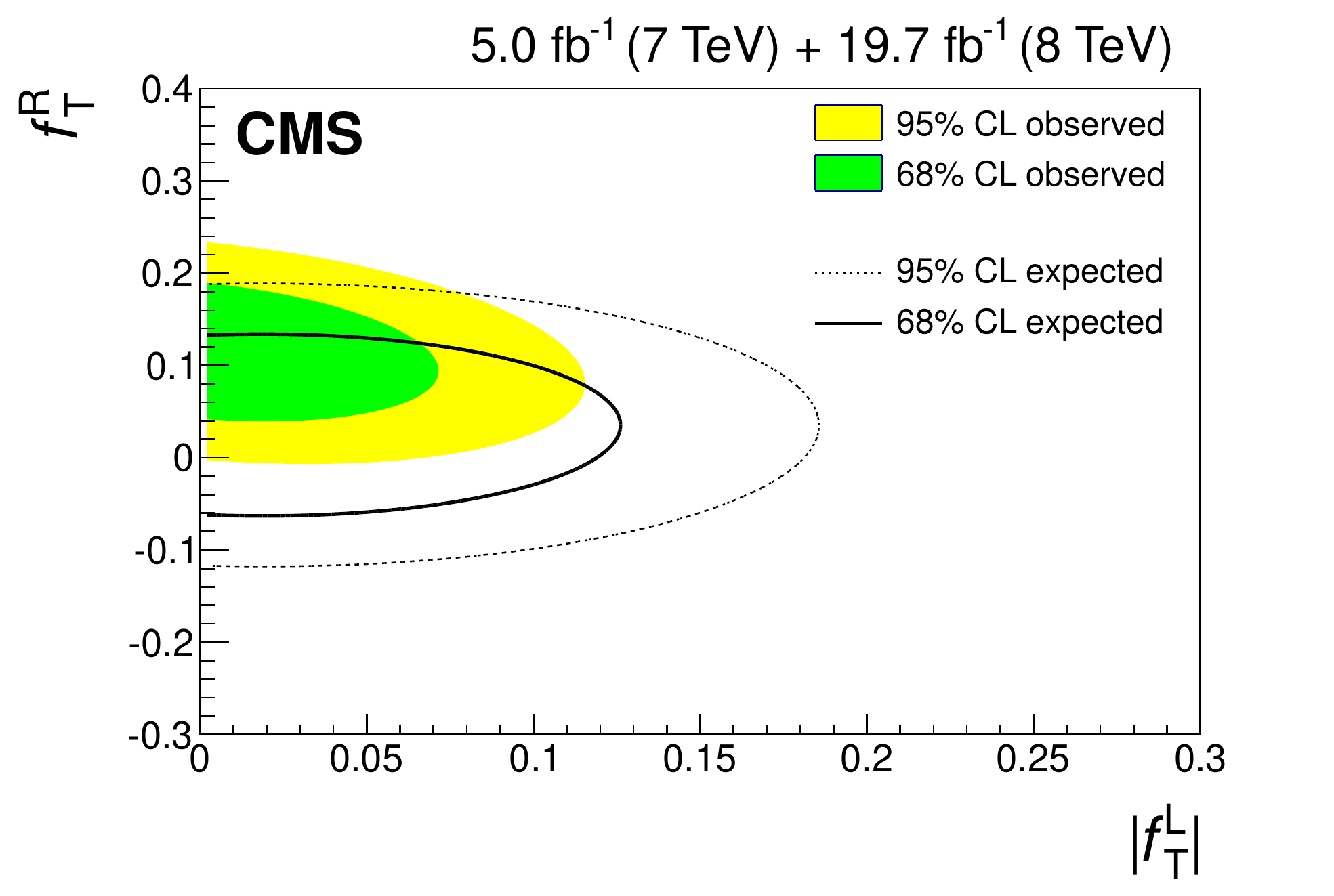}
\hfill
  \includegraphics[width=0.48\textwidth]{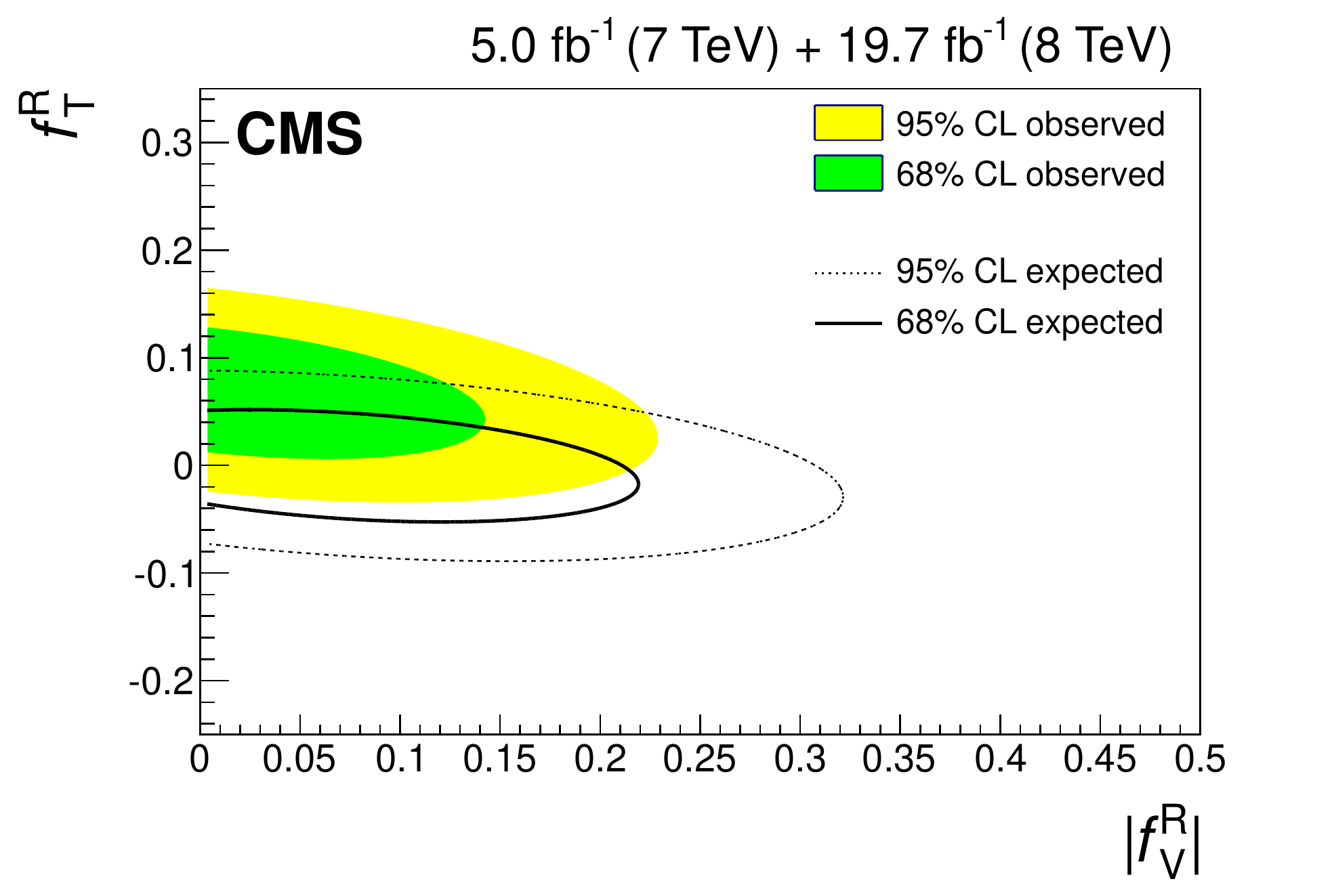}
  \caption{Limits on anomalous \tWb couplings from the CMS analysis combining 7 and 8~TeV, projected onto two dimensions: (left) left- versus right-handed tensorial coupling, and (right) vectorial versus tensorial right-handed coupling (from \textcite{Khachatryan:2016sib}).}
  \label{fig:CMS3ang}
\end{figure}

The CMS analysis that combines 7~and 8~TeV data~\cite{Khachatryan:2016sib} is based on the anomalous couplings model in \textcite{Boos:2016zmp}. The search is for combinations of anomalous couplings similar to the D0 analysis, except that here the limit is set simultaneously on three anomalous couplings: the right-handed vector coupling and the two tensor couplings. A BNN is trained to separate the anomalous signal from the different backgrounds and the SM prediction. The resulting contours projected onto two dimensions are shown in Fig.~\ref{fig:CMS3ang}. The contours are significantly tighter than the two-dimensional limit contours from D0 shown in Fig.~\ref{fig:d0anom}, even though there is is an additional degree of freedom here. Comparing the limits from ATLAS (Fig.~\ref{fig:Atl23ang}) and CMS (Fig.~\ref{fig:CMS3ang}), the graph shows clearly that for the left-handed tensor coupling, the CMS analysis is more sensitive, while for the right-handed tensor coupling, the ATLAS analysis is more sensitive.

\subsection{Searches for Flavor Changing Neutral Currents}
\label{sec:fcnc}

 Models that try to solve the so called ``flavor problem''~\cite{Georgi:1986ku} usually predict a large coupling of new particles to the top quark, and therefore sizable FCNC effects in the top-quark sector, despite the tight constraints in the $B$- and $K$-meson sectors. These are very interesting to look for in single top-quark production, where the effect of a small $u-t$ coupling would be enhanced by the large $u$-quark density~\cite{Tait:2000sh}. 
The same effect would come from a $c-t$ coupling, although with a less spectacular enhancement from the PDF.
Formulations exist where BSM effects in quantum loops are absorbed by effective $tuX$ or $tcX$ couplings, where $X$ can be a gluon, a photon, a $Z$ or $H$ boson (read, for example, \textcite{AguilarSaavedra:2008zc,Zhang:2010dr}). Based on the consideration that higher-order effects mix the effects of different couplings, inducing ambiguities in the interpretation of single signatures, a global approach is advocated in~\textcite{Durieux:2014xla,AguilarSaavedra:2018nen}. However, the results reviewed in this paper make use of leading-order FCNC models.

CDF searched for single top quarks produced by top-gluon FCNC in $W$+1 jet events~\cite{Aaltonen:2008qr}.
The ATLAS collaboration searched for the same exotic signature of a single top quark produced in isolation  (i.e., a $2\to 1$ partonic reaction producing a top quark) with the 7 and 8~TeV data sets~\cite{Aad:2012gd,Aad:2015gea}, to constrain the top-gluon FCNC couplings $tgu$ and $tgc$. The analysis selects events with a single charged lepton, significant \MET and a single jet, passing b-tagging identification. A BNN is applied on the selected events, trained to separate FCNC signals from SM events. 

D0 searched for a single top quark produced together with a light quark, i.e., a $t$-channel signature, created by a top-gluon FCNC~\cite{Abazov:2007ev}. This is also the basis for the CMS top-gluon FCNC search that combines 7~and 8~TeV data~\cite{Khachatryan:2016sib}. Just like for the anomalous couplings search described in the same paper (see Section~\ref{sec:wtb-vertex}), here also a MVA is trained to maximize sensitivity to the $tug$ and $tcg$ interactions.

The CMS collaboration searched for events containing a top quark and a large-\pt photon with the 8~TeV data set~\cite{Khachatryan:2015att}. The semileptonic decay of the top quark is used, and a MVA is performed to discriminate the FCNC signal from the SM backgrounds.
 The dominant $W$+jets and $W+\gamma+$jets backgrounds are estimated from data.
 This statistically-limited analysis makes use of the event counts to set limits on the effective couplings of the \utgamma and \ctgamma types. For the purpose of easy comparison with measurements in \ttbar production, the result is also interpreted in terms of an equivalent branching ratio of top-quark decay into a photon and a quark. CMS also searched for events containing a single top quark and a $Z$ boson decaying to two leptons~\cite{Sirunyan:2017kkr} using the 8~TeV dataset. This analysis not only sets limits on SM $tZ$ production (see Section~\ref{sec:zt}), but also searches for FCNC production of $tZ$. The resulting limit on the $tZq$ coupling is competitive with the sensitivity from top-quark decay searches.

\begin{figure}[!ht]
 \begin{center}
  \includegraphics[width=0.9\textwidth]{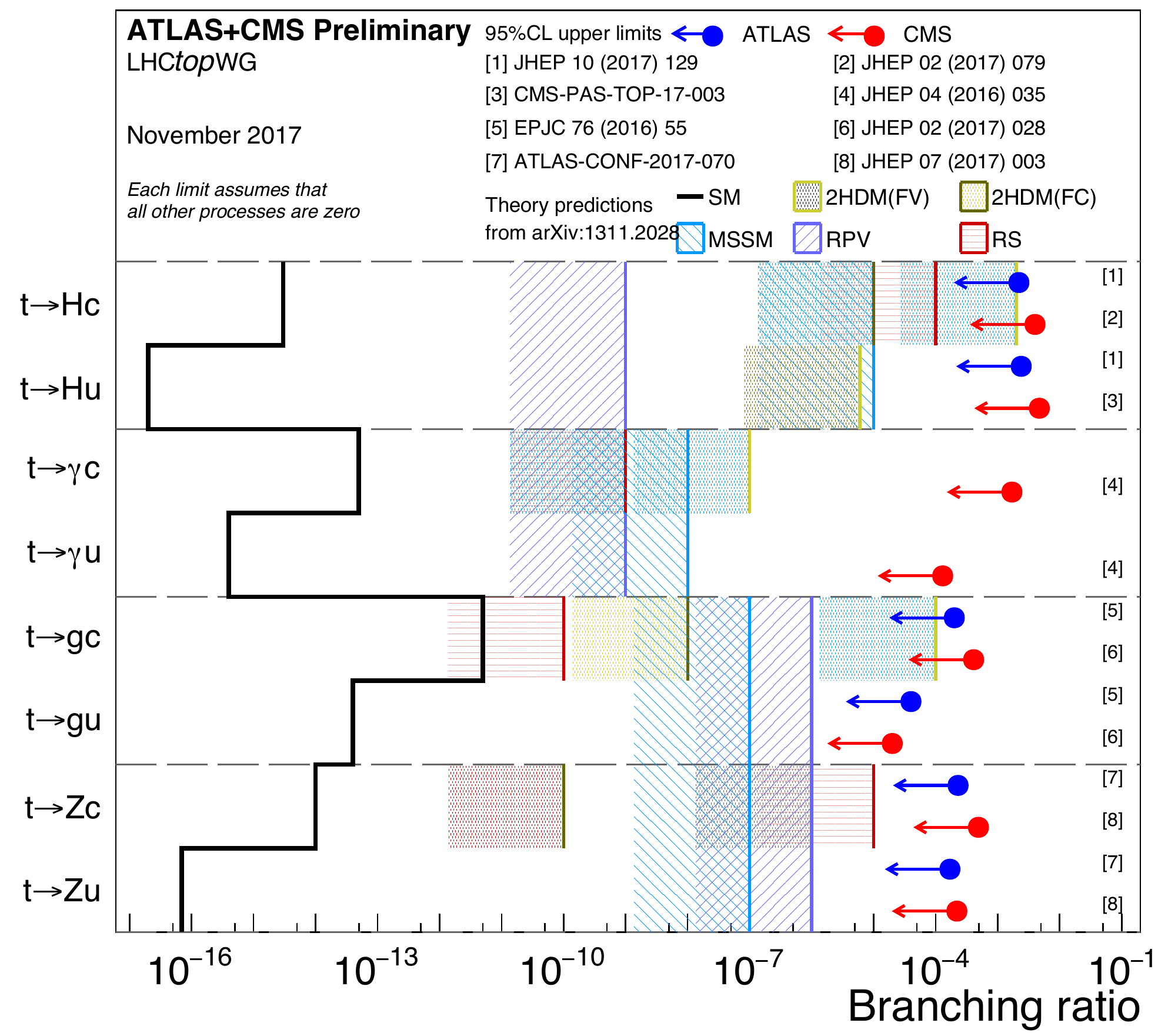}
  \caption{\label{fig:fcnc1}{} Summary of ATLAS and CMS limits on FCNC processes, expressed in equivalent branching ratios and compared with the expectations from the SM and several new physics models. For each FCNC process, the ATLAS limit is shown at the top and the CMS one at the bottom. From \textcite{toplhcwg}, including some preliminary results.}
 \end{center}
\end{figure}

\begin{figure}[!h!tpb]
 \centering
  \includegraphics[width=0.48\textwidth]{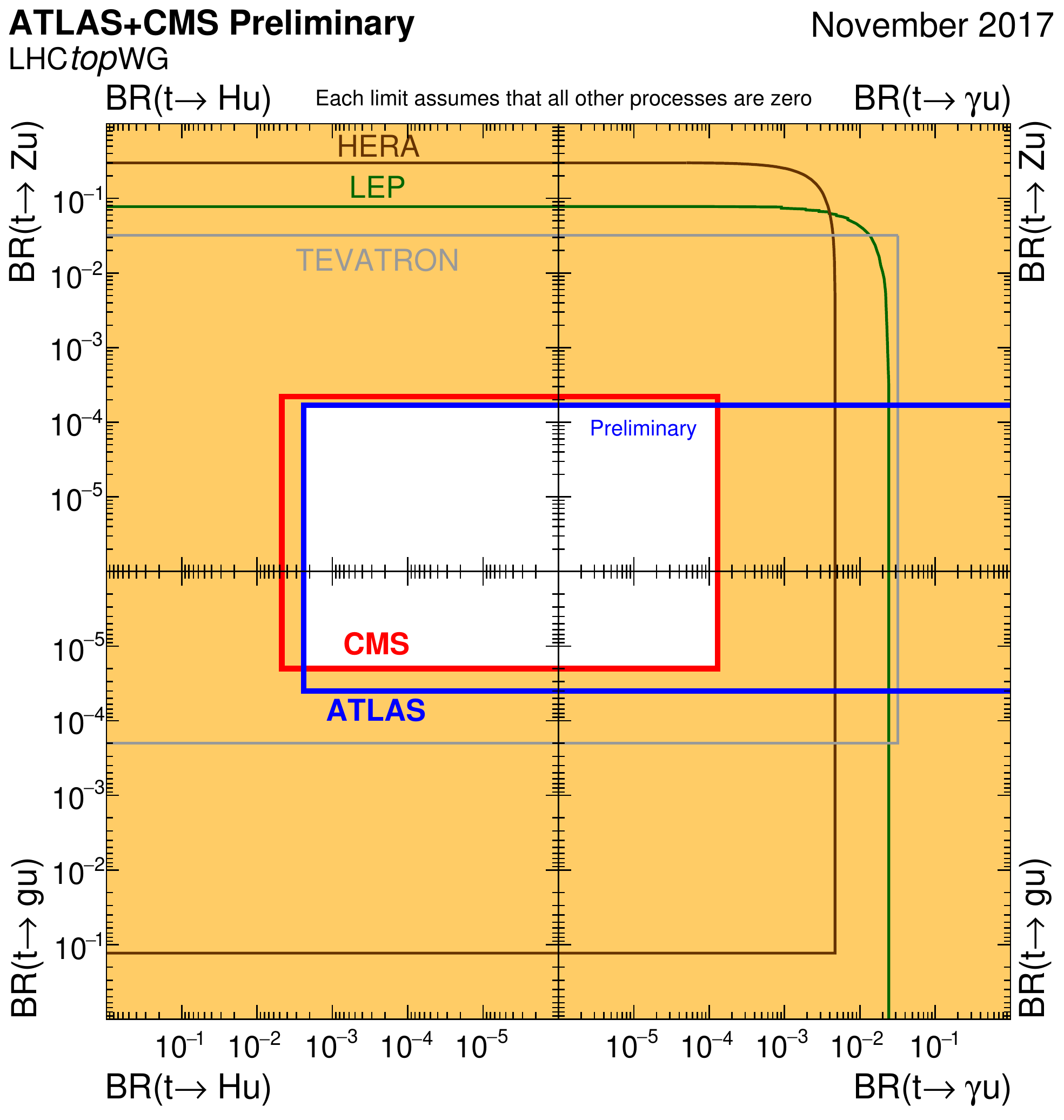}
\hfill
  \includegraphics[width=0.48\textwidth]{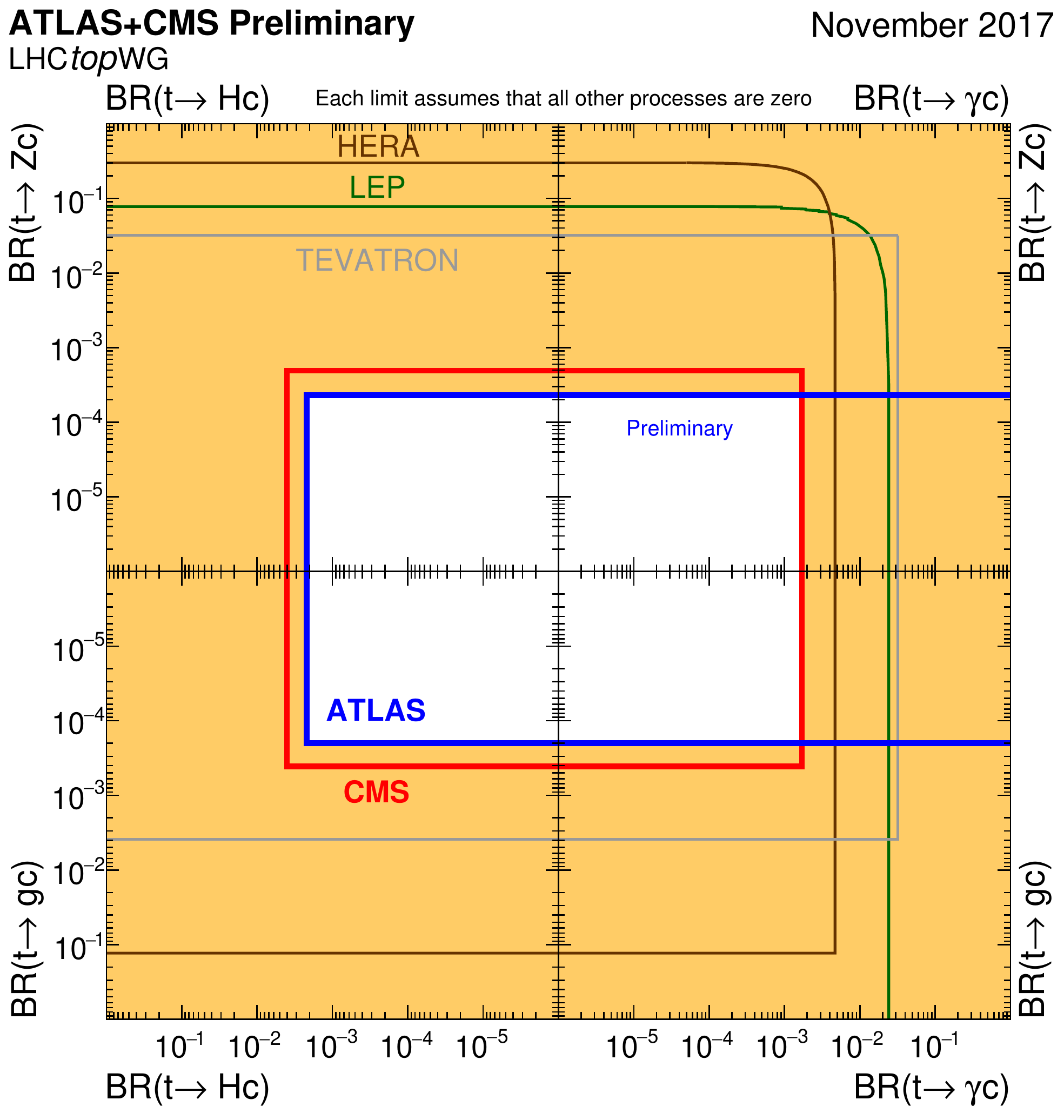}
 \caption{Observed 95\% CL upper limit on the branching ratio of $t\to Zq$ versus the branching of $t\to \gamma q$ ($q = u,c$) as derived directly or indirectly by experiments at LEP, HERA, Tevatron and LHC: search for $e^+e^-\to \gamma^{*}/Z \to t\bar q / \bar t q$ by L3~\cite{Achard:2002vv}, search for $eq\to et$ by ZEUS~\cite{Abramowicz:2011tv} and H1~\cite{Aaron:2009vv}, search for $t\to Zq$ decays in \ttbar events by D0~\cite{Abazov:2011qf}, CDF~\cite{Aaltonen:2008ac}, 
search for $t\to \gamma q$ decays in \ttbar events by CDF~\cite{PhysRevLett.80.2525}. From \textcite{toplhcwg}, including some preliminary results.
}
 \label{fig:tgamma}
\end{figure}

Figure~\ref{fig:fcnc1} summarizes the limits on FCNC interactions from ATLAS and CMS from both top-quark decay searches and single top-quark production searches, expressed in terms of equivalent branching ratios of top-quark decay. Figure~\ref{fig:tgamma} shows a summary that also includes the limits from HERA~\cite{Aaron:2009vv,Abramowicz:2011tv} and LEP~\cite{Achard:2002vv,Abdallah200421,Abbiendi:2001wk,Barate:2000rb}, where the CM energy or the integrated luminosity is not sufficient to produce a measurable number of top-quark events in the SM. At HERA, the FCNC exchange of a photon or $Z$~boson between the electron and the proton leads to a single top quark in the final state. At LEP, the exchange of a photon or $Z$~boson leads to a $tu$ or $tc$ final state. Thus, single top-quark final states are responsible for all HERA and LEP limits in Fig.~\ref{fig:tgamma}, as well as all limits on ${\rm BR}(t\rightarrow gu)$ and ${\rm BR}(t\rightarrow gc)$.

\subsection{$H$-associated single top-quark production (\tH)}
\label{sec:tHq}

\begin{figure}[th!]
        \centering
        \includegraphics[width=0.2\textwidth]{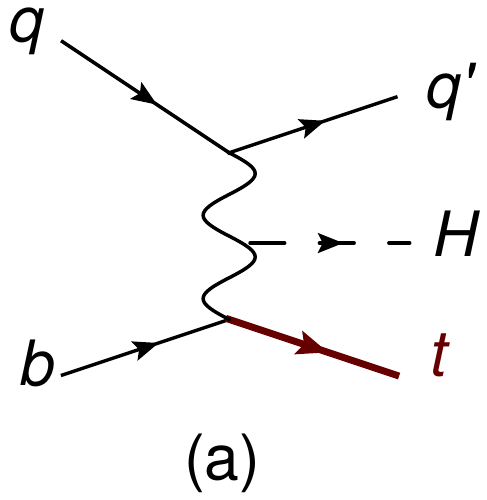}\hspace{1cm}
       \hspace{1cm}
        \includegraphics[width=0.2\textwidth]{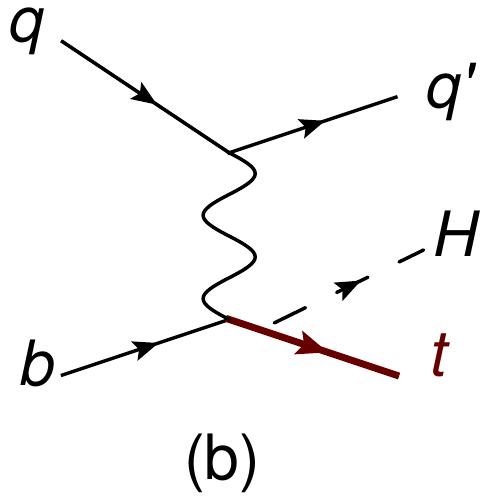}\vspace{.5cm}
        \caption{Dominant Feynman diagrams for the production of \tHq events.}
\label{fig:thq_prod}
\end{figure}

The associated production of a single top quark and a Higgs boson (\tH) provides a complementary experimental view on the interaction of the Higgs boson with the top quark, with respect to the measurement of \ttbar production in association with a Higgs boson (\ttH).
 In particular, while the \ttH process is sensitive to the modulus of \yt,
 \tH production is characterized by a tree-level sensitivity to the relative phase between \yt and the coupling of the Higgs to the gauge bosons~\cite{BORDES1993315}, 
thanks to an accidental numerical similarity of the amplitudes of the diagrams where the Higgs boson is radiated by the $W$~boson and by the top quark (see Fig.~\ref{fig:thq_prod}). In the SM the couplings of the Higgs boson to the $W$ boson and the top quark have opposite sign, leading to destructive interference and very small cross sections, while a significant enhancement is expected if some kind of BSM physics induces a relative phase between these two couplings (more than one order of magnitude in the so called ``inverted top-quark coupling scenario'', or ITC, where $\yt = -1$). 
 In the case of other processes used to set constraints on the \yt phase, like $H\to\gamma\gamma$ and $gg\to HZ$~\cite{Hespel:2015zea}, sensitivity to this phase comes through loop corrections, making their interpretation intrinsically more model-dependent as the particles running in the loop have to be specified. Any analysis of the Higgs-boson couplings that aims at being agnostic about new physics in these loops is unable to use these processes to lift the degeneracy on the sign of \yt~\cite{Ellis:2012rx,Ellis:2013lra}. 

 Single top-quark plus Higgs-boson production proceeds mainly through $t$-channel diagrams (\tHq), as in Fig.~\ref{fig:thq_prod}, and therefore the current searches are optimized for this final state, although the interest of the \tWH signature is similar and it has also been explored in the theoretical literature~\cite{Farina:2012xp,Demartin:2016axk}. The \ttH and \tWH processes feature the same kind of mixing discussed in Section~\ref{sec:wt} in the case of \ttbar and \Wt.
 
 While the SM rate is arguably too low to be observed with available and future LHC data, the large enhancement in the ITC scenario will allow to either observe or exclude this case with the LHC Run 2 data, as has been suggested in a number of phenomenological papers~\cite{Biswas:2012bd,Farina:2012xp,Biswas:2013xva,Chang:2014rfa}.

\begin{figure}[!htbp]
  {\centering
    \includegraphics[width=0.49\textwidth]{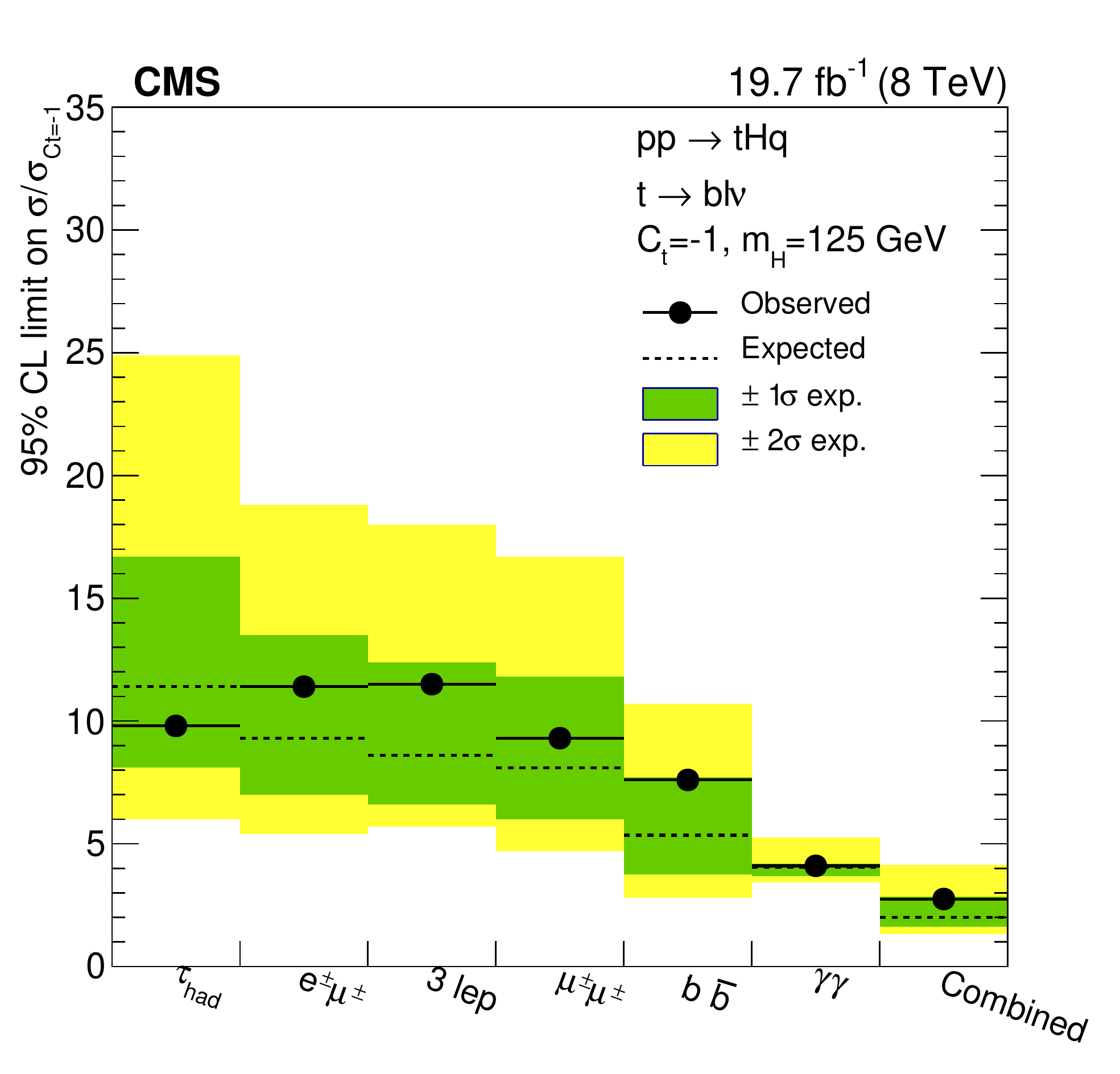}
    \includegraphics[width=0.49\textwidth]{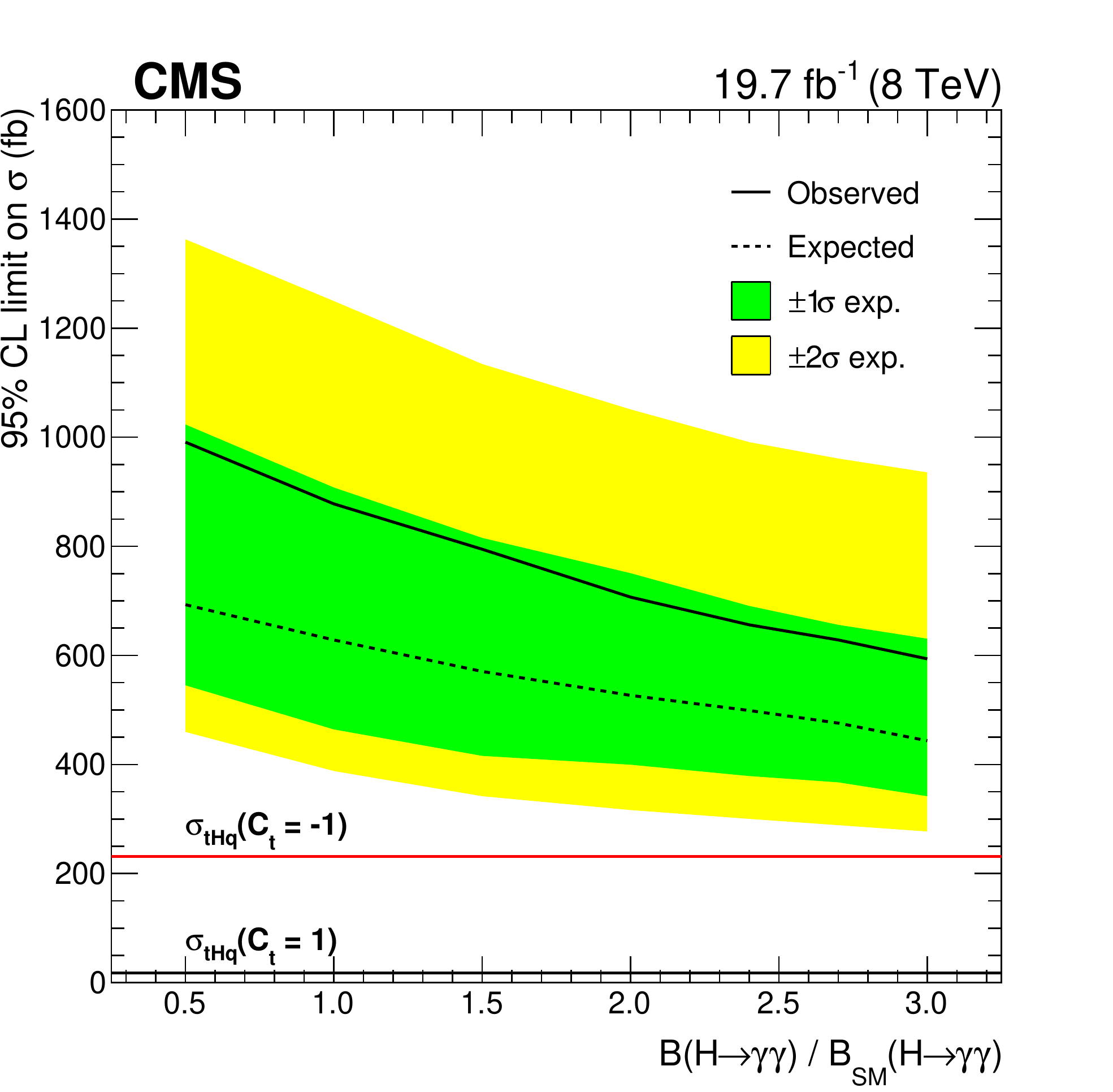}
  \caption{Left: 95\% CL upper limits on the \tHq cross section, divided by its expectation in the $\yt=-1$ scenario, by decay channel and combined. Right: 95\% CL upper limits on the \tHq production cross section versus \BRHgg; the red horizontal line shows the predicted \tHq cross section for the SM Higgs boson with $m_H = 125~{\rm GeV}$ in the $\yt=-1$ scenario, while the black horizontal line shows the predicted \tHq cross section for the SM (i.e., $\yt=+1$) scenario. Figures from~\textcite{Khachatryan:2015ota}.
}
    \label{fig:thq_limits}}
\end{figure}

Using the full 8~TeV data set, the \textcite{Khachatryan:2015ota} performed dedicated searches for \tHq in a variety of signatures: $\gamma\gamma$, $b\bar b$, same-sign leptons, three leptons, and electron or muon plus hadronically-decaying $\tau$. In all Higgs decay channels, the top quark is assumed to decay semileptonically.
The data generally agree with the SM expectations, and limits are set in the individual channels and combined with and without the assumption that the value of \yt affects \BRHgg and $\sigma_{\tHq}$ coherently. When this assumption is made, as shown in Fig.~\ref{fig:thq_limits} (left), the $\gamma\gamma$ channel is the most sensitive as expected from the theory literature~\cite{Biswas:2012bd}.
 The combined limit is also provided with \BRHgg treated as a free parameter, thus facilitating possible reinterpretations in different theoretical frameworks, see Fig.~\ref{fig:thq_limits} (right). 
 The \textcite{Aad:2014lma}, also using the 8~TeV data set, followed a different approach. Instead of a direct search for this process, single top-quark plus Higgs-boson production is included in the signal model in a \ttH-optimised search in the $H\to\gamma\gamma$ decay channel, which allows to set limits on negative values of \yt.

\FloatBarrier

%% file: tex/conclusions.tex
\section{Conclusions and Outlook}
\label{sec:concl}

In the decade that has passed since first experimental evidence for electroweak production of a single top quark was reported, the study of single top-quark production has become a very fertile and mature research direction.
Production rates of processes with a single top quark have been measured in four production modes, at four distinct center-of-mass energies, using five detectors at two accelerators with two different beam particle configurations.
Precision measurements of top-quark properties and searches for new couplings of the top quark utilize single top-quark processes as a powerful probe for new-physics effects.

The groundwork for today's single top-quark studies was laid at the Tevatron, where measurements, searches and analysis techniques that are in use at the LHC today were first established. The single top-quark discovery relied on multivariate approaches, and the first single top-quark samples were used to search for anomalous couplings and new physics.

Thanks to the excellent performance of the LHC during the ongoing Run~2, an integrated luminosity of $\mathcal{O}(100)~\fb$ is expected to be collected at 13~TeV by the end of 2018. 
This large amount of data will have a big impact on several of the analyses described here: 
measurements that so far have been statistics-limited, such as the \tZ cross section and top quark/antiquark cross-section ratios; differential measurements, 
whose power to constrain new physics, SM parameters and MC generator settings will benefit from more bins and more population in the tails of some crucial distributions; and searches for new physics, especially those in clean final states involving neutral bosons.
The interference between \ttbar and \tW will be a point of study in the coming years, both on the theoretical and the experimental side. This effort, and precision measurements in general, rely on improvements in the theoretical modeling of single top-quark processes, not only including off-shell processes but also bringing the theoretical cross-section calculations to NNLO accuracy for single top-quark production channels beyond the $t$-channel. 

At the time of writing, we are still waiting for the first measurement of $s$-channel single top-quark production at 13~TeV. 
The larger amount of available data, by itself, does not make the study of this process easier than it was at 7 and 8~TeV: 
the signal cross section at 13~TeV is only about twice that at 8~TeV~\cite{Kant:2014oha}, while the dominant background, \ttbar, is 
three times larger~\cite{Czakon:2011xx}.
As the Run~1 analyses were already limited by systematic uncertainties, measuring $s$-channel single top at 13~TeV with a useful precision will require significant progress on the theory side, such as to reduce the signal and background modeling uncertainties, and new ideas for an experimental break-through. 
More data can help, for example through a more extended exploitation of auxiliary control regions, to better constrain the modeling of the backgrounds {\it in situ}. 

Single top-quark analyses at Tevatron were among the pioneers for the introduction or broader acceptance of several multivariate analysis techniques in collider physics~\cite{Bhat:2010zz}.  
In spite of a conventional wisdom that, at the time, favored simple cut-and-count methods in the searches for new processes in hadron-hadron collisions, the challenges posed by the search for single top-quark production at Tevatron created a strong incentive for practicing machine-learning methods such as Neural Networks and Boosted Decision Trees, that at the time of writing count among the most popular tools for LHC analysis, and the ME method that had been developed for top-quark physics\cite{Kondo:1988yd,Kondo:1991dw}, although applied until then for different use cases such as top-quark mass measurements.
We are currently witnessing a burst of interest in borrowing even more advanced machine-learning techniques from the larger world outside of High Energy Physics~\cite{HEPML2014}, and it is likely that single top-quark analyses, again, will be among the early adopters.
With regard to the ME method, a recent methodological break-through has been the inclusion of NLO Feynman diagrams in the computation of the dynamical likelihoods~\cite{Martini:2015fsa,Martini:2017fot,Martini:2017ydu}, overcoming the computational challenge by an efficient method to calculate NLO QCD weights for events with jets. This development is expected to reduce the biases in analyses that aim at extracting model parameters, and to improve the sensitivity of the searches for new processes. \textcite{Martini:2017ydu} specifically address the interest of this development in the context of single top-quark studies. 

Apart from pushing the energy and luminosity frontier in its regular proton-proton runs, the LHC continues to advance knowledge by an intense programme of collisions involving heavy ions, complemented by ``reference runs'' of proton-proton collisions at lower energy. 
The \ttbar cross section has already been measured by the CMS collaboration at a CM energy of 5.02~TeV~\cite{Sirunyan:2017ule} using a data set of 26~\pb collected in 2015. With an order-of-magnitude larger data set collected in 2017, the multi-purpose ATLAS and CMS experiments may have the potential to study also single top-quark production at that energy, providing further input to PDF fits.
Recently, top-quark pair production has been observed in proton-lead collisions at $\sqrt{s_{NN}}= 8.16$~TeV~\cite{Sirunyan:2017xku}, and it is expected that single top-quark measurements will also join the physics program with future heavy-ion runs at the LHC~\cite{dEnterria:2015mgr,Baskakov:2015nxa}. The single top-quark production cross section increases by a factor 30 to 40 for heavy ion runs at a possible future circular collider~\cite{dEnterria:2017jyt}, which turns single top-quark events into precise probes. These and \ttbar events will serve as a probe for parton density functions in nuclei at small \xB and large momentum transfer~\cite{Dainese:2016gch}.

At future hadron colliders like the HL-LHC, top-quark measurements will reach high precision~\cite{Agashe:2013hma}, including single top-quark measurements~\cite{Schoenrock:2013jka}. At a possible future 100~TeV hadron collider, single top-quark triggers might be possible, which would allow for unbiased studies of everything produced on the opposite side, including objects at high transverse momenta~\cite{Arkani-Hamed:2015vfh}. 

Top-quark production occurs dominantly through single top-quark events at the future electron-hadron collider~\cite{AbelleiraFernandez:2012cc}, where top-quark pair production (via a neutral current) is suppressed by an order of magnitude. Searches for \tH FCNC interactions are also promising~\cite{Liu:2015kkp}, equivalent to those for $tZ$ and $t\gamma$~\cite{Abramowicz:2011tv,Aaron:2009vv}.

At future lepton colliders, top quarks are produced in pairs through electro-weak interactions. The focus will be on high-precision measurements of the top-quark mass and of the top-quark couplings to the $Z$~boson and the photon~\cite{Agashe:2013hma,Baer:2013cma,Gomez-Ceballos:2013zzn}. Single top-quark production proceeds in an electron-photon collision, with one incoming lepton radiating off a photon and the other incoming lepton radiating off a $W$~boson, resulting dominantly in a final state of a top quark plus a $b$~quark plus a forward lepton~\cite{Boos:2012hi,Penunuri:2011hp}. The cross section for this process is about an order of magnitude smaller than that for $\ttbar$ production. Similar to hadron colliders, single top-quark production at lepton colliders is directly proportional to \vtb and the \vtb precision is limited by the theoretical and experimental understanding of the production process.

\FloatBarrier